\documentclass[a4paper,12pt]{book}
\usepackage[T1]{fontenc}
\usepackage[latin1]{inputenc}
\usepackage{graphics}
\usepackage{eepic}
\usepackage{axodraw}
\usepackage{epsfig}
\usepackage{amssymb}
\usepackage{a4}
\usepackage{rotating}
\usepackage[german,english]{babel}

\makeatletter

\newcommand{\beq}{\begin{equation}}
\newcommand{\eeq}{\end{equation}}
\newcommand{\bqa}{\begin{eqnarray}}
\newcommand{\eqa}{\end{eqnarray}}

\newcommand{\picb}[1]{\;\parbox[c]{48pt}{\begin{picture}(45,30)(-9,0)
\SetWidth{1.0}\SetScale{1.0} #1 \end{picture}}\;}

\def\Lwidth{1}

\def\Agl(#1,#2)(#3,#4,#5){\PhotonArc(#1,#2)(#3,#4,#5){\Lwidth}
{6.283 #3 mul 360 div #4 #5 sub #4 #5 sub mul sqrt mul Ldensity mul}}
\def\Lgl(#1,#2)(#3,#4){\Photon(#1,#2)(#3,#4){\Lwidth}
{#1 #3 sub #1 #3 sub mul #2 #4 sub #2 #4 sub mul add sqrt Ldensity mul}}
\def\Agh(#1,#2)(#3,#4,#5){\DashArrowArc(#1,#2)(#3,#4,#5){1}}
\def\Aagh(#1,#2)(#3,#4,#5){\DashArrowArcn(#1,#2)(#3,#5,#4){1}}
\def\Lgh(#1,#2)(#3,#4){\DashArrowLine(#1,#2)(#3,#4){1}}
\def\Lagh(#1,#2)(#3,#4){\DashArrowLine(#3,#4)(#1,#2){1}}
\def\Ahh(#1,#2)(#3,#4,#5){\DashCArc(#1,#2)(#3,#4,#5){1}}
\def\Lhh(#1,#2)(#3,#4){\DashLine(#1,#2)(#3,#4){1}}
\def\Aqu(#1,#2)(#3,#4,#5){\ArrowArc(#1,#2)(#3,#4,#5)}
\def\Aaqu(#1,#2)(#3,#4,#5){\ArrowArcn(#1,#2)(#3,#5,#4)}
\def\Lqu(#1,#2)(#3,#4){\ArrowLine(#1,#2)(#3,#4)}
\def\Laqu(#1,#2)(#3,#4){\ArrowLine(#3,#4)(#1,#2)}
\def\Aqq(#1,#2)(#3,#4,#5){\CArc(#1,#2)(#3,#4,#5)}
\def\Lqq(#1,#2)(#3,#4){\ArrowLine(#1,#2)(#3,#4)}
\def\Asc(#1,#2)(#3,#4,#5){\ArrowArc(#1,#2)(#3,#4,#5)}
\def\Lsc(#1,#2)(#3,#4){\ArrowLine(#1,#2)(#3,#4)}
\def\DAsc(#1,#2)(#3,#4,#5){\DashCArc(#1,#2)(#3,#4,#5){3}}
\def\DLsc(#1,#2)(#3,#4){\DashLine(#1,#2)(#3,#4){3}}
\def\TAsc(#1,#2)(#3,#4,#5){\SetWidth{2.0}\CArc(#1,#2)(#3,#4,#5)\SetWidth{1.0}}
\def\TLsc(#1,#2)(#3,#4){\SetWidth{2.0}\ArrowLine(#1,#2)(#3,#4)\SetWidth{1.0}}

\begin{document}

\begin{titlepage}
\begin{figure}
\centering
\epsfig{file=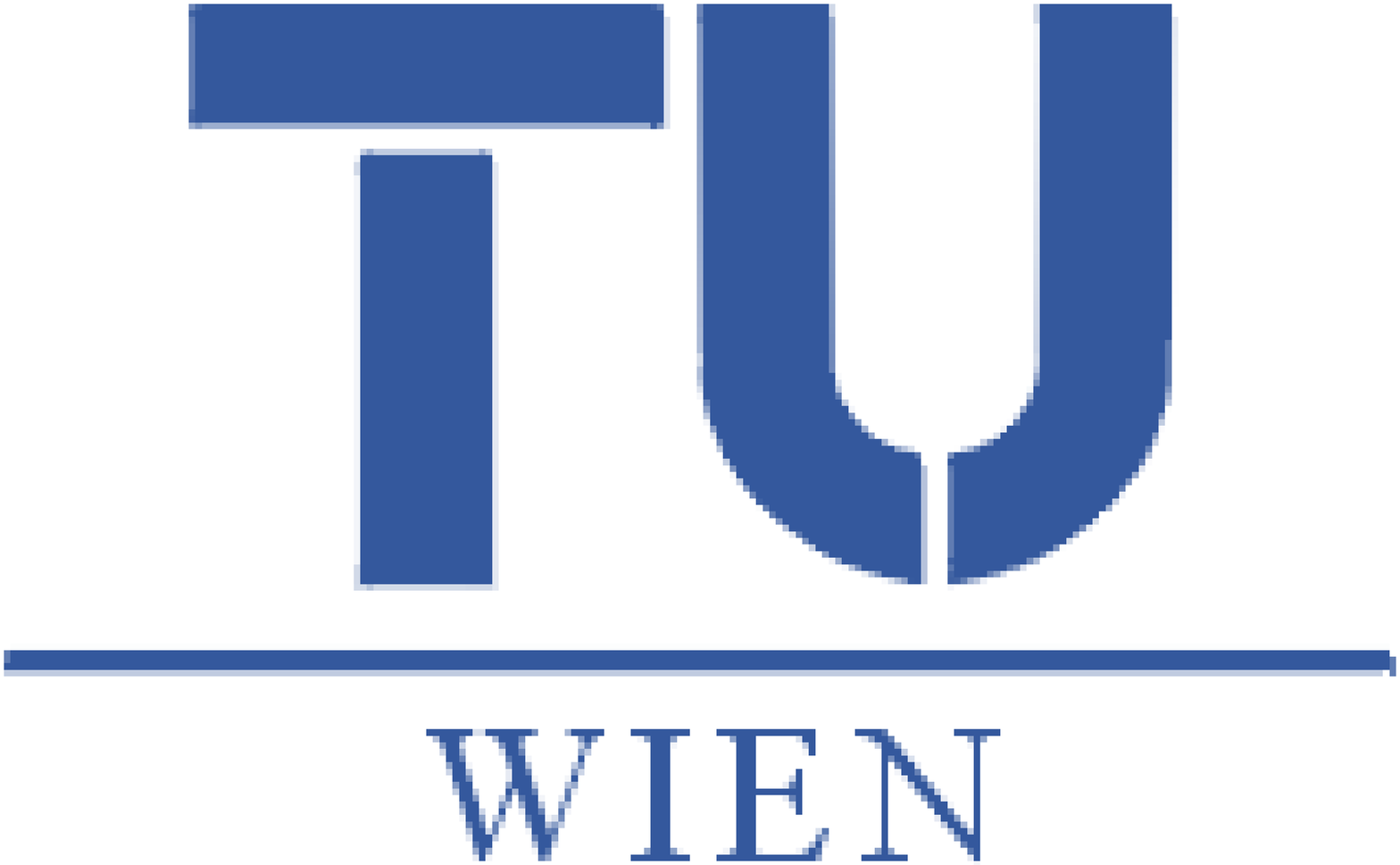,width=5cm}
\end{figure}
\begin{center}
\vfill
\Huge{Dissertation} \\ 
\vfill
\begin{bf}
Quasiparticle description of the hot and dense
quark-gluon plasma\\
\end{bf}
\vfill
\large{ausgef\"uhrt zum Zwecke der Erlangung des akademischen Grades
eines Doktors der technischen Wissenschaften unter der Leitung von\\}
\vfill
\large{Ao. Univ.-Prof. Dr. Anton Rebhan}\\
\large{Institutsnummer: E 136}\\
\large{Institut f\"ur Theoretische Physik} \\
\vfill
\large{eingereicht an der Technischen Universit\"at Wien} \\
\large{Fakult\"at f\"ur Technische Naturwissenschaften und Informatik}\\
\vfill
\large{von}\\
\vfill
\large{Dipl.-Ing. Paul Romatschke}\\
\large{Matrikelnummer: 9625501}\\
\large{Pilgramgasse 13/28, A-1050 Wien}\\
\vfill

\large{\hrulefill \hfill \hfill \hfill \hrulefill}\\
\large{\hfill Datum \hfill \hfill \hfill \hfill \hfill %
\hfill \hfill \hfill \hfill%
\hfill \hfill Unterschrift \hfill}
\end{center}
\end{titlepage}

\quad
\thispagestyle{empty}
\newpage
\newpage

\setcounter{page}{3}
\renewcommand{\thepage}{\roman{page}}
\thispagestyle{empty}
\vspace*{5cm}
\begin{center}
\begin{Large}
For my Parents
\end{Large}
\end{center}
\newpage

\thispagestyle{empty}
\chapter*{Abstract}

\addcontentsline{toc}{chapter}{\protect\numberline{}Abstract}

The collective modes of QCD at temperatures and densities
above its phase-transition are analyzed; models 
for various observables based on the knowledge
gained from this analysis are constructed, and the results derived are
compared to those of other methods, where available.
Specifically, isotropic systems as well as systems having an anisotropy
in momentum-space are investigated using the Hard-Thermal-Loop (HTL) 
approximation. Several observables are calculated, ranging from
thermodynamic quantities and their application in the isotropic case to
the collisional energy loss in the isotropic as well as the anisotropic
case.

For isotropic systems, results at finite densities (or
non-vanishing quark chemical potential) based on a phenomenological description
of lattice calculations at vanishing densities are derived and compared
to those following from different methods. It is shown that results
for the thermodynamic pressure and entropy as well as those for
the quark-number susceptibilities agree well with independent lattice
calculations, and it is demonstrated that the plasmon effect leads to
numerically small contributions, in contrast to what is found in
strictly perturbative approaches. As possible applications of the resulting
equation of state the mass-radius relationship of so-called quark-stars
as well as the hydrodynamic expansion in the Bjorken model
of the quark-gluon plasma created
through a heavy-ion collision are calculated.

It is shown that systems with anisotropic momentum-space distributions
contain unstable modes in addition to the stable quasiparticle modes, which
-- since they correspond to exponentially growing field amplitudes --
may be of great importance for the dynamical evolution of an incompletely
thermalized quark-gluon plasma.
Moreover, the presence of such instabilities and the corresponding
singularities in the propagator lead to divergences of scattering
amplitudes in a perturbative framework, signalling the breakdown of the
latter. However, it is demonstrated that at least one observable, 
namely the collisional energy loss, is protected from these divergencies.
This permits its calculation also for anisotropic systems, as can be
shown analytically both for very weak as well as for extremely
strong anisotropies. A subsequent comparison of results from isotropic
and anisotropic systems exhibits a possibly strong directional dependence
of the energy loss for the latter, which might lead to 
effects that can be verified experimentally.

\chapter*{Acknowledgements}

\addcontentsline{toc}{chapter}{\protect\numberline{}Acknowledgements}

This dissertation in its present form would not have been possible
without the help and support I received from several people, to some
of whom I would like to express my gratitude here.
First of all I want to thank my parents Kurt and Helga who made my
doctoral studies possible in the first place through their continuous
moral and financial support and to whom I have dedicated this dissertation.
Also, I want to warmly thank my supervisor Anton Rebhan and my collaborator
Mike Strickland who's suggestions and help never failed me when 
lengthy equations stubbornly obscured the physics instead of revealing it.
Moreover, I'd like to thank my colleagues and friends
Jean-Paul Blaizot, Andreas Gerhold, Edmond Iancu, 
Andreas Ipp, Urko Reinosa and Dominik Schwarz for countless interesting 
discussions and help on various problems.
Last but not least, 
I'd like to thank my girlfriend Ulli for her love and understanding
during the past months when I was either physically abroad at a conference or 
mentally absent when figuring out an integral.
Finally I want to apologize to all the friends and colleagues not named here
explicitly and assure them of my warmest gratitude, as well as to
all those people I have failed to call back in the last months and who still 
miraculously seem to be talking to me.

\quad\\
\begin{flushright}\textsc{Paul Romatschke}, Vienna 2003 \end{flushright}

\tableofcontents
\newpage
\quad\\ \newpage
\thispagestyle{empty}
\vspace*{5cm}
\begin{center}
\begin{large}
There is a theory which states that if ever anyone\\
discovers exactly what the Universe is for and why it is here,\\
it will instantly disappear and be replaced by\\
something even more bizarre and inexplicable.\\
\end{large}
\end{center}
\newpage
\thispagestyle{empty}
\vspace*{5cm}
\begin{center}
\begin{large}
There is another theory which states\\
that this has already happened.\\
\end{large}
\quad\\
\textsc{Douglas Adams}, {\em The Hitch Hiker's Guide to the
Galaxy II}
\end{center}

\chapter{Introduction}

\setcounter{page}{1}
\renewcommand{\thepage}{\arabic{page}}

\section{Quantum chromodynamics}

Today, quantum chromodynamics (QCD) is accepted to be the 
established theory of strong interactions. It has been formulated
along the lines of the theory of quantum electrodynamics (QED),
which -- as the unification of quantum theory and electrodynamics --
is certainly one 
of the most successful and accurate theories in modern physics.
Consequently, QCD bears several similarities but
also exhibits some striking differences with respect to QED, 
as will be briefly illustrated in the following.
Both theories are gauge field theories, but whereas QED is an Abelian
gauge theory with the gauge group U(1), quantum chromodynamics is 
non-Abelian in nature, having the color group SU(3) as a gauge
group. The gauge boson associated with the U(1) group of QED is the
well-known {\itshape photon}, while for the color group SU(3) there are 8 
associated gauge bosons called \textit{gluons}. 
In contrast to the photon
which is uncharged, the gluons do carry color charges which are 
the QCD equivalents of the electromagnetic charge. For QED,
the matter particles are the {\itshape electron, muon} and {\itshape tauon} 
while for QCD there are the six
quark species {\itshape up, down, strange, charm, beauty} and {\itshape top}, 
which are all spin 1/2 fields or fermions; those quark species
(or flavors) with masses much smaller than the energy
scale under consideration are referred to as {\em active} or light flavors and 
their number is usually denoted by $N_f$. 

Both QED and QCD are renormalizable field theories and as a consequence
their coupling ``constants'' are functions of the energy scale $Q$.
In Abelian gauge theories (as QED) the coupling
increases for larger energy scales, whereas for low energies it turns
out to be small, allowing a valid (and indeed extremely accurate) 
description of our (low energy) everyday world using perturbation theory.

However, unlike in Abelian gauge theories, 
the non-Abelian nature of QCD
makes its coupling constant $\alpha_s(Q)$ {\em decrease} 
as $Q$ becomes large, a property that is known as asymptotic freedom.
Consequently, at very high energies one may hope that 
perturbation theory offers a valid description
of QCD since the coupling constant gets tiny.
On the other hand, when the energy scale gets smaller and the coupling constant
rises, calculating observables in QCD becomes very hard in general,
since one has to deal with a strongly coupled theory and 
perturbation theory breaks down.
Therefore, one obviously needs 
non-perturbative methods to describe the theory and explain 
the experimental fact that at very low energies quarks and gluons cannot
roam freely but are {\em confined} into hadrons 
from which they do not escape.

While at low temperatures quarks and gluons are thus locked up mainly
inside {\itshape protons} and {\itshape neutrons},
one expects them to propagate freely in a state that has been dubbed the
{\em quark-gluon plasma} at very high temperatures. The exact nature of
this transition between the confined phase and the quark-gluon plasma phase
(whether it is a first or second order phase transition or a crossover)
is still a matter of active research.

Several theoretical methods have been proposed to study the physics of QCD,
each most apt for a certain energy range, the most popular being:
\begin{itemize}
\item{Effective models based on the QCD Lagrangian that allow to calculate
the low-energy behavior of the theory. Perhaps the most interesting of
these is chiral perturbation theory, based on the effective chiral Lagrangian
(see \textsc{Gasser}'s and \textsc{Leutwyler}'s 
original articles on this subject 
\cite{Gasser:1984yg,Gasser:1985gg} or \cite{Scherer:2002tk} for a 
pedagogical review).}
\item{Monte-Carlo simulations of QCD on a lattice, which so far is the 
only method to provide 
quantitative results for the intermediate energy range near the deconfinement
transition. Since simulations with dynamical quarks are much more time
consuming than those without, the situation best studied in lattice QCD
is that of a pure gauge theory. Nevertheless, there has also been considerable
progress for systems with quarks and recently even for non-vanishing 
chemical potential. The latter in general poses a major problem for
lattice calculations because of the {\em sign problem}, which prohibits
the use of conventional numerical algorithms (see \textsc{Karsch}
\cite{Karsch:2001cy} for a pedagogical review).}
\item{Perturbative approaches, which are expected to be most accurate for very
high energies. For instance, there exist several 
methods to calculate the equation
of state for the quark-gluon plasma phase; a short review on this subject
will be given in section 1.4.}
\end{itemize}
On the experimental side, the main earth-based tool 
to investigate QCD near or above 
the phase transition is ultrarelativistic particle collisions. The study of
these makes huge collider facilities necessary, such as those of CERN
({\em Centr\'e Europ\'eenne pour la Recherche Nucl\'eaire}) in Switzerland/France
and BNL ({\em Brookhaven National Laboratory}) in the U.S.A.
While the SPS ({\em Super Proton Synchrotron})
collider at CERN produced
the first experimental indication of the existence of the quark-gluon plasma
phase, it has currently been dismantled to make way for the next generation
collider LHC ({\em Large Hadron Collider}), which will allow to probe energies
well above the phase transition and is due to become operational in 2007.
Since 1999, RHIC ({\em Relativistic Heavy Ion Collider}) is the collider
operational at BNL, which has already produced a wealth of new data bringing
the study of nucleus-nucleus collisions carried out at SPS to a new
energy regime. 

Finally, indirect studies of QCD may be possible through astrophysical 
observations and cosmology \cite{Grigorian:2002ih}.

\section{The deconfinement transition}

Our current understanding of QCD is such that whenever the system 
under consideration becomes
sufficiently hot and/or dense, hadrons dissolve into a gas of
almost free quarks and gluons, the quark-gluon plasma. More precisely,
either the temperature $T$, or the quark chemical potential $\mu$, or
a combination of both, has to be big enough so that the system undergoes
a transition to the deconfined state of matter. In a slight abuse of
terminology, one refers to the temperature and chemical potential, for
which the system undergoes the transition, as critical temperature $T_c$ and 
critical 
chemical potential $\mu_c$, regardless of the exact nature of the transition
(first/second order phase transition or crossover). 

At vanishing chemical potential, lattice studies of QCD tell us that the
critical temperature is on the order of a few hundred MeV, which is 
much smaller than the masses of the charm, beauty and 
top quark. For this
reason, those heavy quarks do not play a role in the description of
the physics near the deconfinement transition, so it is useful to limit
the number of active (=light) 
quark flavors to $N_f=3$. Moreover, since the up and
down quark masses are so small compared to the energy scale of the 
deconfinement transition, these quarks may be taken to be massless. The 
strange quark,
finally, is neither much heavier nor much lighter than the deconfinement
scale, so the number of active quark flavors near the deconfinement 
transition is between two and three, also sometimes denoted as
$N_f=2+1$.

\begin{figure}
\begin{center}
\includegraphics[width=0.6\linewidth]{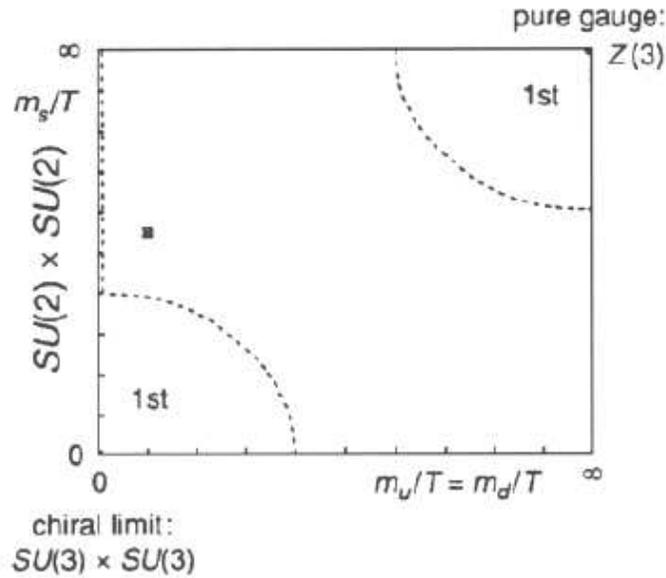}
\caption{Character of the transition as a function of the light quark masses
$m_u=m_d$ and the strange quark mass $m_s$. The dot represents
the physical values of the parameters. From \cite{LeBellac:1996}.}
\label{fig:qcd-phtr}
\end{center}
\end{figure}

Based on universality arguments, it is possible to show that for a pure
gauge theory ($N_f=0$, corresponding to infinitely heavy up, down and
strange quarks) as well as for three and more active flavors
($N_f \ge 3$, corresponding to zero up, down and strange
 quark masses)
the deconfinement transition should be a first-order phase transition.
For $N_f=2$, the transition is probably a second order phase transition,
but the situation is not entirely clear \cite{LeBellac:1996}. 
Fig.~\ref{fig:qcd-phtr} shows a cartoon of our current knowledge 
about the nature
of the deconfinement transition at vanishing chemical potential, indicating
that for light up and down quarks and a moderately heavy
strange quark
the transition is probably a cross-over \cite{Fodor:2001pe}, though
a weak first-order transition cannot be ruled out either (if the physical
point in Fig~\ref{fig:qcd-phtr} moves down).

The value of the critical temperature has been studied 
intensively in lattice QCD 
calculations at $\mu=0$, giving \cite{AliKhan:2000iz,Karsch:2001cy}
\begin{itemize}
\item{$T_c = 271\pm 2$ MeV for $N_f=0$}
\item{$T_c = 171\pm 4$ MeV and $T_c = 173\pm 8$ MeV for $N_f=2$, 
depending on whether one uses Wilson or staggered fermion lattice
implementations}
\item{$T_c = 154\pm 8$ MeV for $N_f=3$.}
\end{itemize}

\section{The quark-gluon plasma}

The quark-gluon plasma phase is unlike any of the other phases 
we know of in nature:
it is a new state of matter, on equal footing with solid, fluid, gaseous
and the electromagnetic plasma, which makes its study interesting from
a conceptual point of view already. Furthermore, from standard cosmology 
we expect that the quark-gluon plasma was the prevailing form of matter
until about $10^{-5}$ seconds after the Big Bang, which could have 
interesting implications on the universe. Nowadays, it is expected to
be produced only in ultrarelativistic heavy-ion collisions, and perhaps
reside in the core of heavy neutron stars, which might have interesting
astrophysical consequences.
For these reasons, one would like to know more about the properties of
this state of matter, especially since ultrarelativistic heavy-ion 
collision measurements in the next decade may allow to verify these 
predictions experimentally.

One of these properties is the equation of state (EOS) of the quark-gluon
plasma; once the latter is created through the collision of 
two ultrarelativistic nuclei, the expansion of the ``fireball'' is 
controlled mainly by the EOS. Also, the EOS specifies the mass-radius
relation of so-called quark-stars through the Tolman-Oppenheimer-Volkoff
equations.

The EOS itself follows from the pressure $p(T,\mu)$ (which is 
related to the thermodynamic potential $\Omega=-pV$) 
through the standard thermodynamic equations
\beq
s=\frac{dp}{dT}, \quad n=\frac{dp}{d\mu}, \quad \epsilon=-p+sT+n\mu,
\eeq
where $s$, $n$ and $\epsilon$ are the entropy, number and energy density,
respectively; from these, the energy density is given as a function of
the pressure only, $\epsilon=\epsilon(p)$, which is one representation of
the EOS.
As a consequence, knowledge of the pressure for arbitrary temperature
and chemical potential automatically implies the EOS.

\section{The QCD pressure}

Since the QCD coupling becomes small because of asymptotic freedom, 
one could expect that weak coupling calculations
at high temperature $T$ or chemical potential $\mu$ 
lead to reasonable estimates of the thermodynamic pressure of QCD. To
zeroth order in the coupling (where the quark-gluon plasma would 
correspond to an ideal gas of quarks and gluons), the 
pressure takes the form
\beq
p_{0}=\frac{(N^2-1) \pi^2 T^4}{45}+N N_f%
\left(\frac{7 \pi^2 T^4}{180}+\frac{\mu^2 T^2}{6}+\frac{\mu^4}{12 \pi^2}%
\right),
\label{psb}
\eeq
where $N_f$ is the number of massless quark-flavors and $N=3$ is the number
of colors \cite{Kapusta:1989}.
The first correction to this result (which itself corresponds to black-body
radiation in electrodynamics) is of first order in $\alpha_s$ and given by
\beq
p_{2}=-N (N^2-1) \frac{\alpha_s \pi T^4}{36} - (N^2-1) N_f%
\frac{\alpha_s (N^2-1)\pi}{144}
\left(5 T^4+18 \frac{mu^2 T^2}{\pi^2}+\frac{9 \mu^4}{\pi^2}\right).
\label{pLO}
\eeq

\subsection{Vanishing chemical potential}

To simplify the argument, consider the chemical potential to be zero 
for the moment, so that
the pressure is proportional to $T^4$ and all the 
non-trivial temperature behavior of $p/T^4$ resides
in the temperature dependence of the strong coupling $\alpha_s$.
The standard procedure is to take the solution for the running coupling
from the renormalization group equation, which to lowest (1-loop) order
is given by
\beq
\alpha_s(\bar{\mu})=\frac{12 \pi}{(11N-2 N_f)\ln{\bar{\mu}^2/%
\Lambda^2_{\overline{\hbox{\footnotesize MS}}}}},
\label{alpha1loop}
\eeq
and put the renormalization point $\bar{\mu}$ (not to be confounded
with the chemical potential $\mu$) proportional 
to the first Matsubara frequency,
$\bar{\mu}=c_{\bar{\mu}} 2 \pi T$, which for massless quarks is the 
only dimensionfull quantity inherent to the theory. 
The parameter $c_{\bar{\mu}}$ expresses our ignorance about the ideal
renormalization point and is usually varied by a factor of two to 
obtain an estimate of how strongly the result depends on this
chosen renormalization point; obviously, one would prefer the result
to have as little dependence as possible on 
the arbitrary parameter $c_{\bar{\mu}}$. The strong coupling $\alpha_s$
also depends on the scale parameter of the modified minimal subtraction
scheme, $\Lambda_{\overline{\hbox{\footnotesize MS}}}$, which cannot be
fixed by theoretical considerations but has to come from experiments.
However, $\alpha_s$ is not easily measured at energy scales near
the deconfinement transition, so one has to extrapolate measurements
down from higher energies. Currently, the official extrapolation value
at $\bar{\mu}=2$ GeV is given by \cite{Hagiwara:2002fs}
\beq
\alpha_s(\bar{\mu}=2 {\rm GeV})=0.2994;
\eeq
since from the choice of $\bar{\mu}\simeq 2 \pi T$ this corresponds to
a temperature where one expects up, down and strange
 quarks to be effectively massless ($N_f=3$), this would fix 
$\Lambda^{(3)}_{\overline{\hbox{\footnotesize MS}}}\simeq194$MeV. However, by
using a higher order correction to the running coupling (2-loop),
\beq
\alpha_s(\bar{\mu})=\frac{12 \pi}{(11N-2 N_f) {\bar L}(\bar{\mu})}\left(%
1-\frac{(34 N^2-13 N N_f+3N_f/N) \ln{\bar{L}(\bar{\mu})}}{6 (11N-2N_f)^2 %
\bar{L}(\bar{\mu})}\right)
\label{alpha2loop}
\eeq
where $\bar{L}(\bar{\mu})=\ln{\bar{\mu}^2/%
\Lambda^2_{\overline{\hbox{\footnotesize MS}}}}$, the value of 
$\Lambda_{\overline{\hbox{\footnotesize MS}}}$ changes to 
$\Lambda^{(3)}_{\overline{\hbox{\footnotesize MS}}}\simeq378$MeV and adopting a
3-loop running coupling, one finds 
$\Lambda^{(3)}_{\overline{\hbox{\footnotesize MS}}}\simeq344$ MeV.
One could still improve on this result by taking into account 
matching conditions between the different quark mass scales 
\cite{Rodrigo:1993hc} as 
well higher order (e.g. 4-loop \cite{vanRitbergen:1997va})
beta-functions for the running coupling; however, being 
interested in a rough estimate only, one can 
naively use the critical temperature $T_c\sim 154$MeV for $N_f=3$
to obtain the ratio
$T_c/\Lambda^{(3)}_{\overline{\hbox{\footnotesize MS}}}\sim 0.45$,
which turns out to be reasonably close to the value for two active flavors 
advocated by \textsc{Gupta}
\cite{Gupta:2000hr},
\beq
\frac{T_c}{\Lambda^{(2)}_{\overline{\hbox{\footnotesize MS}}}}=0.49.
\eeq

Therefore, I will adopt this ratio for all the two-flavor calculations 
in the following. Clearly, if a different value is adopted,
some quantitative results in this work will change somewhat, but
the qualitative picture remains the same and the modifications 
to the results should
be straightforward (I will also give the trend of the modified result
in chapter \ref{qpmodels2-chap}).

\begin{figure}
\begin{center}
\includegraphics[width=0.6\linewidth]{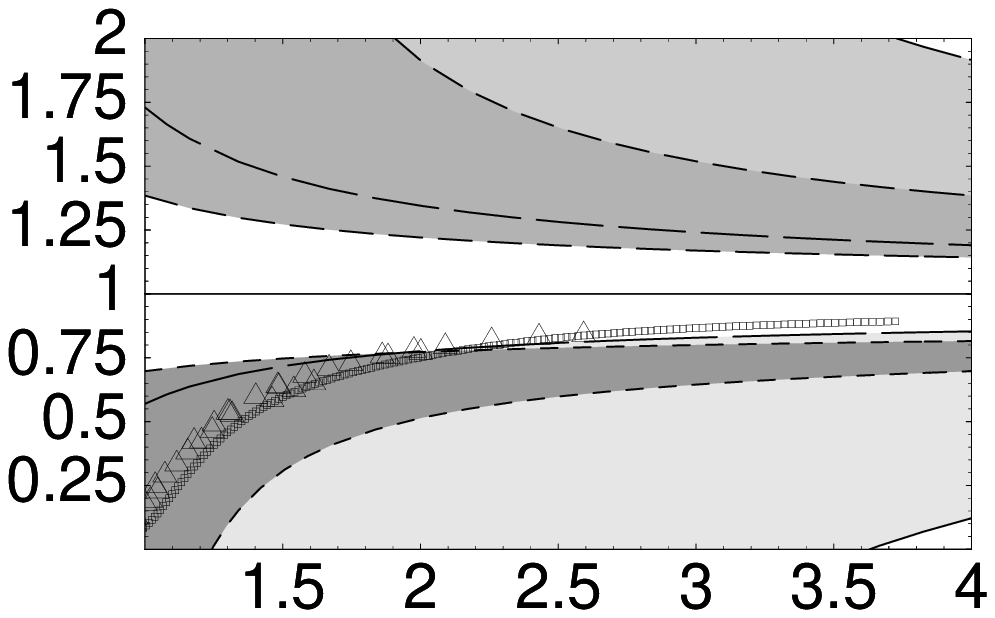}
\setlength{\unitlength}{1cm}
\begin{picture}(6,0)
\put(4,0.1){\makebox(0,0){\begin{large} $T/T_{c}$ \end{large}}}
\put(-1.5,3.5){\makebox(0,0){\begin{large}  $p/p_0$ \end{large}}}
\put(4.5,2){\makebox(0,0){$\alpha_s^{5/2}$}}
\put(4.5,5){\makebox(0,0){$\alpha_s^{2}$}}
\put(2,4.7){\makebox(0,0){$\alpha_s^{3/2}$}}
\put(3,2.7){\makebox(0,0){$\alpha_s$}}
\end{picture}
\caption{Strictly perturbative results for the thermal pressure of 
$N_f=2$ QCD, normalized to the ideal-gas value $p_0$, as a function of
$T/T_c$ (assuming $T_c/\Lambda_{\overline{\hbox{\footnotesize MS}}}=0.49$).
The various gray bands bounded by differently dashed lines show the 
perturbative results to order $\alpha_s$, $\alpha_s^{3/2}$, $\alpha_s^{2}$
and $\alpha_s^{5/2}$, with the $\overline{\hbox{\small MS}}$ 
renormalization point $\bar{\mu}$ varied between $\pi T$ and $4 \pi T$ 
\cite{Blaizot:2003iq}.
The boxes and triangles show the estimated continuum-extrapolated lattice
results from \cite{AliKhan:2001ek,Karsch:2000ps}.}
\label{fig:Spert}
\end{center}
\end{figure}

Having determined the temperature dependence of the coupling one can 
proceed to
evaluate the perturbative pressure; indeed, 
explicit calculations of higher order contributions to the pressure
have been pushed to $\alpha^{5/2}_{s}$ \cite{
Arnold:1995eb,Zhai:1995ac} and even to 
$\alpha_s^{3}\log{\alpha_s}$ in a recent heroic effort \cite{Kajantie:2002wa}.
The result, however, is rather depressing: instead of showing any sign
of convergence, adding successive orders gives a total pressure that 
is sometimes below, sometimes above the free pressure $p_0$, jumping 
in an nearly unpredictable fashion (see Fig.~\ref{fig:Spert}). 
Perhaps worse, the ambiguity introduced
by the constant $c_{\bar{\mu}}$ seems to increase for every order added
to the pressure, signalling a complete loss of predictive power.

Various mathematical extrapolation techniques have been tried, such as 
Pad\'e approximants and Borel resummation 
\cite{Kastening:1997rg,Hatsuda:1997wf,%
Cvetic:2002ju,Parwani:2000rr,Parwani:2000am}.
The resulting expressions are 
indeed better behaved than polynomial approximations
truncated at order $\alpha_s^{5/2}$ or lower, showing a weaker dependence
on the renormalization scale. However, while these methods do improve the
situation somewhat, it is fair to say that they offer little physical
insight on the source of the difficulty.

However, in recent years resummations based on
the hard-thermal-loop (HTL) effective action 
\cite{Braaten:1992gm
} have been proposed,
alternatively in the form of so-called HTL perturbation theory
\cite{Andersen:1999fw,
Andersen:2002ey
}
or based upon the 2-loop $\Phi$-derivable approximation 
\cite{Blaizot:1999ip
,Blaizot:2000fc,Blaizot:2003tw}
(see also \textsc{Peshier} \cite{Peshier:2000hx})%
\footnote{For recent progress
with higher-loop $\Phi$-derivable approximations in scalar field
theory see Refs.~\cite{Braaten:2001en,vanHees:2001ik,
Blaizot:2003br}.}.
The latter approach, which assumes weakly interacting
quasiparticles as determined by the HTL propagators
and NLO corrections thereof, leads to results which
agree remarkably well with lattice data\footnote{%
This is in fact not the case for 2-loop HTL perturbation theory
\cite{Andersen:2002ey
}. However a dimensionally reduced variant \cite{Blaizot:2003iq} of this form
of improved perturbation theory (which is perturbatively
equivalent and moreover much simpler) has recently been shown
to agree well with the results of Refs.~\cite{Blaizot:1999ip,%
Blaizot:2000fc,Blaizot:2003tw}
and correspondingly also with lattice data.}; more details
about this approach will be given later on.
Recently, it has also been shown that the results from the HTL quasiparticle
models are in fact
consistent with those obtained in high-order perturbation theory if the
latter is organized through an effective (dimensionally reduced)
field theory according to \cite{Braaten:1996jr} and effective-field-theory
parameters are kept without further expanding in powers of the
coupling \cite{Blaizot:2003iq}, as already 
advocated in Ref.~\cite{Kajantie:2002wa}.

\subsection{Finite chemical potential}

In contrast to vanishing chemical potential, where lattice studies seem
to be quite capable of calculating the EOS for a quark-gluon plasma near
the deconfinement transition, this task is by far more difficult
at finite (and especially large) chemical potential. 
The reason for this is that the quark determinant becomes complex at
finite chemical potential $\mu$, 
so Monte-Carlo simulations cease to be directly 
applicable \cite{Allton:2002zi}; this is known as the sign problem. 
Recently however, there has been important progress also for lattice 
calculations\footnote{A perturbative expansion of the pressure
for small chemical potential has also been calculated recently
\cite{Vuorinen:2003fs}.} at non-vanishing $\mu$ 
\cite{Fodor:2001au,deForcrand:2002ci,Fodor:2002km,Allton:2002zi,D'Elia:2002gd,
Gavai:2003mf,Allton:2003vx},
which circumvent the problem of a complex quark determinant by either
a Taylor expansion in $\mu$ 
of the observables at vanishing chemical potential or an analytic continuation
of observables calculated at purely imaginary $\mu$. 
Unfortunately, all these calculations are limited to the case of
small chemical potential ($\mu/T\ll 1$) for now, so an EOS for 
cold dense matter, 
which is of importance in astrophysical 
situations \cite{Peshier:1999ww,Peshier:2002ww,Fraga:2001xc,%
Andersen:2002jz})
is currently beyond the reach of lattice calculations.
Moreover, a recent result for a model of QCD with an infinite number of 
flavors ($N_f\rightarrow \infty$) indicates that extrapolation of results
obtained at $\mu=0$ breaks down rather abruptly for $\mu/T \gtrsim \pi$,
pointing presumably towards a generic obstruction for extrapolating
data from small to large chemical potential \cite{Ipp:2003zr}.

As a remedy for this situation, \textsc{Peshier} 
\emph{et al.} \cite{Peshier:1999ww,Peshier:2002ww} proposed a method which
can be used to map the available lattice data for $\mu=0$ to finite $\mu$
and small temperatures by describing the interacting plasma as a system
of massive quasiparticles. Since this procedure does not involve
any direct extrapolation, the model proposed by \textsc{Pehsier} 
and its hard-thermal-loop (HTL)
extension which will be considered later 
on are generally capable of calculating
the EOS for the quark-gluon plasma for any (i.e. also large) value
of $\mu/T$, given that the models stay reliable for all temperatures
and densities.
Accordingly, such quasiparticle models\footnote{%
See also the model in Ref.~\cite{Schneider:2001nf} 
which has been extended to finite chemical 
potential in Ref.~\cite{Thaler:2003uz}.} are 
a valuable complementary method to e.g. lattice calculations 
at small chemical potential, and they turn out to give quite robust 
quantitative predictions on the EOS for cold dense QCD matter as well as 
on the critical chemical potential.

I will give a detailed discussion of the results for the EOS obtained by
these quasiparticle models and also compare with results from other 
approaches \cite{Baluni:1978mk,Freedman:1977ub,Fraga:2001xc,%
Andersen:2002jz}.

\section{Anisotropic systems}

So far, all the discussion has been limited to homogeneous and isotropic
systems. However, considering that the system created through a collision
of two heavy ions propagating essentially at the speed of light is certainly
not homogeneous and isotropic, one has to
ask whether a description through {\em equilibrium} field theory actually
constitutes a valid treatment of this system. This question can be restated
in the form of whether the system {\em thermalizes} fast enough so that
at least at a local scale it can be described by equilibrium field theory.
The advent of ``bottom-up'' thermalization by 
\textsc{Baier, M\"uller, Schiff} and \textsc{Son} \cite{Baier:2000sb}
was a big step towards answering this question,
although there are indications that this scheme
probably will have to be modified because of 
more recent developments discussed below.

Despite continuous progress \cite{Berges:2002wr,Berges:2000ur,%
Baacke:2002ee,Cooper:2002qd}, 
an out-of equilibrium study of QCD has not been achieved yet;
however, one may try to learn more about out-of-equilibrium situations 
by studying systems that are close to equilibrium. For instance, one can
consider the effects that an anisotropic momentum-space distribution
function has on the thermalization, or more precisely isotropization of
a quark-gluon plasma. As has been pointed out first by 
\textsc{Mr\'owczy\'nski}\
\cite{Mrowczynski:1993qm,Mrowczynski:1994xv,Mrowczynski:1997vh}, 
such a system possesses instabilities that 
correspond to so-called Weibel or filamentation instabilities in 
electrodynamics \cite{Weibel:1959}.  
\textsc{Weibel} showed that
unstable transverse modes exist in electrodynamic plasmas with anisotropic 
momentum distributions and derived their growth rate in linear response theory.
In plasma physics, it has been shown
that these instabilities generate strong
magnetic fields resulting in the filamentation of the electron current 
by simulations and recently also experimentally \cite{Wei:2002}.
The effects of these instabilities are much less clear, but they may 
potentially be very important for
the quark-gluon plasma evolution at RHIC or LHC due to the large 
amount of momentum-space anisotropy in the 
gluon distribution functions at $\tau \sim 1\;{\rm fm/c}$.  

\textsc{Mr\'owczy\'nski} and \textsc{Randrup} recently performed 
phenomenological estimates of the growth
rate of the instabilities for two types of anisotropic distribution 
functions \cite{Randrup:2003cw}, finding 
that the degree of amplification of the Weibel instability 
is not expected to dominate the
dynamics of a QGP, but instead is comparable to the contribution from 
elastic Boltzmann collisions.  
However, if a large number of the unstable 
modes could be excited 
then it is possible that their combined effect on the overall dynamics 
could be significant.

Moreover, \textsc{Arnold, Lenaghan} and \textsc{Moore} 
\cite{Arnold:2003rq} argued that the presence of instabilities giving
rise to large gauge fields necessarily change the first stage of
the bottom-up thermalization scenario qualitatively, making a revision
of the original scheme unavoidable.

In this work, I will perform a detailed study of the HTL quasiparticles 
and unstable modes 
in an anisotropic quark-gluon plasma and discuss their effects
on the properties of such a system with respect to an isotropic quark-gluon
plasma.

\section{Outline of this work}

In chapter \ref{peshier-chap}, which is still of introductory character,
 I will give a definition of what I will refer to 
as quasiparticles and also introduce a simple quasiparticle model for
the QCD pressure of the isotropic quark-gluon plasma. In chapter 
\ref{sewm-chap},
I will introduce resummed perturbation theory based on the 2PI effective
action and an extension of the simple quasiparticle model that follows
from it. In chapter \ref{qpmodels-chap}, 
I will discuss the results one obtains 
from mapping lattice data from zero to arbitrary chemical potential potential,
including two possible applications.
In chapter \ref{qpmodels2-chap}, 
I will investigate to what extent these results are modified
when one limits the input to only the value of 
$T_c/\Lambda_{\overline{\hbox{\footnotesize MS}}}$.
In chapter \ref{asi-chap}, I will calculate the HTL resummed gluon
self-energy of an anisotropic quark-gluon plasma, identify all stable
and unstable collective modes of the system, and consider their effect
on the partonic energy loss with respect to an isotropic system in
chapter \ref{eloss-chap}.
Finally, I will give my conclusions in chapter \ref{conc-chap}.

\chapter{QCD Quasiparticles}
\label{peshier-chap}

When elementary particles propagate through a medium, their vacuum properties
are changed through the effect of their interactions, which is also 
referred to as ``dressing''. In the simplest case, a quasiparticle is then
just an elementary particle which had e.g. its vacuum mass changed to an
effective mass by the medium.

However, in general a medium is characterized by a whole 
set of collective modes or
quasiparticles (I will use the terms interchangeably),
 some of which correspond to elementary particles
in the vacuum and others which do not. Loosely speaking, a quasiparticle
is an in-medium excitation that behaves as if it was an elementary particle,
although it might not correspond to one in the vacuum.

Generally, quasiparticles possess a dispersion law $E(k)$ that gives
their energy $E$ as a function of their momentum $k$ and --
since their lifetime may not be infinite -- a decay or damping rate 
$\gamma(k)$. Both the dispersion relation as well as the decay rate are
linked to the peak and width of the spectral densities; in the simple
cases considered in the following the spectral densities can 
be assumed to have a Breit-Wigner form so that the dispersion relations
and decay rate are linked to a simple pole of the propagator.


\section{The gluon propagator in the HTL approximation}

The gluon propagator $G^{\mu \nu}$ is related to the gluon self-energy
$\Pi^{\mu \nu}$ through the Dyson-Schwinger equation
\beq
G^{-1}_{\mu \nu}=G^{-1}_{\mu \nu,0}+\Pi_{\mu \nu}.
\label{Dysonglue}
\eeq
Using a tensor decomposition of the self-energy and restricting to
the HTL approximation \cite{Rebhan:2001wt,Blaizot:2000fc} and the 
temporal axial gauge\footnote{%
This particular gauge is chosen for convenience only; a proof of 
the gauge independence of the dispersion relations below can be found
in \cite{Kobes:1991dc}.} one finds
\bqa
&\Pi_{ij}(\omega,k)=\left(\delta_{ij}-\frac{k_i k_j}{k^2}\right) \Pi_T(%
\omega,k)-\frac{k_i k_j}{k^4} \Pi_L(\omega,k),&\nonumber\\
&\Pi_{00}(\omega,k)=-\Pi_L(\omega,k), \quad \quad \Pi_{0i}(\omega k)=-%
\frac{\omega k_i}{k^2} \Pi_L(\omega,k),
\eqa
where the longitudinal and transversal parts of the self-energy are given
by
\beq
\Pi_L(\omega,k)=m_D^2\left[1-\frac{\omega}{2 k}\ln{\frac{\omega+k}{%
\omega-k}}\right],\quad\Pi_T(\omega,k)=\frac{1}{2}\left[m_D^2+%
\frac{\omega^2-k^2}{k^2}\Pi_L\right],
\label{HTLPiL}
\eeq
and 
\beq
m_D^2=4 \pi \alpha_s\left(\frac{2 N + N_f}{6} T^2+\frac{N_f \mu^2}{2\pi^2}\right)
\label{mD}
\eeq
is the QCD Debye mass squared.
In the temporal axial gauge the gluon propagator then reads
\beq
G_{ij}(\omega,k)=\left(\delta_{ij}-\frac{k_i k_j}{k^2}\right) G_T(\omega,k)
+\frac{k_i k_j}{k^2} \frac{k^2}{\omega^2} G_L (\omega,k),
\eeq
where
\beq
G_L(\omega,k)=\frac{-1}{k^2+\Pi_L(\omega,k)}, \quad %
G_T(\omega,k)=\frac{-1}{\omega^2-k^2-\Pi_T(\omega,k)}.
\eeq

\begin{figure}
\begin{center}
\includegraphics[width=0.6\linewidth]{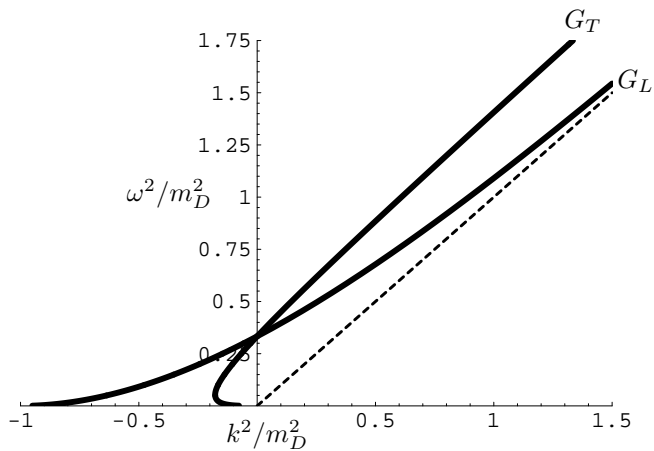}
\setlength{\unitlength}{1cm}
\begin{picture}(6,0)
\put(2.3,0.5){\makebox(0,0){\footnotesize $k^2/m_D^2$}}
\put(1,3.7){\makebox(0,0){\footnotesize $\omega^2/m_D^2$}}
\put(7.2,5.2){\makebox(0,0){\footnotesize $G_L$}}
\put(6.5,6){\makebox(0,0){\footnotesize $G_T$}}
\end{picture}
\caption{Dispersion relations for 
the transversal ($G_T$) and longitudinal ($G_L$) part
of the gluon propagator in quadratic scales. Positive values of $k^2$
correspond to propagating modes (mass hyperboloids show as straight lines
parallel to the light cone indicated by the dashed line) whereas negative
values show screening phenomena.}
\label{fig:gluonmodes1}
\end{center}
\end{figure}

The poles $\omega$ of the gluon propagator then fall into two branches, 
corresponding to a longitudinal 
($G_L^{-1}(\omega_L(k),k)=0$) and a transversal 
\hbox{($G_T^{-1}(\omega_T(k),k)=0$)}
mode. The dispersion relations $E_L(k), E_T(k)$ 
and associated damping rates \linebreak $\gamma_L(k), \gamma_T(k)$ are simply
given by the real and imaginary part of the poles, respectively,
\beq
\omega_L(k)=E_L(k)-i \gamma_L(k), \quad \omega_T(k)=E_T(k)-%
i\gamma_T(k),
\eeq
or more specific for the above modes,
\bqa
&0=k^2+{\rm Re} \Pi_T(E_L(k),k), \quad%
E_T^2(k)=k^2+{\rm Re} \Pi_T(E_T(k),k),&\nonumber\; \;\; \; \\
&0={\rm Im} \Pi_L(E_L(k)+i\gamma_L(k),k),%
\quad \gamma_T(k)=-\frac{1}{2 E_T(k)} {\rm Im} \Pi_T(E_T(k)+i\gamma_T(k),k).
& \; \;\; \;
\eqa
In Fig.~\ref{fig:gluonmodes1}, a plot of the dispersion relations
in quadratic scale is shown: there are propagating modes above a
common plasma frequency $\omega>\omega_{pl}=\sqrt{m_D^2/3}$, 
which for
the transverse branch tend to a mass hyperboloid with asymptotic mass
squared
\beq
\label{minfty}
m_{\infty}^2=\frac{m_D^2}{2},
\eeq
whereas the longitudinal branch approaches the light-cone exponentially
with a residue that is exponentially suppressed. Indeed, this last mode
(which is sometimes called the longitudinal plasmon)
does not have a vacuum analogue and disappears from the spectrum 
at $k\rightarrow \infty$. The transverse mode, however, represents an
in-medium version of the physical gauge boson polarization.

For $\omega<\omega_{pl}$, $k$ turns out to be imaginary which corresponds
to the phenomenon of screening; in QCD, there is magnetic (transversal
mode) as well as electric (longitudinal mode) screening with a screening
length given by $1/|k|$ as long as $\omega>0$.

In the static limit $\omega=0$, the self-energies become 
$\Pi_L(0,k)=m_D^2$ and $\Pi_T(0,k)=0$, respectively. Accordingly,
whereas the longitudinal branch predicts screening in the electrostatic
sector with inverse screening length $|k|=m_D$, the solution
to the transverse branch, $k=0$, signals the absence of magnetostatic 
screening. In fact, in Abelian gauge theories there are strong arguments 
that the absence of a magnetic screening mass is required by gauge invariance
of the theory \cite{Blaizot:1995kg}. However, in non-Abelian gauge theories
this absence is not guaranteed and indeed lattice QCD simulations
do find a screening behavior in the transverse sector 
\cite{Cucchieri:2001tw}. Note that also analytically one can find a nontrivial
magnetostatic behavior if one no longer requires the system to be isotropic;
the result for the magnetic screening mass, however, is somewhat
unexpected and this case will be described in more detail in chapter 
\ref{asi-chap}.

To leading order, the imaginary part of the gluon self-energy
vanishes for $\omega^2>k^2$. Accordingly, the damping rates of the
quasiparticles are zero to this order of the coupling. However, 
there is a contribution at order $\alpha_s$, which has been calculated 
by \textsc{Braaten} and \textsc{Pisarski} \cite{Braaten:1990it} 
in the zero momentum limit after several
incomplete attempts by various authors in the 1980's 
(see \cite{Rebhan:2001wt} for a collection
of published values). For nonzero momentum the damping rates have an
apparent logarithmic singularity that is cut off by the magnetostatic mass
\cite{Pisarski:1993rf}, which has to be calculated non-perturbatively;
only the zero momentum result seems to be protected from this effect,
although there may be different problems arising at next-to-leading order 
\cite{Flechsig:1995sk}.

Finally, in the region with $\omega^2<k^2$ the imaginary part of the 
gluon self-energy is large ($\sim \alpha_s T^2$); this corresponds to the
possibility of Landau damping, which is the dissipation of energy in
the plasma \cite{LeBellac:1996} (see also 
chapter \ref{eloss-chap} where the energy loss of
a heavy parton is calculated).

\section{The fermion propagator in the HTL approximation}

\begin{figure}
\begin{center}
\includegraphics[width=0.6\linewidth]{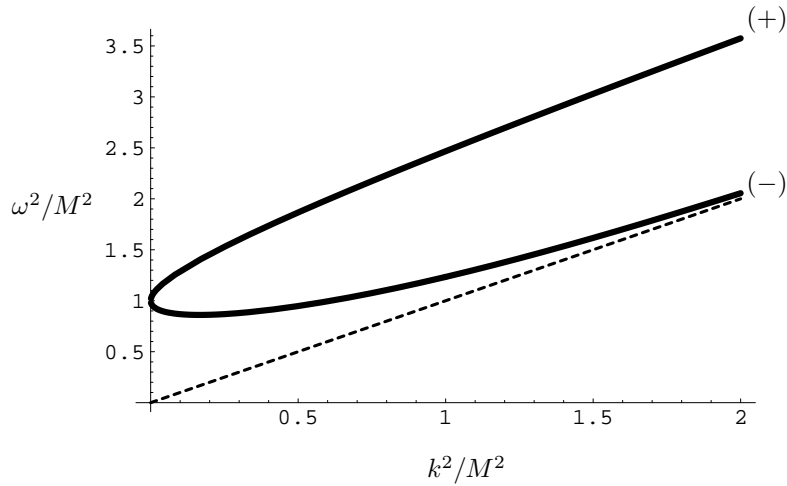}
\setlength{\unitlength}{1cm}
\begin{picture}(6,0)
\put(3.5,0){\makebox(0,0){\footnotesize $k^2/M^2$}}
\put(-2,3.5){\makebox(0,0){\footnotesize $\omega^2/M^2$}}
\put(7.5,6){\makebox(0,0){\footnotesize $(+)$}}
\put(7.5,3.8){\makebox(0,0){\footnotesize $(-)$}}
\end{picture}
\caption{Dispersion relations for the fermionic quasiparticles
in quadratic scale. Shown are 
the regular (+) and plasmino (-) branch, together with the light cone
(dashed-line).}
\label{fig:fermionmodes1}
\end{center}
\end{figure}

The fermion propagator $S$ (suppressing Lorentz indices) 
fulfills a Dyson-Schwinger equation similar to the gluon propagator
\beq
S^{-1}=S^{-1}_0+\Sigma,
\label{Dysonferm}
\eeq
where $\Sigma$ is the fermionic self-energy. The most general form of
$\Sigma$ compatible with the rotational and chiral symmetries is
\beq
\Sigma(\omega,k)=a(\omega,k)\Gamma^{0}+b(\omega,k){\bf \hat{k}}\cdot%
{\bf \Gamma},
\eeq
where $\Gamma$ are the Dirac matrices. This can be rewritten as
\beq
\Gamma_0 \Sigma(\omega,k)=\Sigma_+(\omega,k)\Lambda_{+}({\bf \hat{k}})-%
\Sigma_-(\omega,k)\Lambda_{-}({\bf \hat{k}}),
\eeq
where $\Sigma_{\pm}(\omega,k)=b(\omega,k)\pm a(\omega,k)$ and the spin
matrices
\bqa
&\Lambda_{\pm}({\bf \hat{k}})\equiv\frac{1\pm \Gamma^{0} {\bf \Gamma}\cdot%
{\bf \hat{k}}}{2},\quad  \Lambda_+ +\Lambda_-=1, \quad %
\Lambda_{\pm}^2=\Lambda_{\pm}, &\\
&\Lambda_+ \Lambda_-=\Lambda_- \Lambda_+=0,\quad {\rm Tr}\Lambda_{\pm}=2,&
\eqa
project onto spinors whose chirality is equal ($\Lambda_+$), or
opposite ($\Lambda_-$), to their helicity. Using $S_0^{-1}=-/\!\!\!k+\Sigma$
one can decompose the propagator 
\beq
S(\omega,k)\Gamma_0=\Delta_+(\omega,k)\Lambda_+ + \Delta_-%
(\omega,k) \Lambda_-,
\eeq
with $\Delta_{\pm}^{-1} \equiv-[\omega \mp(k+\Sigma_{\pm})]$. Accordingly,
there are also two fermionic pole branches corresponding to
$\Delta_+^{-1}=0$ and $\Delta_-^{-1}=0$, where the latter one is occasionally
referred to as plasmino branch. Using the result for the fermion self-energy
in the HTL approximation,
\beq
\Sigma_{\pm}(\omega,k)=\frac{M^2}{k}\left(1-\frac{\omega\pm k}{2k}%
\ln{\frac{\omega+k}{\omega-k}}\right),
\label{HTLSigmapm}
\eeq
where $M^2$ is the plasma frequency for fermions,
\beq
M^2=\frac{4 \pi \alpha_s (N^2-1)}{16 N}\left(T^2+\frac{\mu^2}{\pi^2}\right), 
\eeq
one can proceed to plot the dispersion relations $E_+(\omega), E_-(\omega)$
for the fermionic modes, shown in Fig.~\ref{fig:fermionmodes1}.

The plasmino branch shows an interesting minimum of $\omega$ at 
$\omega/M\simeq 0.93$, $k/M\simeq 0.41$ and approaches 
the light-cone for large momenta with exponentially vanishing residue;
the regular branch tends to a mass hyperboloid with asymptotic mass squared
\beq
\label{Minfty}
M^2_{\infty}=2 M^2.
\eeq

The damping rates also vanish to leading order but there is a contribution
to order $\alpha_s$, as was the case for the gluonic modes \cite{Pisarski:1993rf}.

\section{Simple quasiparticle model for the EOS}

The main building block one uses in the derivation of the HTL approximation
is that the principal
contribution for the loop integral in e.g. the self-energies
comes from momenta which are of order $k\sim T$, dubbed ``hard'' (hence 
the name hard-thermal-loop). Therefore, for weak coupling $\alpha_s$ 
when momenta of order $k\sim \alpha_s^{1/2} T$ 
(dubbed ``soft'') are much smaller than
hard momenta, the thermodynamic behavior of the system is dominated by 
the hard excitations. However, the relevant collective modes at large momenta
have already been identified in the previous sections: while the longitudinal
plasmon as well as the plasmino mode are exponentially suppressed, the 
transverse gluonic and positive fermionic branches represent particle 
excitations that propagate predominantly on simple mass shells with 
dispersion relations 
\beq
E_T(k)^2=m_{\infty}^2+k^2,\quad E_{+}(k)^2=M_{\infty}^2+k^2,
\eeq
with the asymptotic masses $m_{\infty}$ and $M_{\infty}$ given above.
It is therefore straightforward to model the thermodynamic pressure as 
a gas of weakly interacting massive particles with residual 
mean-field interaction $B$ \cite{Gorenstein:1995vm,Peshier:1999ww},
\beq
p(T,\mu)=p_{g}(T,m_{\infty}^2)+p_{q}(T,\mu,M_{\infty}^2)-B(m_{\infty}^2,M_%
{\infty}^2),
\label{simP}
\eeq
where
\bqa
p_{g}(T,m_{\infty}^2)&=&- 2 (N^2-1) \int\frac{d^3 k}{(2\pi)^3} %
\ln{\left(1-e^{-E_T(k)/T}\right)}, \\
p_{q}(T,\mu,M_{\infty}^2)&=& 2 N N_f \int \frac{d^3 k}{(2\pi)^3} %
\ln{\left(1+e^{-(E_+(k)-\mu)/T}\right) + (\mu\rightarrow -\mu)}
\eqa
represent the partial pressure from gluons (g) and quarks (q). Using
the stationarity of the thermodynamic potential under variation of the
self-energies one finds
\beq
\frac{\partial B}{\partial m_{\infty}^2}=\frac{\partial p_{g}(T, m_{\infty}^2)%
}{\partial  m_{\infty}^2}, \quad %
\frac{\partial B}{\partial M_{\infty}^2}=\frac{\partial p_{q}(T, \mu, %
M_{\infty}^2)}{\partial  M_{\infty}^2},
\label{Bdet1}
\eeq
implying 
\beq
s=\frac{dp}{dT}=\frac{\partial [p_g(T,m_{\infty}^2)+p_q%
(T,\mu,M_{\infty}^2)]}{\partial T}, \quad%
n=\frac{dp}{d\mu}=\frac{\partial p_q%
(T,\mu,M_{\infty}^2)}{\partial \mu}
\eeq
for the entropy and number density, $s$ and $n$, respectively.

If one expands Eq.(\ref{simP}) in powers of the coupling $\alpha_s$, the 
leading order perturbative result $p_0+p_2$ given in Eqs.(\ref{psb},\ref{pLO}),
and part of the higher order correction is reproduced. Unexpanded,
however, Eq.(\ref{simP}) represents a thermodynamically consistent
resummation of terms in all orders of $\alpha_s$, suggesting also a possible
applicability of the model in the strong coupling regime. Indeed, by 
introducing an ansatz for the strong coupling (inspired by the running
coupling Eq.(\ref{alpha1loop})) that contains two phenomenological 
parameters $\lambda$ and $T_s$ at $\mu=0$,
\beq
\alpha_{s,{\rm eff}}%
(T,\mu=0)=\frac{12 \pi}{(11N-2 N_f)\ln{\left[\lambda(T+T_s)/%
T_c\right]^2}},
\label{alphaeff}
\eeq
it turns out that one can accurately describe lattice data for the entropy 
and eventually the pressure (see chapter \ref{sewm-chap} for details).

Moreover, for non-vanishing chemical potential, one obtains a flow
equation for the coupling: since the quasiparticle masses depend on $T$
and $\mu$ explicitly as well as implicitly through the coupling 
$\alpha_s(T,\mu)$, Maxwell's relation $ds/d\mu=dn/dT$ implies a partial
differential equation for $\alpha_s$ which takes the form
\beq
a_T\frac{\partial \alpha_s}{\partial T}+a_{\mu}\frac{\partial \alpha_s}{%
\partial \mu}=b
\label{floweq},
\eeq
where the coefficients $a_T,a_{\mu}$ and $b$ depend on $T,\mu$ and 
$\alpha_s$. Given a valid boundary condition, a solution for 
$\alpha_s(T,\mu)$ can be found by solving the above flow equation
by the method of characteristics; once $\alpha_s$ is thus known
in the $T,\mu$ plane, the quasiparticle partial pressures $p_g$ and 
$p_q$ are also fixed completely. Finally, the residual interaction
$B$ is then given by the integral
\beq
B=\int \sum_{i=g,q}%
 \frac{\partial p_{i}}{\partial m_{i}^2}\left(\frac{\partial %
m_{i}^2}{\partial \mu} d\mu + \frac{\partial m_{i}^2}{\partial T} dT\right)%
+B_{0},
\label{Bdet}
\eeq
where $B_0$ is an integration constant that has to be fixed either through
lattice data or through matching the pressure 
on to a low-energy effective theory. In chapter \ref{sewm-chap}
the details of this calculation and also an interpretation
of the results will be given.

\subsection{Properties of the flow equation}

When calculating the coefficients $a_T, a_{\mu}$ and $b$
of Eq.(\ref{floweq}), one finds that in particular $a_T$ and $a_{\mu}$
obey \cite{Peshier:2002ww}
\beq
a_T(T,\mu\rightarrow 0)=0,\quad a_{\mu}(T\rightarrow 0,\mu)=0,
\eeq
while the respective other coefficient does not vanish in these limits.
The characteristic equations following from Eq.(\ref{floweq}),
\beq
a_{\mu}dT=a_T d{\mu}, \quad a_{\mu} d\alpha_s=b d\mu,
\eeq
then imply that the characteristics are perpendicular to both the $T$ and
the $\mu$ axes. Therefore, specifying the coupling $\alpha_s$ on some
interval on the $T$ or $\mu$ axes represents a valid boundary condition
for the flow equation (\ref{floweq}).

In the limit of vanishing coupling $\alpha_s\rightarrow 0$ it can be shown
that the coefficient $b$ vanishes. The coupling is then constant along 
the characteristics, which become ellipses in the variables $T^2$ and
$\mu^2$ \cite{Peshier:1999ww}, given by 
\beq
\frac{4 N+5 N_f}{9 N_f} T^4+2 T^2 \frac{\mu^2}{\pi^2}+\left(\frac{\mu}{\pi}%
\right)^4={\rm const.}
\eeq
Note that for nonzero coupling this is still true approximately.

However, the solution to
the flow equation is not forcibly unique everywhere, since the characteristics
may turn out to be intersecting for certain regions in the $T,\mu$-plane.
This ambiguity as well as its resolution will be treated in more detail
in chapter \ref{sewm-chap}.

\section{Summary}

In this chapter I have defined and characterized 
the leading-order 
gluonic and fermionic quasiparticle excitations relevant for QCD. From the 
respective self-energies the dispersion relations were calculated
and it was 
found that for hard momenta two of these excitations vanish exponentially
while the remaining two propagate approximately on simple mass shells.
Following \textsc{Gorenstein} and 
\textsc{Peshier} \cite{Gorenstein:1995vm,Peshier:1999ww}, 
I argued that the thermodynamic pressure of QCD can be modeled by that
of a free gas of massive particles together with 
some residual interaction, which itself
is determined by requiring stationarity of the pressure with 
respect to the self-energies.
Anticipating that by using a simple 2-parameter ansatz for an 
effective strong coupling this model can be used to accurately 
describe lattice data
for the entropy and pressure (which will be done in chapter \ref{sewm-chap}),
I showed that by using Maxwell's relation one obtains a flow equation
for the coupling, which, once solved, will give an equation of state 
of deconfined QCD for
arbitrary temperature and chemical potential.
Finally, I derived the equations for the characteristics of 
the coupling flow equation and anticipated that they will resemble
ellipses in the variables $T^2$ and $\mu^2$, 
with endpoints perpendicular to the $T$
and $\mu$ axes.

\chapter{The HTL quasiparticle model}
\label{sewm-chap}

In the last chapter I have set up a simple quasiparticle model of the 
QCD pressure that only takes into account the hard excitations, ignoring
longitudinal plasmon and plasmino quasiparticles. A comparison
between the strictly perturbative pressure known to order $\alpha_s^{5/2}$
and a perturbative expansion of the simple quasiparticle model shows
that -- while the leading orders are correctly reproduced -- only
$1/4/\sqrt{2}$ or about $18$\% of the order $\alpha_s^{3/2}$ term are
included in the simple quasiparticle model.
However, it is precisely the inclusion of this term 
(also called the plasmon term) that makes the result for the pressure
go ``wild'' in the first place,
spoiling the convergence of the strictly perturbative expansion
as was shown in Fig.~\ref{fig:Spert}.
Consequently, neither does the simple quasiparticle model incorporate
a big part of the plasmon effect (which could mean quantitative results
are incorrect by an unknown amount) nor is it a priori
clear whether an extension including the {\em full} plasmon term would not
go astray as violently as in the strict perturbative case (implying
that also qualitatively the results from the simple quasiparticle model
would be dubious). Therefore, it is important to consider refinements
of the simple quasiparticle model that incorporate more of the physically
important plasmon term, so that the above caveats can be settled.

One such refinement,
namely that of a quasiparticle model for the 
QCD thermodynamic pressure that is based on the HTL-resummed entropy
\cite{Blaizot:1999ip,Blaizot:1999ap,Blaizot:2000fc},
will be considered in the following.

\section{The entropy of the QCD plasma}

\begin{figure}
\begin{center}
\includegraphics[width=0.6\linewidth]{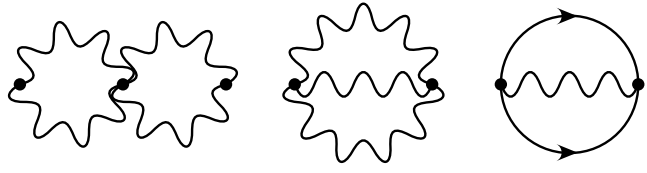}
\caption{Diagrams for $\Phi$ at 2-loop order in QCD in the temporal axial
gauge. Wiggly lines refer to gluon, plain lines refer to fermions.}
\label{fig:2Ldiags}
\end{center}
\end{figure}

As a starting point, one uses the result that the thermodynamic potential 
$\Omega=-p V$ can be written as the following functional of the full gluon
and fermion propagators $G$ and $S$ \cite{Blaizot:1999ip}:
\beq
\frac{1}{T} \Omega[G,S]=\frac{1}{2} {\rm Tr} \ln{G^{-1}}-{\rm Tr}\ln{S^{-1}}%
-\frac{1}{2}{\rm Tr} \Pi G+{\rm Tr} \Sigma S+\Phi[G,S],
\label{OmegaQCD}
\eeq
where $\Pi$ and $\Sigma$ are the gluon and quark self-energies of 
chapter \ref{peshier-chap} (related 
to the full propagators by Dyson's equation (\ref{Dysonglue}, 
\ref{Dysonferm})) and
${\rm Tr}$ includes traces over color indices and also over Lorentz
and spinor indices when applicable. $\Phi[G,S]$ is the sum of the 2-particle-%
irreducible ``skeleton'' diagrams, which to 2-loop order are
shown in Fig.~\ref{fig:2Ldiags}.

The essential property of the functional $\Omega[G,S]$ is to be 
stationary under variations of $G$ and $S$ (at fixed $G_0$ and $S_0$),
\beq
\frac{\delta \Omega}{\delta G}=0,\quad \frac{\delta \Omega}{\delta S}=0,
\eeq
so the physical pressure is then obtained as the value of $\Omega$
at its extremum. Note that the stationarity conditions imply
\beq
\frac{\delta \Phi}{\delta G}=\frac{1}{2} \Pi,%
\quad \frac{\delta \Phi}{\delta S}=\Sigma,
\label{selfcon}
\eeq
which together with the Dyson's equations define the physical propagators
and self-energies in a self-consistent way. Accordingly, by selecting
a class of skeletons in $\Phi[G,S]$ and calculating $\Pi$ and $\Sigma$
from the above relations one obtains a self-consistent approximation
of the physical pressure.

Using the temporal axial gauge which preserves rotational invariance
and also makes the Faddeev-Popov ghosts decouple from the theory, the gluon
and quark propagators can be decomposed as in the previous chapter, so that
one can evaluate Eq.(\ref{OmegaQCD}) using standard contour integration 
techniques. Accordingly, the entropy density $s=-\frac{d \Omega/V}{d T}$ is
evaluated to be
\beq
s=s_g+s_q+s^{\prime},
\label{entropyexpr}
\eeq
where 
\bqa
s_g&=&- (N^2-1)\int \frac{d^4 k}{(2\pi)^4} \frac{\partial n(\omega)}{%
\partial T}\left\{2 \left[%
{\rm Im} \ln{(-\omega^2+k^2+\Pi_T)}-{\rm Im}\Pi_T {\rm Re}D_T\right]\right.%
\nonumber\\%
&&\hspace*{6cm}\left.+{\rm Im}\ln{(k^2+\Pi_L)}+{\rm Im} \Pi_L {\rm Re} D_L%
\right\}, \nonumber\\
s_f&=&%
-2 N N_f \int \frac{d^4 k}{(2\pi)^4} \frac{\partial f(\omega)}{\partial T}%
\left\{{\rm Im}\ln \Delta_+^{-1}+{\rm Im}\ln{\left(-\Delta_-^{-1}\right)}%
\right.\nonumber\\
&&\hspace*{6cm}\left.
-{\rm Im}\Sigma_+ {\rm Re} \Delta_+ +{\rm Im}\Sigma_- {\rm Re} \Delta_+%
\right\},\nonumber \\
s^{\prime}\!\!\!&=&\!\!\!-%
\left.\frac{\partial (T/V \Phi[G,S])}{\partial T}\right|_{D,S}\!\!\!\!+%
(N^2\!-\!1)\! \int\! \frac{d^4 k}{(2\pi)^4} \frac{\partial n(\omega)%
}{\partial T} \left(%
2 {\rm Re} \Pi_T {\rm Im} D_T\!-\!%
{\rm Re} \Pi_L {\rm Im} D_L \right) \nonumber\\
&&\hspace*{2cm}%
+2 N N_f\int \frac{d^4 k}{(2\pi)^4} \frac{\partial f(\omega)}{\partial T}%
\left\{%
{\rm Re}\Sigma_+ {\rm Im} \Delta_+ -{\rm Re}\Sigma_- {\rm Im} \Delta_+%
\right\}. \;\;\;
\label{entropyprec}
\eqa
Here $n(\omega)$ and $f(\omega)$ are the Bose-Einstein and Fermi-Dirac
distributions,
\beq
n(\omega)=\frac{1}{\exp{\omega/T}-1},\quad%
f(\omega)=\frac{1}{\exp{(\omega-\mu)/T}+1}+\frac{1}{\exp{(\omega+\mu)/T}+1},
\eeq
and decomposition of the propagators from
chapter \ref{peshier-chap} has been used.

When restricting to an approximation where $\Phi[D,S]$ contains only
2-loop skeletons (shown in Fig.~\ref{fig:2Ldiags}), 
one finds the remarkable property that
\beq
s'=0,
\label{sprimezero}
\eeq
which has been shown in Refs.~\cite{Vanderheyden:1998ph,Blaizot:2000fc} 
for QED and QCD, respectively.
Since in this approximation
the self-energies and propagators are to be determined self-consistently
by solving ``gap-equations'' of the form
\beq
D_T^{-1}=-\omega^2+k^2+\Pi_T[D_T,D_L,\Delta_+,\Delta_-],
\label{gapeq}
\eeq
where $\Pi_T$ is given by one-loop self-energy expressions only, the 
resulting expression for the entropy (\ref{entropyexpr}) is also effectively
one-loop. 

This has important consequences: firstly, one has reduced the problem
of calculating a two-loop quantity (the free energy or thermodynamic pressure)
to a simple integration of an effectively one-loop quantity
(the entropy expression from above); secondly, since the term $s'$ has 
been interpreted as the contribution from the residual interaction of
the quasiparticles \cite{Blaizot:2000fc}, the result $s'=0$ means that
this contribution vanishes for the QCD entropy in the 2-loop approximation.
I will make use of this interpretation in the following.

\subsection{Approximately self-consistent solutions}

Unfortunately, ``gap equations'' like Eq.(\ref{gapeq}) are non-local,
which makes their exact solution prohibitively difficult. Moreover,
the gap equations contain UV divergencies and therefore require
a renormalization scheme compatible with the self-consistent structure 
of this non-local equation. Whereas this daunting task has been at 
least formally resolved in the scalar $\lambda \phi^4$ model
\cite{vanHees:2001ik,Blaizot:2003br}, the richer structure of QCD 
along with unresolved problems
concerning gauge-dependencies of the two-particle-irreducible scheme
at higher orders \cite{Arrizabalaga:2002hn} 
render a self-consistent renormalization of the
gap-equations in QCD an open problem.

Therefore, it is convenient to construct so-called ``approximately 
self-consistent'' solutions, which maintain the properties 
of Eq.(\ref{selfcon}) together with the Dyson's equations up to and including
a given order of $\alpha_s$, and which are manifestly gauge 
independent and UV finite. For example, using the HTL results for 
the self-energies and the resulting full propagators (given in chapter
\ref{peshier-chap}) constitutes just one 
such approximately self-consistent solution; therefore, using the HTL
approximation for the self-energies and propagators in 
Eqs.(\ref{entropyexpr},\ref{entropyprec},\ref{sprimezero}) 
gives a gauge-invariant, non-perturbative approximation to the full entropy.

\section{The HTL quasiparticle model}

Similar to the simple quasiparticle model, one can construct a model
for the thermodynamic pressure that incorporates the relevant HTL quasiparticle
excitations, which I will call the {\em HTL quasiparticle model} in the 
following \cite{Romatschke:2002pb}. 
Since this model should be a refinement of the simple quasiparticle
model, two requirements have to be fulfilled: firstly, the HTL model 
has to incorporate more of the physically important plasmon effect, and
secondly, it should constitute a natural extension of the simple 
quasiparticle model, including its structure in some sense.

The key to the construction of the HTL quasiparticle model lies in the 
observation that similar to the results in the simple quasiparticle model 
the entropy in the two-loop $\Phi$-derivable approximation
of the previous section is given in terms of HTL quasiparticle excitations
only, while the contribution from residual interactions vanishes.
One therefore {\em defines} the HTL quasiparticle model 
entropy to be given by Eq.(\ref{entropyprec}), which 
is known to be a fairly 
good approximation to the full entropy of the quark-gluon plasma
\cite{Blaizot:2000fc}.
Taking all this into account, one is led to the following
ansatz for the thermodynamic QCD pressure:
\beq
p(T,\mu)=p_{g,HTL}(T,m_D^2)+p_{q,HTL}(T,\mu,M^2)-%
B_{HTL}(m_D^2,M^2),
\label{HTLPmodel}
\eeq
with
\bqa
p_{g,HTL} & = & - 2 (N^2-1) \int \frac{d^3 k}{(2\pi)^3} \int _{0}^{\infty}%
 \frac{d\omega}{2\pi} n(\omega)%
\left[2 \rm{Im} \ln{\left(-\omega^2+k^2+\Pi_{T}\right)}%
\right. \nonumber \\ %
&& \left. -2\rm{Im} \Pi_{T} \rm{Re} D_{T} %
+ \rm{Im} \ln{\left(k^2+\Pi_{L}\right)} +%
\rm{Im} \Pi_{L} \rm{Re} D_{L}\right] \label{pg} \\
p_{q,HTL} & = & - 4 N N_f \int \frac{d^3 k}{(2\pi)^3} \int_{0}^{\infty} %
\frac{d \omega}{2 \pi}f(\omega)%
\left[\rm{Im}\ln{\left(k-\omega+\Sigma_{+}\right)} \right. \nonumber\\
& & \left.-\rm{Im}\Sigma_{+}\rm{Re}\Delta_{+}%
+ \rm{Im}\ln{\left(k+\omega+\Sigma_{-}\right)}+%
\rm{Im}\Sigma_{-}%
\rm{Re}\Delta_{-}\right],
\label{HTLQPs}
\eqa
obeying the stationarity conditions
\beq
\frac{\partial p}{\partial m_D^2}=\frac{\partial p}{\partial M^2}%
=0.
\eeq
Clearly, the model (\ref{HTLPmodel}) fulfills all of the above requirements:
the pressure is described by a gas of weakly interacting 
HTL quasiparticles, which have momentum-dependent
dispersion relations that were given in chapter \ref{peshier-chap}. 
Therefore, the integrations in Eq.(\ref{HTLQPs}) pick up contributions
equal to those of the simple quasiparticle model for large momenta, while 
for soft momenta the richer structure of the quasiparticle excitations
are taken into account. Consequently, when doing a perturbative expansion
of Eq.(\ref{HTLQPs}) at $\mu=0$ in powers of the coupling $\alpha_s$, one finds
that $25$\% of the plasmon effect is included in the HTL model, in contrast
to $18$\% for the simple quasiparticle model.
Furthermore, the stationarity condition ensures that there is no
residual interaction to the entropy ($s'=0$) and the resulting model
entropy is indeed given by $s=s_g+s_f$.

In its setup, the HTL quasiparticle model is very similar to the 
simple quasiparticle model, so the general discussion of 
this model in chapter \ref{peshier-chap} remains valid to a large extent
also for the HTL model. Indeed, to tell whether the plasmon effect has
a destabilizing effect in these quasiparticle models similar to what
happens in strict perturbation theory, one has to compare 
the results from both models in a standardized procedure, so I will
follow the steps from  Eq.(\ref{alphaeff}) onwards for both models, 
filling in the details and pointing out the differences.

\section{Modeling the lattice entropy and pressure}

\begin{figure}
\begin{center}
\includegraphics[width=0.6\linewidth]{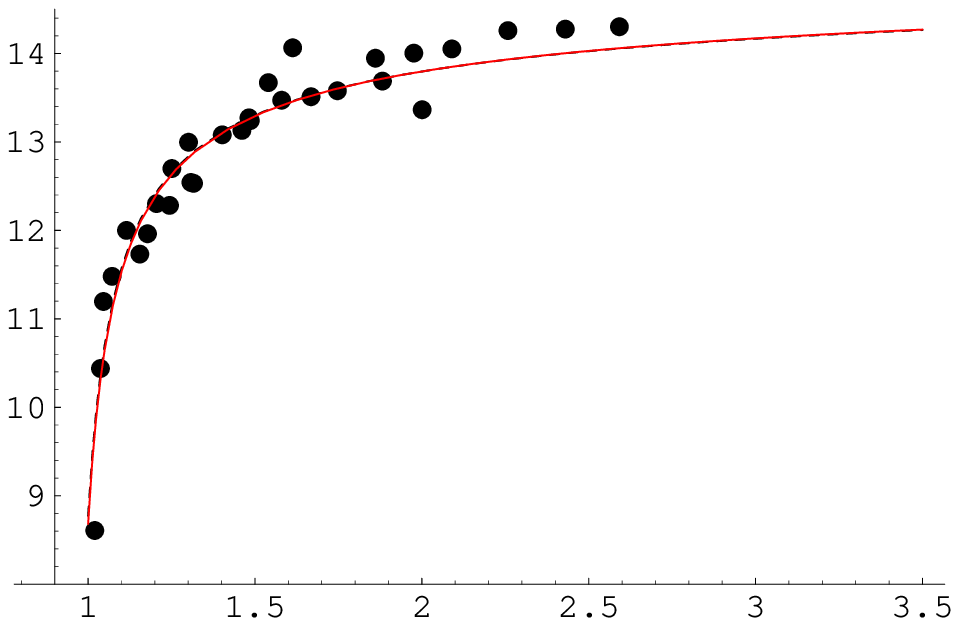}
\setlength{\unitlength}{1cm}
\begin{picture}(6,0)
\put(4,0){\makebox(0,0){\footnotesize $T/T_c$}}
\put(-1.5,3.5){\makebox(0,0){\footnotesize $s/T^3$}}
\end{picture}
\caption{%
Entropy data generated from Ref.~\cite{AliKhan:2001ek} vs. fitted model
entropy. The HTL and simple quasiparticle model fits are nearly
indistinguishable to the naked eye.}
\label{fig:Smu01}
\end{center}
\end{figure}

Restricting $\mu=0$ and using the ansatz 
(\ref{alphaeff}) for the strong coupling $\alpha_s$, 
the entropy of the HTL and simple quasiparticle models becomes a function
of the temperature $T$ and two fit parameters $T_s$ and $\lambda$, 
$s=s(T,T_s,\lambda)$. Using lattice data for $N_f=2$ and $N=3$ from 
Ref.~\cite{AliKhan:2001ek} together with an estimated continuum extrapolation
from \cite{Peshier:2002ww}, these two parameters $T_s$ and $\lambda$ are
determined by a least square fit to the data. The numerical values
for these fit parameters (which are reproduced in chapter 
\ref{qpmodels-chap}) turn out to be quite similar, and the respective fits
to the lattice data (shown in Fig.~\ref{fig:Smu01}) are nearly
indistinguishable.

\begin{figure}
\begin{center}
\includegraphics[width=0.6\linewidth]{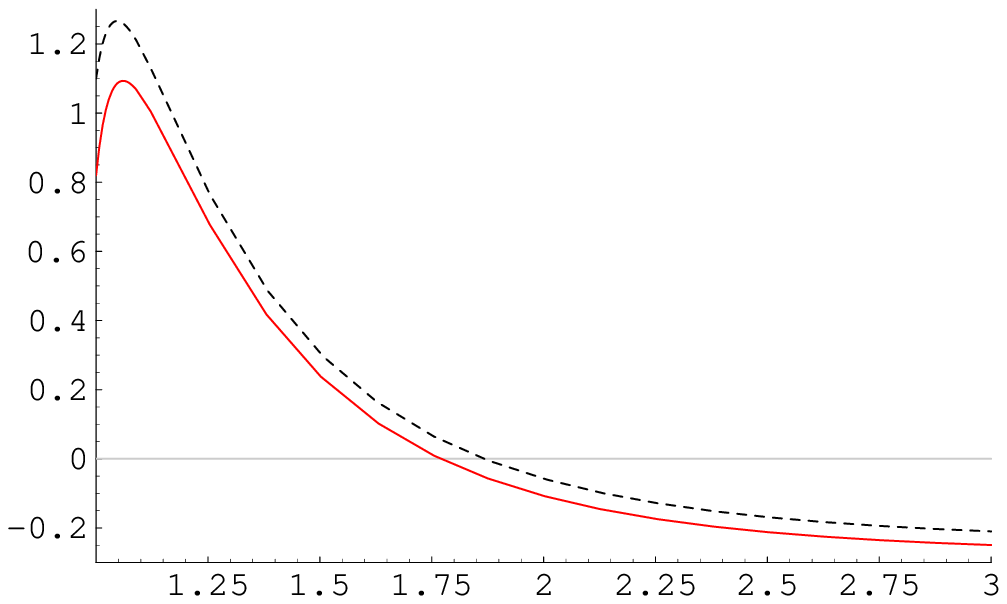}
\setlength{\unitlength}{1cm}
\begin{picture}(6,0)
\put(4,0){\makebox(0,0){\footnotesize $T/T_{c}$}}
\put(-1.7,3){\makebox(0,0){\footnotesize $B/T^4$}}
\end{picture}
\caption{%
Residual interaction $B$ for the HTL (full line)
and simple model (dashed-line), respectively.}
\label{fig:B1}
\end{center}
\end{figure}

\begin{figure}
\begin{center}
\includegraphics[width=0.6\linewidth]{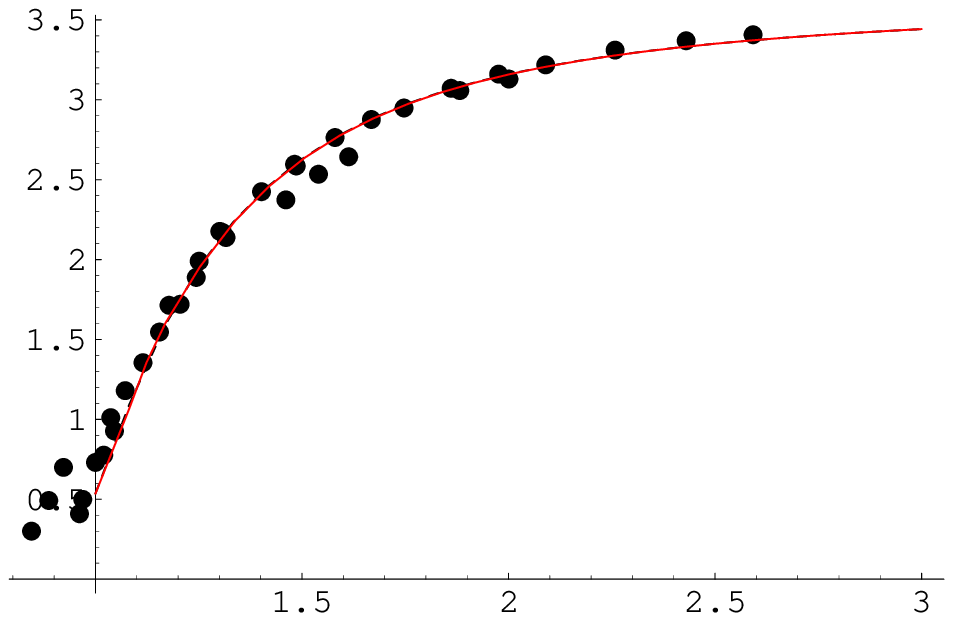}
\setlength{\unitlength}{1cm}
\begin{picture}(6,0)
\put(4,0.1){\makebox(0,0){\footnotesize $T/T_{c}$}}
\put(-1.7,4){\makebox(0,0){\footnotesize $p/T^4$}}
\end{picture}
\caption{Pressure data from Ref.~\cite{AliKhan:2001ek} vs. model pressure.
HTL and simple model results are indistinguishable for the naked eye.}
\label{fig:P1}
\end{center}
\end{figure}

Once the fit parameters have been fixed, so has $\alpha_{s,{\rm eff}}$ 
at vanishing chemical potential,
and consequently also the quasiparticle masses. One can then
use Eq.(\ref{Bdet}) together with $d\mu=0$, 
\beq
B=\int \sum_{i=g,q}%
 \frac{\partial p_{i}}{\partial m_{i}^2}%
\frac{\partial m_{i}^2}{\partial T} dT+B_{0},
\eeq
to determine
the residual interaction as a function of the temperature up to an
integration constant $B_0$. This integration constant has been determined
by a fit of the lattice pressure with the simple quasiparticle model 
in \cite{Peshier:2002ww}, obtaining
\beq
B_0=B|_{T_c}=1.1\ T_c^4,
\eeq
so that the resulting model pressure at $T_c$ becomes
\beq
p(T_c,\mu=0)= 0.536(1) T_c^4.
\eeq
To facilitate comparison, I fix the HTL model value of $B_0$ by requiring
\linebreak
\hbox{$p_{HTL}(T_c,\mu=0)\simeq 0.536\ T_c^4$}, 
obtaining $B_{0,HTL}=0.82\ T_c^4$.
The resulting residual interactions for the HTL and simple quasiparticle
models are plotted in Fig.~\ref{fig:B1}; finally, the overall
model pressure is compared to the lattice data in Fig.~\ref{fig:P1}.

Clearly, both the HTL and the 
simple quasiparticle model can be used to describe
the lattice data very accurately once a 2-parameter fit for the effective
coupling is adopted, even close to the deconfinement transition. 
Furthermore, since the fit parameters turn out to be only slightly different
and the resulting fits to the lattice data are nearly indistinguishable, one
concludes that (at least at $\mu=0$) the inclusion of the plasmon effect
does not seem to destabilize the quasiparticle models; on the contrary,
even though the effective coupling gets rather large near $T=T_c$ (see chapter 
\ref{qpmodels-chap} for details), the effects of changing
the percentage of the included plasmon effect are rather small.
Therefore, one gains some confidence that a phenomenological description
of the thermodynamic QCD pressure through quasiparticle models is also 
applicable in the non-perturbative regime, and in turn may be used to 
derive results that lie parametrically far from the initial (lattice) data.

\section{Solving the flow equation}

At non-vanishing chemical potential, the models are required 
to fulfill Maxwell's relation
\beq
\frac{d s}{d \mu}=\frac{d n}{d T}.
\eeq
Since the entropy $s$ as well as the number density $n$ depend on
$T$ and $\mu$ both explicitly through the Bose-Einstein and Fermi-Dirac
distribution functions as well as implicitly through the quasiparticle masses,
one has
\beq
\frac{d s}{d \mu}-\frac{d n}{d T}=\sum_{i=g,q}%
\frac{\partial s}{\partial m_{i}^2}\frac{d m_{i}^2}{d \mu}-%
\frac{\partial n}{\partial m_{i}^2}\frac{d m_{i}^2}{d T},
\label{Maxepl}
\eeq
because the explicit derivations of $s$ and $n$ with respect to
$\mu$ and $T$ cancel, respectively. The quasiparticle masses themselves
also depend on $T$ and $\mu$ both explicitly and implicitly through 
the coupling $\alpha_{s,{\rm eff}}$, so that e.g. for the Debye mass
one finds
\beq
\frac{d m_D^2}{d \mu}=\frac{N_f}{2\pi^2} 2 \mu 4 \pi \alpha_s(T,\mu)+%
\frac{\partial \alpha_s(T,\mu)}{\partial \mu} 4\pi%
\left(\frac{2 N + N_f}{6} T^2+\frac{N_f \mu^2}{2\pi^2}\right).
\eeq
Therefore, Eq.(\ref{Maxepl}) represents a partial differential equation
for the coupling that can be written in the form already anticipated
in Eq.(\ref{floweq}),
\beq
a_T\frac{\partial \alpha_s}{\partial T}+a_{\mu}\frac{\partial \alpha_s}{%
\partial \mu}=b;
\eeq
since it takes a considerable amount of algebra to derive 
the coefficients $a_{T},%
a_{\mu}$ and $b$ in the HTL model, their explicit form may be of use
in further studies.
However, since their appearance is rather
unwieldy, I refrain from reproducing these coefficients in the main text
and have relegated them to appendix \ref{app1-chap}.

\begin{figure}
\begin{center}
\includegraphics[width=.6\linewidth]{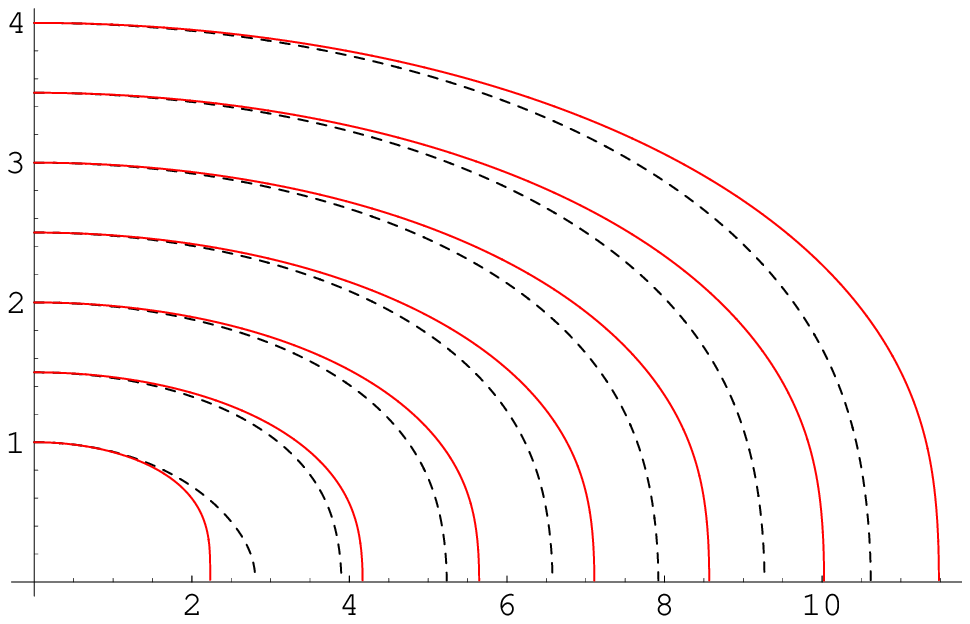}
\setlength{\unitlength}{1cm}
\begin{picture}(6,0)
\put(4,0.1){\makebox(0,0){\footnotesize $\mu/T_{c}$}}
\put(-1.7,3.5){\makebox(0,0){\footnotesize $T/T_c$}}
\end{picture}
\caption{Comparison of the shape of the characteristics for
the HTL (full lines) and simple quasiparticle model (dashed lines).
$T_c$ denotes $T_c|_{\mu=0}$.}
\label{fig:charcomp1}
\end{center}
\end{figure}

\begin{figure}
\begin{center}
\includegraphics[width=.6\linewidth]{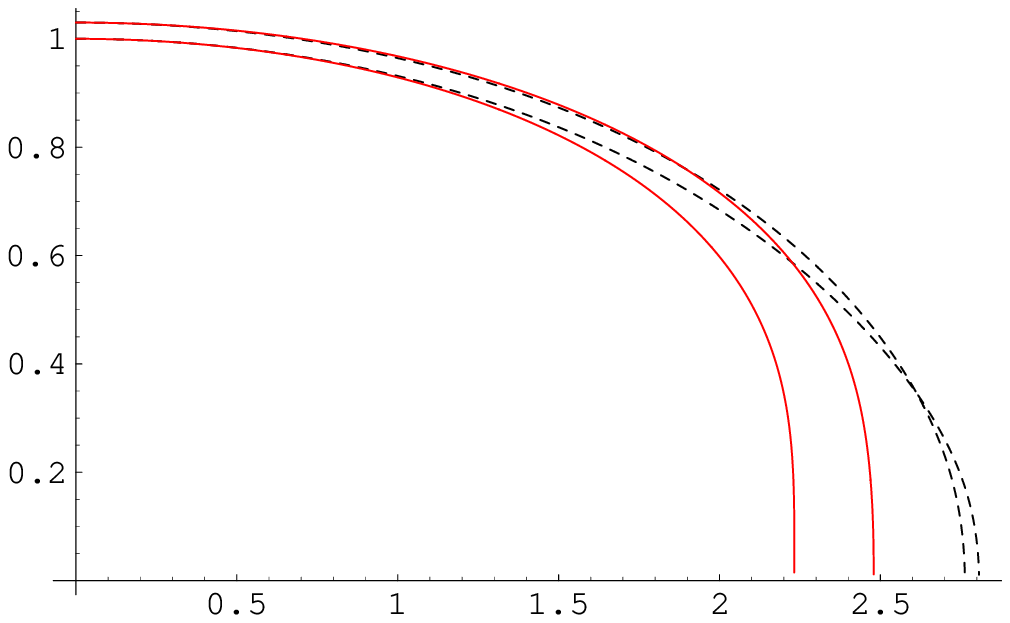}
\setlength{\unitlength}{1cm}
\begin{picture}(6,0)
\put(4,0.1){\makebox(0,0){\footnotesize $\mu/T_{c}$}}
\put(-1.7,3.5){\makebox(0,0){\footnotesize $T/T_c$}}
\end{picture}
\caption{Characteristics starting very near $T_c$: for the simple
quasiparticle model (dashed-lines) intersections are occurring, while
for the HTL model (full lines) this is not the case.}
\label{fig:charcomp2}
\end{center}
\end{figure}

When evaluating the coefficients numerically and comparing their value
for the HTL and simple quasiparticle model, one finds that while 
$a_{T}$ and $a_{\mu}$ assume nearly equal values, the coefficient $b$
turns out to differ noticeably which has interesting consequences, as will
become clear in the following.
The solution to the flow equation (\ref{floweq}) for the coupling is
found by using 
the initial value condition $\alpha_{s,{\rm eff}}(T,\mu=0)$
and then numerically solving the characteristics of Eq.(\ref{floweq}),
\beq
a_{\mu}dT=a_T d{\mu}, \quad a_{\mu} d\alpha_s=b d\mu.
\eeq
The shapes of the characteristics are shown in Fig.~\ref{fig:charcomp1}
for both the HTL and the simple quasiparticle model;
as anticipated in chapter \ref{peshier-chap}, the characteristic curves 
resemble ellipses in $T^2$ and $\mu^2$, 
which are perpendicular to both the $T$ and the $\mu$ axis.
Furthermore, for the simple quasiparticle model there is a set of 
characteristics (those which start at $\mu=0$ in the interval 
$[T_c,1.06 T_c]$) that intersect in a narrow region, indicating that 
there the solution to the flow equation is not unique (see 
Fig.~\ref{fig:charcomp2}). However, when also calculating the pressure
along the characteristics 
it turns out that it becomes negative 
in the intersection region; it has been argued \cite{Peshier:2002ww}
that this implies that a transition to another phase with positive
pressure has been occurring 
somewhere outside this region, which therefore means that the quasiparticle
description had broken down and the ambiguity would be of no
physical relevance. 

Interestingly, for the HTL model, 
the difference in the flow equation parameter $b$ seems to completely
lift the ambiguity problem: the characteristic curves are well behaved
and do not intersect anywhere, as can be seen in Fig.~\ref{fig:charcomp2}.
Unfortunately, the pressure still turns negative for some region
in the $T,\mu$ plane for the HTL model, so that clearly also this
model breaks down for certain $T,\mu$. On the other hand, one can
turn this breakdown of the quasiparticle models into a virtue by assuming
that the deconfinement transition occurs near the line $T=T(\mu)$ 
where the pressure vanishes. Therefore, an estimate for 
the ``critical'' transition
line for arbitrary $\mu$ becomes calculable within the quasiparticle models,
(see chapter \ref{qpmodels-chap} for details).

\section{The pressure at finite chemical potential}

\begin{figure}
\begin{center}
\includegraphics[width=.8\linewidth]{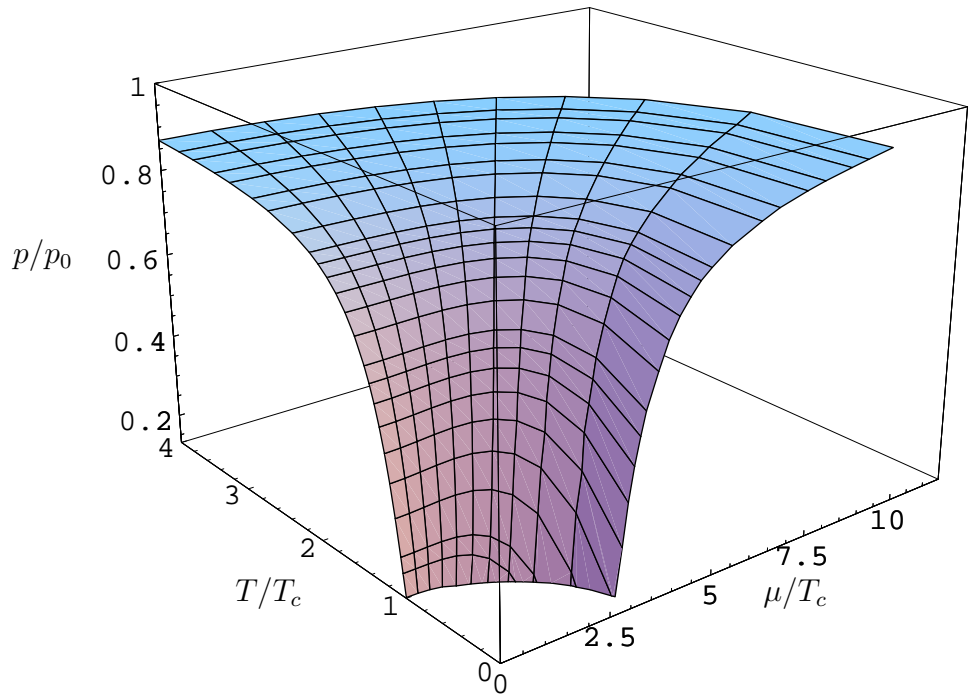}
\setlength{\unitlength}{1cm}
\begin{picture}(5,0)
\put(6,2){\makebox(0,0){$\mu/T_{c}$}}
\put(-4,6.5){\makebox(0,0){$p/p_0$}}
\put(-1,2){\makebox(0,0){$T/T_c$}}
\end{picture}
\end{center}
\caption{%
The full pressure of the HTL quasiparticle model normalized to 
the free pressure $p_0$ in the $T,\mu$ plane.}
\label{fig:Palongchar}
\end{figure}

Solving the flow equation provides one with an effective coupling 
in the whole $T,\mu$ plane, so that -- via the explicit form of the
quasiparticle masses -- the contributions $p_g$ and $p_q$ to the
full pressure (\ref{simP},\ref{HTLPmodel}) are fixed.
It remains to calculate the residual interaction $B$, which is most
easily done by integrating Eq.(\ref{Bdet}) along the characteristic curves
of the flow equation (\ref{floweq}). The result for the full pressure 
is shown in a 3D-plot in Fig.~\ref{fig:Palongchar}.


As will be discussed in more detail in chapter \ref{qpmodels-chap}, this
result is in fairly good agreement with various lattice approaches.
On the other hand, \textsc{Ipp} and \textsc{Rebhan} \cite{Ipp:2003jy} 
have found
that in the exactly solvable large-$N_f$ model of QCD the pressure
has a conspicuous ``kink'' near $\mu \simeq \pi T$, signalling an abrupt
change of the pressure for larger chemical potentials. This kind of
behavior is not reproduced in the quasiparticle model approach
for 2 flavors; on the contrary, it turns out that the quasiparticle
model pressure is well described by its small-$\mu$ behavior even for
large values of the chemical potential, as will be discussed in more detail
in chapter \ref{qpmodels-chap}.
However, it is by no means clear that results obtained in the large-$N_f$
case are also relevant in general 
for the physically interesting situations of $N_f=2$ or $N_f=3$, although 
for special cases their effect may well be considerable, as has 
recently been demonstrated \cite{Ipp:2003cj}.


\section{Summary}

In this chapter I have introduced the entropy of the quark gluon plasma 
through the so-called $\Phi$-derivable formalism. 
A 2-loop approximation
for the functional $\Phi$ that allows the entropy to be written 
as an effectively 
one-loop quantity was presented, and it was pointed out that 
because of problems concerning
renormalization and gauge-dependencies it is useful to introduce the notion
of {\em approximate self-consistency}. 
Since using the HTL self-energies
together with the respective Dyson's equations for the propagators 
introduced in chapter \ref{peshier-chap} corresponds to such an
approximately self-consistent approximation to the full quark-gluon 
plasma entropy, this expression was used as a basis
for an HTL quasiparticle model, and it was shown that this model corresponds
to an extension of the simple quasiparticle model introduced in 
chapter \ref{peshier-chap}.

Fitting two flavor lattice data for the pressure it was found that at vanishing
chemical potential the HTL model constitutes only a small correction to
the simple quasiparticle model, so it seems that the plasmon effect which 
plays such a devastating role in strict perturbation theory is not 
destabilizing the quasiparticle implementation. Moreover, an explicit
solution to the flow equation for the coupling shows that while 
the simple quasiparticle model solution becomes ambiguous in some region 
in the $T,\mu$ plane, one does not encounter any such problems for
the HTL model.

Finally, I investigated the pressure at finite chemical potential 
and discussed its general behavior in view of recent exact results for
the large-flavor limit of QCD.

\chapter{The QCD EOS for arbitrary $\mu$ and completion of plasmon
effect}
\label{qpmodels-chap}
\chaptermark{The QCD EOS for arbitrary $\mu$}

Having demonstrated in the last chapter
that lattice data at vanishing chemical
potential can be accurately described by quasiparticle models with 
an effective coupling, I now concentrate on the quantitative predictions one
can make by mapping lattice data to nonzero chemical potential.
More precisely, first the quasiparticle model results 
at low chemical potential will be compared 
to those obtained through lattice calculations
and strictly perturbative methods, and then the case of 
high chemical potential and low temperatures, where nearly all other 
methods eventually break down, will be investigated. 
However, I will
ignore the effect of color superconductivity, which should become
important at very small temperatures only, assuming that because of 
the comparatively small energy gap it constitutes only a minor modification
to the thermodynamic pressure in the region of interest \cite{Fraga:2001xc}.
The results in this chapter are based on 
Ref.~\cite{Rebhan:2003wn}; however, here I will be able to go into
more detail and also add some new results.

In the previous chapters, I have introduced the HTL quasiparticle
model that includes $1/4$ of the full plasmon effect at vanishing chemical
potential, 
which is a factor of $\sqrt{2}$ more than in the simple quasiparticle model
that uses only the asymptotic HTL masses.
Motivated by the rather successful generalization of the simple 
to the HTL quasiparticle model I will try to improve the accuracy of 
the HTL quasiparticle model further by setting up a model that 
includes even more of the perturbative plasmon effect.
In principle, for this task one would need the complete
(momentum-dependent) next-to-leading order corrections to the self-energies,
which up to now no one had the strength to calculate;
however, it has been shown in Ref.~\cite{Blaizot:1999ip,Blaizot:2000fc} 
that -- as far as the plasmon effect is concerned -- the contribution to 
the thermodynamic quantities of these next-to-leading order corrections
may be approximated by their averaged contribution to the asymptotic 
masses. Therefore, this opens the possibility of setting up a model
that contains the {\em full} plasmon effect at $\mu=0$, albeit at
the price of introducing another unknown parameter, as will be
discussed in the following.

\section{NLO quasiparticle models}

Denoting the momentum-dependent next-to-leading ($\alpha_s^{3/2}$)
order corrections to the self-energies as $\delta \Pi_T(\omega,k)$ and
$\delta \Sigma_{+}$ (the other branches do not receive contributions
at this order), respectively, the averaged corrections to the asymptotic
masses (\ref{minfty},\ref{Minfty})
have been calculated to be \cite{Blaizot:1999ip,Blaizot:2000fc}
\beq
\label{deltamasav}
\bar\delta m_\infty^2={\int dk\,k\,n'(k) {\rm Re}\, \delta\Pi_T(\omega=k)
\over \int dk\,k\,n'(k)}=-2 \alpha_s NT m_D
\eeq
and similarly
\beq
\label{deltaMasav}
\bar\delta M_\infty^2={\int dk\,k\,(f'_+(k)+f'_-(k)) {\rm Re}\,
2k \delta\Sigma_+(\omega=k)
\over \int dk\,k\,(f'_+(k)+f'_-(k))}=-{(N^2-1)\over N} \alpha_s T m_D\;.
\eeq

For the values of the coupling $\alpha_s$ considered here, 
these corrections are so large that they 
give tachyonic masses when treated strictly perturbatively.
In Ref.~\cite{Blaizot:2000fc} it has been proposed to incorporate
these corrections through a quadratic gap equation
which works well as an approximation
in the exactly solvable scalar O($N\to\infty$)-model, where
strict perturbation theory would lead to identical difficulties.
However, for the fermionic asymptotic masses,
in order to have the correct scaling of Casimir factors in
the exactly solvable large-$N_f$ limit of QCD
\cite{Moore:2002md
}, a corresponding gap equation has
to remain linear in the fermionic mass squared. 
Choosing the correction
therein to be determined by the solution to the gluonic gap
leads to \cite{Rebhan:2003fj,Blaizot:2003tw}
\begin{eqnarray}
\label{minftyb2}
\bar{m}_{\infty}^2& =& m_\infty^2-2 \sqrt{2} \alpha_s N T %
\bar{m}_{\infty} \\
\label{Minftyb2}
\bar{M}_{\infty}^{2}&=& M_\infty^2-\frac{\sqrt{2} \alpha_s (N^2-1)T }{N}
\bar{m}_{\infty} ,
\label{NLOmassgaps}
\end{eqnarray}
where $m_\infty^2$ and $M_\infty^2$ are the leading-order
gluonic and fermionic asymptotic
masses as given in (\ref{minfty}) and (\ref{Minfty}).
This in fact avoids tachyonic masses for the fermions
as long as $N_f\le 3$.
\footnote{%
For $N_f=3$ the solutions to the above approximate gap equations happen to
coincide with those obtained in
the original version of two independent quadratic
gap equations of Ref.~\cite{Blaizot:2000fc}. For the case $N_f=2$
considered here, the differences are fairly small.
For $N_f>3$, however, the necessity to avoid tachyonic masses would
restrict the range of permissible coupling strength $\alpha_s$.}

Finally, since the averaged quantities $\bar m_\infty^2$
and $\bar M_\infty^2$ are the effective masses at
hard momenta only, a cutoff scale 
$\Lambda=\sqrt{2 \pi T m_{D} c_{\Lambda}}$
is introduced that separates soft from hard momenta. 
The quasiparticle pressure for this model, 
which in the following will be referred to
as NLA-model, then separates into a soft and a hard component for both
gluons and fermions. The soft contributions are given by expressions similar to
Eq.(\ref{HTLQPs}), but with $\Lambda$ as upper limit for 
the momentum integration. For the hard contributions, the momentum 
integrations run from $\Lambda$ to $\infty$ and the mass pre-factors
in the HTL self-energies (\ref{HTLPiL},\ref{HTLSigmapm})
are replaced by their asymptotic counterparts, 
$m_{D}^2\rightarrow 2 \bar m_{\infty}^2$ and $M^2\rightarrow %
\frac{1}{2} \bar M_{\infty}^2$.

The single free parameter $c_{\Lambda}$ in $\Lambda$ can be varied around
$1$ to obtain an idea of the ``theoretical error'' of the model. In the 
following the range
$c_{\Lambda}=\frac{1}{4}$ to $c_{\Lambda}=4$ will be considered; note that 
$c_{\Lambda}=\infty$ corresponds to the HTL-model 
(which has been used as cross-check) since all hard corrections are ignored.
On the other hand, $c_{\Lambda}=0$ would assume that 
(\ref{minftyb2}) and (\ref{Minftyb2}) represent good approximations
for the NLO corrections to the spectral properties of soft
excitations. However, the few existing results, in particular on
NLO corrections to the Debye mass \cite{Rebhan:1993az
} and 
the plasma frequency \cite{Schulz:1994gf}, appear to be rather different
so that it seems safer to leave the soft sector unchanged by keeping
a finite $c_\Lambda$.

\section{Fitting lattice data at \protect$\mu=0$}

\begin{figure}
\begin{center}
\includegraphics[width=0.6\linewidth]{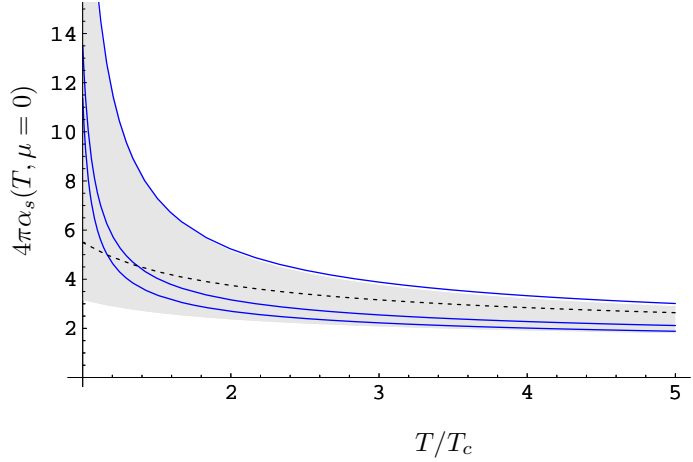}
\setlength{\unitlength}{1cm}
\begin{picture}(6,0)
\put(4,0){\makebox(0,0){\footnotesize $T/T_c$}}
\put(-1.5,2.5){\makebox(0,0){\begin{rotate}{90}%
\footnotesize $4 \pi\alpha_{s}(T,\mu=0)$
\end{rotate}}}
\end{picture}
\caption{%
Effective coupling: NLA model results
for $c_{\Lambda}=$ $4$,$1$ and $1/4$ (full lines from
bottom to top), 2-loop perturbative coupling in 
\protect$\overline{\hbox{MS}}$ (gray band) and result 
from Ref.~\protect{\cite{Schneider:2003uz}} (dashed line).}
\label{fig:G1}
\end{center}
\end{figure}

Going through the program of fitting the entropy expressions 
from the models under consideration to lattice data \cite{AliKhan:2001ek}
for $N_{f}=2$ as described in chapter \ref{sewm-chap}, one finds
the following values for the fit parameters $T_s$ and $\lambda$:\\
\begin{center}
\begin{tabular}[c]{|c|c|c|c|c|c|}
\hline
& simple & HTL & $c_{\Lambda}=4$ & $c_{\Lambda}=1$ & $c_{\Lambda}=1/4$ \\
\hline
$T_{s}/T_{c}$ & -0.89 & -0.89 & -0.89 & -0.84 & -0.61 \\
\hline
$\lambda$ &17.1 & 19.4 & 18.64 & 11.43 & 3.43 \\
\hline
\end{tabular}\\
\end{center}
It can be seen that the results for the 
simple, HTL and NLA quasiparticle model with $c_{\Lambda}=4$ are very close.
The fits to the entropy data all lie in
a narrow band for all the models considered \cite{Rebhan:2003wn}, resembling
the fits for the simple and HTL quasiparticle models shown in 
Fig.~\ref{fig:Smu01}.
The fitted effective coupling $\alpha_s$ for the various models is shown in 
Fig.~\ref{fig:G1}; for comparison, also the 2-loop perturbative 
running coupling in $\overline{\hbox{MS}}$ 
is shown, 
where the renormalization scale is varied between $\pi T$ and $4 \pi T$
and
$T_c=0.49 \Lambda_{\overline{\hbox{\scriptsize MS}}}$. 
As can be seen from the plot, the results for the effective 
coupling are well within the range of the 
2-loop perturbative running coupling (for the case $c_{\Lambda}=1/4$
and renormalization scale $\pi T$ the results even seem to be identical,
which is, however, probably only a coincidence).
The result for the coupling obtained
in the semi-classical approach of Ref.~\cite{Schneider:2003uz}\footnote{%
For finite $\mu$ and constant temperatures,
 however, the coupling obtained in \cite{Schneider:2003uz}
rises while the quasiparticle model results indicate a decrease of
the coupling (which is consistent with the standard QCD running coupling
with renormalization scale proportional to $\sqrt{T^2+(\mu/\pi)^2}$).}
is also shown in Fig.~\ref{fig:G1}.

In general, one can see that the effective coupling becomes bigger when
$c_{\Lambda}$ gets smaller; this is because the hard masses (for equal
values of the coupling) are 
smaller than the soft masses, which makes the entropy increase when the
hard parts become more important. Accordingly,
the coupling has to rise in order for the entropy to match the data 
(therefore, for the extreme case $c_{\Lambda}=0$ one finds huge values
of the effective coupling constant).

The fact that the fitted effective coupling agrees so well with the band of
the 2-loop perturbative running coupling suggests that also simply using
the perturbative running coupling as an input might lead to results
for the thermodynamic pressure that are not very different;
this approach will be tested in \hbox{chapter \ref{qpmodels2-chap}}.

Once the effective coupling is known for $\mu=0$ one 
can proceed to fix the remaining integration constant $B_0$ as indicated
in chapter \ref{sewm-chap}. Setting $p(T_{c})=0.536(1)\ T_c^4$ one finds \\
\begin{center}
\begin{tabular}[c]{|c|c|c|c|c|c|}
\hline
&simple & HTL & $c_{\Lambda}=4$ & $c_{\Lambda}=1$ & $c_{\Lambda}=1/4$ \\
\hline
$B_{0}/T_{c}^4$ &1.1 & 0.82 & 0.73 & 0.6 & 0.47 \\
\hline
\end{tabular}\\
\end{center}

The fits for the model pressure for the various models also turn out
to lie in a narrow band \cite{Rebhan:2003wn} 
resembling the HTL and simple quasiparticle model
result shown in Fig.~\ref{fig:P1}. Therefore, also the NLA quasiparticle
model can be used to accurately describe lattice data for the entropy
and pressure at vanishing chemical potential. In contrast to what was 
found from the comparison between the HTL and simple quasiparticle model,
the fit parameters indicate that at least for $c_{\Lambda}<1$
the NLA model differs considerably from the other models. Unfortunately,
it is hard to say if this difference is due to the full inclusion
of the plasmon effect in the NLA model and thereby represents
a better approximation of the underlying physics than the HTL model,
or rather to the fact that
only the {\em averaged} NLO corrections to the self-energies were taken
into account and the deviations 
of this model signal the onset of the breakdown of
applicability, which probably occurs\footnote{%
This hypothesis is strengthened by the fact that although the 
NLA model with $c_{\Lambda}=0$ can be made to describe the
lattice data at $\mu=0$, the values of the fit parameters for
this model do not resemble anymore 
those of the other quasiparticle models given above, and its subsequent
extension to finite chemical potential gives results that are at odds
with respect to the other models as well as lattice
calculations.}
 at $c_{\Lambda}\ll 1$.
In any case, the results for physical quantities from the NLA models
with $1/4<c_{\Lambda}<4$ turn out to fulfill the requirement that they
represent again small corrections to the HTL and simple quasiparticle
model results, giving further confidence that the stability of 
quasiparticle models is unaffected by the plasmon effect.

\section{Results for small chemical potential}

The NLA model flow equations for the effective coupling can be solved
along the lines of the HTL and simple model in chapter \ref{sewm-chap};
the shape of the characteristics resembles those of the other models 
and the crossing of characteristics found 
in the simple quasiparticle model also does not occur in the NLA models
(similar to what has been found for the HTL-model). 

\subsection{Susceptibilities}

\begin{figure}
\begin{minipage}[t]{.48\linewidth}
\includegraphics[width=\linewidth]{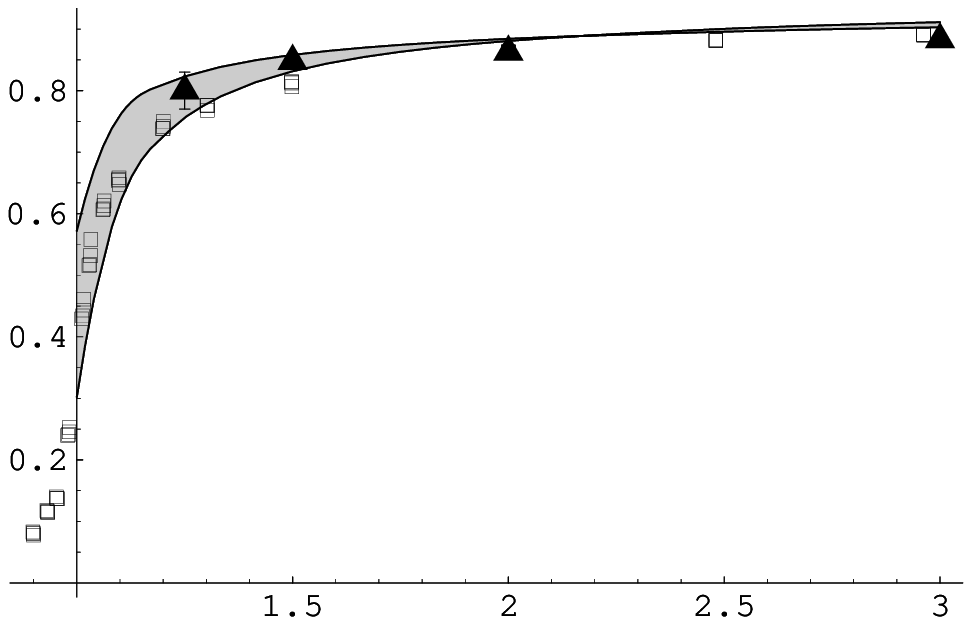}
\setlength{\unitlength}{1cm}
\begin{picture}(6,0)
\put(4,0){\makebox(0,0){\footnotesize $T/T_{c}$}}
\put(-0.5,3.5){\makebox(0,0){\footnotesize $\chi/\chi_{SB}$}}
\end{picture}
\caption{Susceptibilities from NLA models (light gray band), 
from lattice data for 
$N_{f}=2$ \cite{Gavai:2001ie} (triangles) and 
for 2+1 flavors \cite{Fodor:2002km} (boxes).}
\label{fig:Suscep1}
\end{minipage}\hfill
\begin{minipage}[t]{.48\linewidth}
\includegraphics[width=\linewidth]{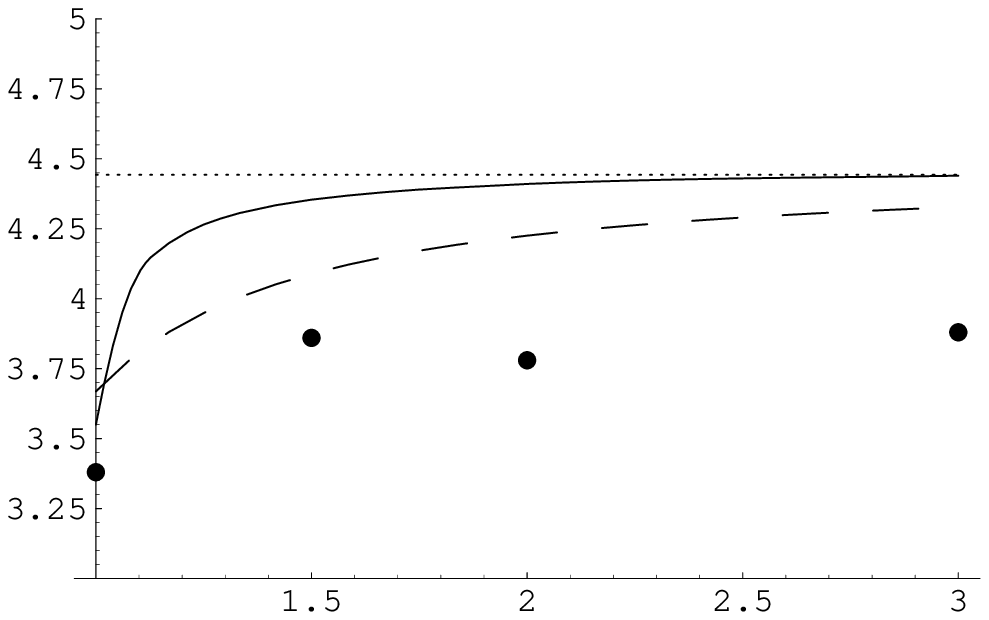}
\setlength{\unitlength}{1cm}
\begin{picture}(6,0)
\put(3.5,0){\makebox(0,0){\footnotesize $\mu/T_{c}$}}
\put(0,2.8){\makebox(0,0){\footnotesize $x(T)$}} 
\end{picture}
\caption{The tree-level result $\sqrt{2} \pi$ (dotted line) 
for the quantity $x(T)$ compared to
the HTL model result (full line),  lattice results from 
Ref.~\cite{Gavai:2003mf}
as well as results from strict perturbation theory \cite{Vuorinen:2003fs}
(dashed line).}
\label{fig:Suszep4HTL}
\end{minipage} \hfill
\end{figure}

Once the flow equations of the various models have been solved, it
is straightforward to calculate the quark-number susceptibilities at
$\mu=0$,
\beq
\chi(T)=\left.\frac{\partial^2 p}{\partial \mu^2}\right|_{\mu=0}.
\label{suscepsdef}
\eeq
The result -- normalized to the tree level result 
$\chi_{0}=\frac{N N_f}{3} T^2$ --
is compared to lattice data for $N_f=2$ \cite{Gavai:2001ie} and $N_f=2+1$ 
\cite{Fodor:2002km} in Fig.~\ref{fig:Suscep1}. As can be seen,
there is a very good agreement between the model predictions and the
lattice data (which is completely independent from the one used in
determining the model fit parameters).
The result from a strictly perturbative calculation \cite{Vuorinen:2002ue}
(not shown) also is consistent with the results of Fig.~\ref{fig:Suscep1}
at larger temperatures, although the uncertainty in the latter is much bigger
than that of the quasiparticle model results (see also chapter 
\ref{qpmodels2-chap}).

Calculating the pressure at finite chemical potential $p(T,\mu)$
as described
in chapter \ref{sewm-chap}, one is able to numerically extract the 
higher order susceptibility 
\beq
\bar{\chi}(T)=\left.\frac{\partial^4 p}{\partial \mu^4}\right|_{\mu=0}
\eeq
by doing a least-square fit to the model pressure
\beq
p_{\rm m}(T,\mu)= p(T,0)+\frac{\chi(T)}{2} \mu^2+\frac{\bar{\chi}(T)}{4!}\mu^4.
\label{modelP}
\eeq
Parameterizing
\beq
\bar{\chi}(T)=\frac{\chi(T)}{2 T^2} \frac{4!}{x(T)}
\eeq
one obtains the result shown in Fig.~\ref{fig:Suszep4HTL}. In this figure,
also the lattice results for the corresponding quantity in 
Ref.~\cite{Gavai:2003mf} (which however have not been extrapolated to 
the continuum limit) and strictly perturbative results 
\cite{Vuorinen:2003fs,Vuopriv}
(for the specific renormalization scale $\bar{\mu}=8.112 T$ 
\cite{Kajantie:1997tt}) are shown,
indicating general agreement between the different approaches.

\begin{figure}
\begin{center}
\includegraphics[width=0.6\linewidth]{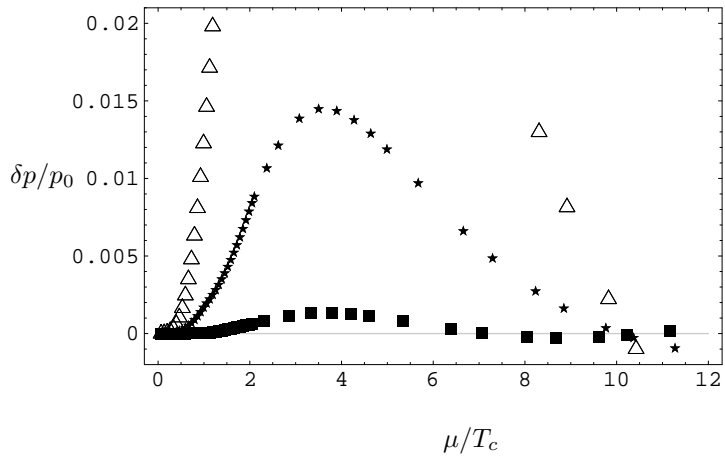}
\setlength{\unitlength}{1cm}
\begin{picture}(6,0)
\put(4,0){\makebox(0,0){\footnotesize $\mu/T_{c}$}}
\put(-1.7,3.5){\makebox(0,0){\footnotesize $\delta p/p_0$}}
\end{picture}
\caption{Deviations of the quasiparticle pressure from the model pressure
Eq.(\ref{modelP}) as a function of $\mu$ scaled with the free pressure $p_0$: 
shown is HTL model data for $T=T_c$ (triangles),
$T=1.1 T_c$ (stars) and $T=1.3 T_c$ (boxes). For higher
temperatures, the deviations are even smaller than those for $T=1.3 T_c$. 
The result for
$T=T_c$ reaches to a maximum of about $8$\%. }
\label{fig:scaling}
\end{center}
\end{figure}

A measure of how well Eq.(\ref{modelP}) describes
the quasiparticle result for the pressure can be obtained by considering
the quantity $\delta p=p(T,\mu)-p_{\rm m}(T,\mu)$, shown in 
Fig.~\ref{fig:scaling}: in the region considered ($\mu<10 T_c$)
the biggest deviations occur for temperatures near $T_c$; at $T=T_c$ 
there is a maximum deviation of $8$\% near $\mu\simeq 3.5\ T_c$, while 
for temperatures $T>1.3\ T_c$ the deviations stay below $3\times 10^{-3}$.

In fact, this agrees with the result of \textsc{Fodor} and \textsc{Katz}
\cite{Fodor:2002km},
who found that a scaling relation Eq.(\ref{modelP}) without 
the $\mu^4$ coefficient works rather well for small $\mu\lesssim T_c$.

\subsection{Lines of constant pressure}

\begin{figure}
\begin{center}
\includegraphics[width=0.6\linewidth]{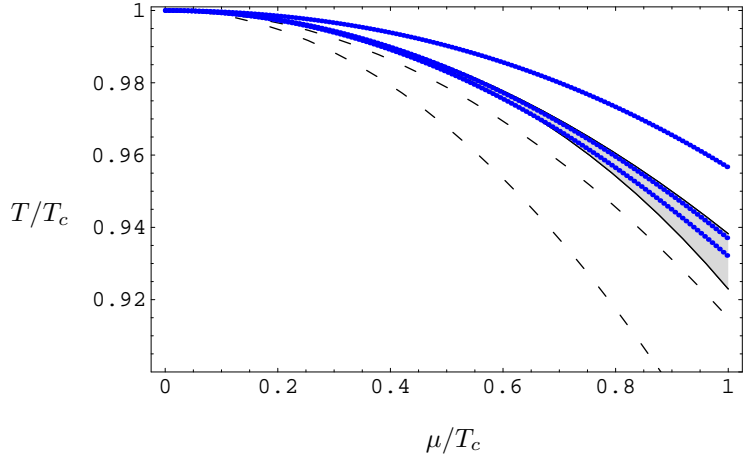}
\setlength{\unitlength}{1cm}
\begin{picture}(6,0)
\put(3.5,0){\makebox(0,0){\footnotesize $\mu/T_c$}}
\put(-2,3){\makebox(0,0){\footnotesize $T/T_c$}}
\end{picture}
\caption{%
Lines of constant pressure: NLA models for $N_{f}=2$
from $c_{\Lambda}=4$, 
$c_{\Lambda}=1$ and $c_{\Lambda}=1/4$ (dotted lines from lowest to highest);
lattice data for $N_{f}=2+1$ \cite{Fodor:2002km} (light gray band)
and $N_{f}=2$ \cite{Allton:2002zi} (long dashed-lines).}
\label{fig:constP}
\end{center}
\end{figure}

Once the data for the pressure at finite chemical potential has been 
calculated, it is also straightforward to extract lines of constant
pressure $p(T,\mu)=p(T_0,0)$. Clearly, 
the line $p(T,\mu)=p(T_c,0)$ is of special interest, 
since it is believed that close to this line the transition
from the hadronic to the quark-gluon plasma phase is occurring. For small
$\mu$, a quantity that can be compared to recent lattice results
is the slope $T_c \frac{dT}{d\mu^2}$
of this constant pressure line, which is readily calculated for the 
quasiparticle models:
\begin{center}
\begin{tabular}[c]{|c|c|c|c|c|c|}
\hline
& simple &HTL & $c_{\Lambda}=4$ & $c_{\Lambda}=1$ & $c_{\Lambda}=1/4$ \\
\hline
$T_c \frac{dT}{d\mu^2}$ &-0.0634(1) & -0.06818(8) & -0.06810(6) 
& -0.06329(34) & -0.041(9) \\
\hline
\end{tabular}\\
\end{center}
One finds that -- as anticipated -- 
the simple and HTL quasiparticle models as well as the
NLA models for $c_{\Lambda}>1$ give results that lie rather close to each
other whereas for $c_{\Lambda}=1/4$ the NLA model result is already somewhat
off the other values. In Fig.~\ref{fig:constP} the quasiparticle model
results are compared to recent lattice data for $2+1$ flavors 
\cite{Fodor:2002km} and $2$ flavors \cite{Allton:2002zi}. One can see
that the band for the $2+1$ flavor lattice calculation (resulting from
fitting the data with a second-order (upper limit) and fourth-order
(lower limit) polynomial in $\mu$ as proposed in \cite{Szabo:2003kg})
is in very good agreement with the simple, HTL, and NLA $c_{\Lambda}>1$
calculations, whereas for the NLA $c_{\Lambda}=1/4$ model the slope
is somewhat flatter. 
The lattice study for $2$ flavors (dashed lines in 
Fig.~\ref{fig:constP}) predicts a slope of $T_c \frac{dT}{d\mu^2}=-0.107(22)$, 
which is significantly steeper.

Although there is no perfect match between 
all the various lattice and quasiparticle
model results for the constant pressure slope, the deviations are small 
enough that there is overall consistency between the different approaches.
Indeed, it is plausible that the remaining disagreement is due
to the fact that none of the lattice data have been rigorously extrapolated
to the continuum limit, so that both the data and the fits of the
quasiparticle models are still likely to change somewhat when this will
be done eventually.

\section{Large chemical potential}

Extending the lines of constant pressure from the different 
models to very small temperatures one 
obtains a crude estimate of the deconfinement transition line in
this region of phase space (assuming that color superconductivity has
only a minor effect, as argued above).
Denoting with $\mu_{c}$ 
the chemical potential where (for vanishing temperature) the pressure 
at vanishing temperature
equals the pressure at $\mu=0$, $p(0,\mu_{c})=p(T_{c},0)$, one finds
(assuming $T_{c}=172$ MeV for easier comparison)
\begin{center}
\begin{tabular}[c]{|c|c|c|c|c|c|}
\hline
& simple & HTL & $c_{\Lambda}=4$ & $c_{\Lambda}=1$ & $c_{\Lambda}=1/4$ \\
\hline
$\mu_{c}$& 548 MeV &533 MeV& 536 MeV & 558 MeV & 584 MeV \\
\hline
$\mu_{0}$& 525 MeV &509 MeV& 511 MeV & 537 MeV & 567 MeV \\
\hline
\end{tabular}\\
\end{center}
Here $\mu_0$ denotes the value of $\mu$ where the pressure
vanishes, which may be taken as a definite lower bound for the
critical chemical potential within the respective models.

In general, these results are in agreement with the estimates 
for $\mu_{c}$ from Refs.~\cite{Szabo:2003kg,Peshier:2002ww}; for the QP models
considered, the lowest and highest results for $\mu_{c}$ are obtained
for the HTL model and the NLA model with $c_{\Lambda}=1/4$, respectively,
while the $\mu_{c}$ of the simple QP model lies between the  
NLA $c_{\Lambda}=4$ and $c_{\Lambda}=1$ model values.
However, even the lowest result for $\mu_{c}$ turns out to exceed
the value for the critical chemical potential expected in
Ref.~\cite{Fraga:2001id,Fraga:2001xc}. 

\begin{figure}
\begin{minipage}[t]{.48\linewidth}
\includegraphics[width=\linewidth]{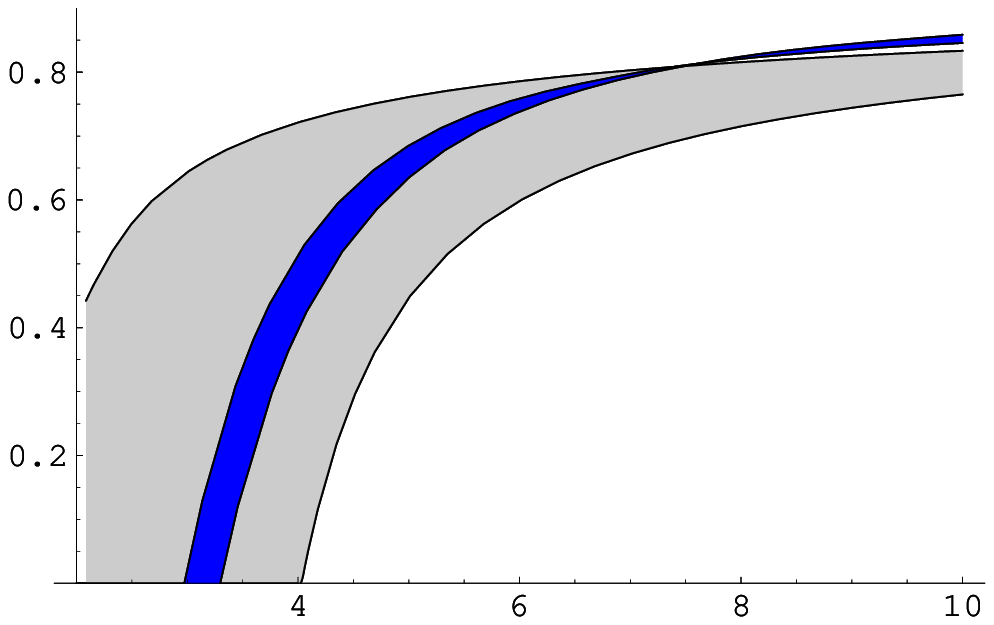}
\setlength{\unitlength}{1cm}
\begin{picture}(6,0)
\put(4,0){\makebox(0,0){\footnotesize $\mu/T_{c}$}}
\put(-0.3,2.8){\makebox(0,0){\footnotesize $\delta p/p_0$}}
\end{picture}
\caption{The pressure at vanishing $T$: shown are perturbative
results (light gray band) and NLA models for $c_{\Lambda}$ from $1/4$
to $4$ (dark band).}
\label{fig:PT0comp1}
\end{minipage}\hfill
\begin{minipage}[t]{.48\linewidth}
\includegraphics[width=\linewidth]{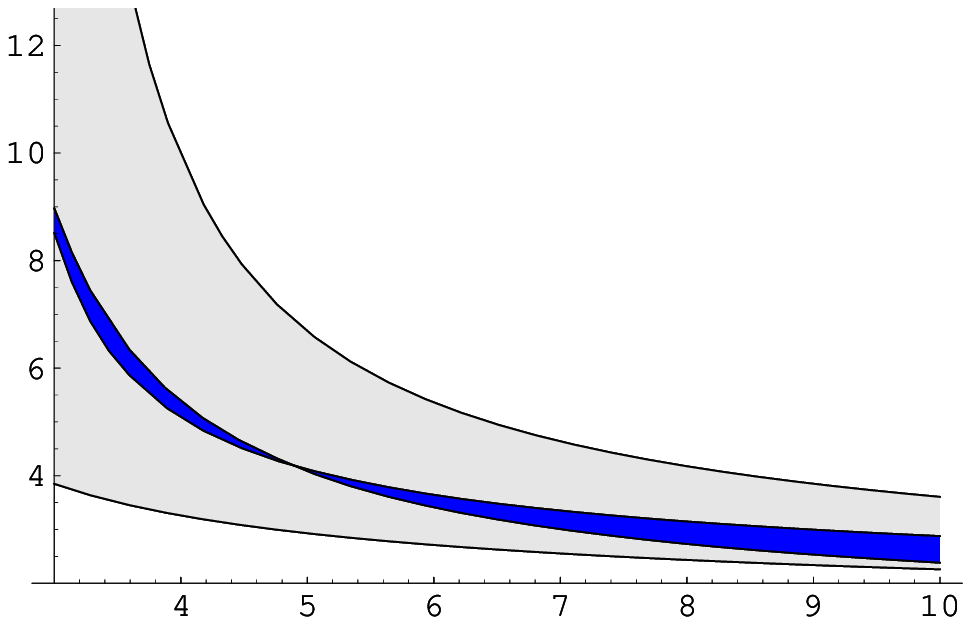}
\setlength{\unitlength}{1cm}
\begin{picture}(6,0)
\put(4,0){\makebox(0,0){\footnotesize $\mu/T_{c}$}}
\put(0,2.5){\makebox(0,0){\footnotesize \begin{rotate}{90}
$4 \pi \alpha_s$ \end{rotate}}}
\end{picture}
\caption{Effective coupling from NLA models with $c_{\Lambda}=4$ to $1/4$
(dark band) and perturbative 2-loop running coupling (light gray band).} 
\label{fig:G2T0comp}
\end{minipage}\hfill
\end{figure}

The quasiparticle results for the pressure at $T/T_{c}\simeq 0.01$
has been compared to the result of the perturbative pressure at vanishing
temperature in Fig.~\ref{fig:PT0comp1}. The latter is originally
due to \textsc{Freedman, McLerran} and \textsc{Baluni} 
\cite{Freedman:1977ub,Baluni:1978mk} but has been recalculated more
precisely by \textsc{Vuorinen} \cite{Vuorinen:2003fs}, 
\bqa
p(T=0,\mu)&=&\frac{\mu^4 N_f}{4 \pi^2}\left\{1-2\frac{\alpha_s}{\pi}-\left[%
18-11 \ln 2-0.53583 N_f +N_f \ln \frac{N_f \alpha_s}{\pi}\right.\right.%
\nonumber \\
&&\left.\left.+(11-\frac{2}{3}%
N_f)\ln{\frac{\bar{\mu}}{\mu}}\right]\left(\frac{\alpha_s}{\pi}\right)^2
+O(\alpha_s^3 \ln \alpha_s)\right\},
\eqa
where in this \cite{Kraemmer:2003gd} 
form $\bar{\mu}$ again denotes the renormalization scale
in the $\overline{\hbox{MS}}$ renormalization scheme.
The perturbative result has been evaluated using the standard 
two-loop running coupling Eq.(\ref{alpha2loop}) with
$\Lambda_{\overline{\hbox{\scriptsize MS}}}={T_{c}}/{0.49}$ 
\cite{Gupta:2000hr} and renormalization scale varied from $\mu$ to $3\mu$,
as has been considered in Ref.~\cite{Fraga:2001id}.
It can be seen in Fig.~\ref{fig:PT0comp1} that while in general 
the quasiparticle model and perturbative results 
are consistent, the variation in 
the latter (due to the uncertainty in the renormalization
scale $\bar{\mu}$) is much larger than the narrow band obtained 
from the evaluation of the different quasiparticle models.
The perturbative two-loop running coupling at vanishing temperature itself 
is shown in Fig.~\ref{fig:G2T0comp} as a function of 
chemical potential, compared to 
the result of the effective coupling for the quasiparticle models.

\subsection{EOS for cold deconfined matter}

By calculating the number density at $T/T_{c}\simeq 0.01$ and using
the results for the pressure one obtains 
an equation of state for cold deconfined matter. As is the case for 
the simple quasiparticle model \cite{Peshier:2002ww}, one finds that also for 
the HTL and NLA models the energy density $\epsilon$ is well fitted
by the linear relation 
$$
\epsilon(p)=4 \tilde{B}+ \alpha p
$$
with
\begin{center}
\begin{tabular}[c]{|c|c|c|c|c|c|}
\hline
& simple & HTL & $c_{\Lambda}=4$ & $c_{\Lambda}=1$ & $c_{\Lambda}=1/4$ \\
\hline
$4 \tilde{B}/T_{c}^4$& 12.3 (6) & 11.1(8)&12.3(8)& 14.7(9) & 19.2(1.6) \\
\hline
$\alpha $ & 3.26(5) &3.23(5)&3.22(4)& 3.22(4)& 3.17(4)\\
\hline
\end{tabular}\\
\end{center}
The value of $\alpha\simeq 3.2$ for $N_{f}=2$ 
seems to be model independent, in contrast to the bag 
constant $\tilde{B}^{1/4}$, which varies between $314$ and $360$ MeV.

\section{Application 1: Quark stars}

\begin{figure}
\begin{center}
\includegraphics[width=0.6\linewidth]{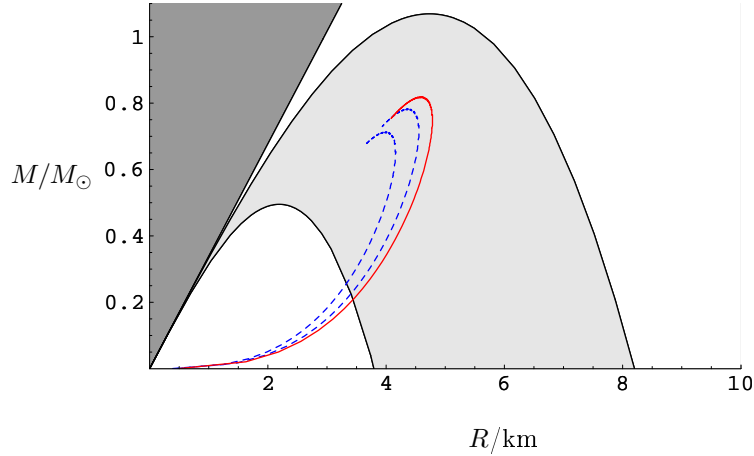}
\setlength{\unitlength}{1cm}
\begin{picture}(6,0)
\put(4,0){\makebox(0,0){\footnotesize $R$/km}}
\put(-2,3.5){\makebox(0,0){\footnotesize $M/M_{\odot}$}}
\end{picture}
\caption{%
Mass-radius relations of non-rotating quark-stars: shown are the results using
the EOS from the HTL (red full line) and NLA $c_{\Lambda}=1/4$ and 
$c_{\Lambda}=1$ models (blue dashed lines). 
Also shown are the 
inferred mass-radius relations of RX J1856.5-3754 from 
Ref.~\cite{Drake:2002bj,Kohri:2002hf} (light gray region), 
a possible candidate for a quark star. The dark shaded
region is excluded by the condition $R>2 G M$.}
\label{fig:quarkstar}
\end{center}
\end{figure}

As a first application I will use the above equation of state for 
cold deconfined matter to determine the mass-radius relations of non-rotating
quark-stars. If these stars exist, they would differ from ordinary
neutrons stars by being more compact and therefore 
considerably smaller than current neutron-star models allow 
\cite{Fraga:2001xc}, as I will show in the following.

Starting from Einstein's equations
\beq
R_{\mu \nu}-\frac{1}{2} g_{\mu \nu} R = 8 \pi T_{\mu \nu}(\epsilon(p),p)
\eeq
where $R_{\mu \nu},g_{\mu \nu}$ and $R$ denote the Ricci and metric tensor and 
Ricci scalar, respectively. Using the Schwarzschild solution for the metric
of a spherically symmetric and static star one readily finds the familiar
Tolman-Oppenheimer-Volkoff equations 
\cite{Oppenheimer:1939ne,Heiselberg:1999mq,Andersen:2002jz},
\beq
\frac{dp(r)}{d r}=%
-\frac{G}{r^2} \frac{\left[\epsilon(r)+p(r)\right]\left[M(r)+4 \pi r^3 p(r)%
\right]}{1-\frac{2 G M(r)}{r}},
\label{TOV}
\eeq
where
\beq
M(r)=4 \pi \int_{0}^{r} \epsilon(r) r^2 dr
\eeq
and $G$ is Newton's constant. The equations are easily solved by specifying
a central pressure or density and calculating the size and mass of the 
corresponding star using the equations of state from the previous section.
The results are shown in Fig.~\ref{fig:quarkstar}, where it can 
be seen that the quasiparticle model EOS allows for quark-stars
with masses of $\sim 0.8 M_{\odot}$ and radii of less than 5 km
(which is similar to the results found in 
Refs.~\cite{Fraga:2001xc,Peshier:2002ww,Blaschke:1998hy}). Current
neutron star models -- which are consistent with the pulsar data
\cite{Fraga:2001xc} -- predict a maximum mass in the range of $1.4$ to
$2$ solar masses with radii ranging from $10$ to $15$ km, so if pure
quark-stars exist, they should in principle be distinguishable
from ordinary neutron stars by observation. 
Interestingly, several observations of neutron star
candidates possessing properties that are in apparent contradiction 
with current neutron star models exist, most prominently the 
object RX J1856.5-3754, whose inferred mass-radius relation 
(see Ref.~\cite{Drake:2002bj,Kohri:2002hf}) is also shown in
Fig.~\ref{fig:quarkstar}. The exciting possibility that with the observation
of RX J1856.5-36754 one has a direct confirmation of the existence 
of quark-stars is still a matter of ongoing debate, as there are
propositions claiming that the available data can be 
described by an ordinary neutron star; future observations are therefore
needed to settle the issue of the exact nature of RX J1856.5-36754.

Finally, it should be kept in mind that the outermost
layers of such a possible quark star consist of hadronic matter, giving
rise to what is usually named a ``hybrid star''. The details of the 
star structure will depend sensitively
on the hadronic equation of state \cite{Peshier:2002ww}, which 
unfortunately does not match on naturally to the EOS of the quark-gluon
plasma phase calculated above \cite{Fraga:2001xc}. Therefore,
to make clear predictions for the phenomenology of compact stars one
has to find a way to describe the intermediate range of densities in
the equation of state, possibly by considering effective field theory
models \cite{Fraga:2003uh}.

\section{Application 2: Expansion of the quark-gluon plasma}

As a second application, 
I shall now describe the longitudinal expansion of the matter produced
in a central collision of identical nuclei, using Bjorken's relativistic
hydrodynamic model and the quasiparticle EOS for the 
quark-gluon plasma calculated above.
The starting point is the energy-momentum tensor of a relativistic
fluid with local energy density $\epsilon(x)$ and pressure $p(x)$,
\beq
T_{\mu \nu}=(\epsilon+p) U_{\mu} U_{\nu}-g_{\mu \nu} p,
\label{EMtensor}
\eeq
where $U^{\mu}$ is the four-velocity of the fluid obeying $U^2=1$.
Energy, momentum as well as the baryon number $N^{\mu}=n U^{\mu}$ 
of the system should be conserved, therefore
\beq
\partial^{\mu} T_{\mu \nu}=0, \quad %
\partial^{\mu} N_{\mu}=0.
\label{conseveq}
\eeq
Introducing proper time $\tau$ and space-time rapidity $\eta$ defined as
\beq
\tau=\sqrt{t^2-z^2},\quad \eta=\frac{1}{2} \ln{\frac{t+z}{t-z}}
\eeq
and neglecting the effect of the transverse expansion as well as
assuming that the properties of the fluid are invariant under Lorentz
boosts in the $z$-direction this implies
\cite{Bjorken:1983qr}
\beq
\epsilon=\epsilon(\tau),\quad p=p(\tau),\quad T=T(\tau),\quad
U_{\mu}=\frac{1}{\tau}\left(t,0,0,z\right).
\eeq
With the use of the expressions
\beq
\partial_{\mu} \tau = U_{\mu},\quad \partial_{\nu} U_{\mu}=\frac{1}{\tau}%
(\tilde{g}_{\mu \nu}-U_{\mu} U_{\nu})
\eeq
and $\tilde{g}_{\mu \nu}=(1,0,0,1)$, the conservation equations
can be brought into the form
\beq
U^{\mu} \partial_{\mu} \left(\frac{s}{n}\right)=0,\quad %
\partial_{\mu} N^{\mu}=0,
\label{conseq}
\eeq
implying that the entropy density $s$ per number density $n$ 
is conserved by the longitudinal expansion.

\begin{figure}
\begin{center}
\includegraphics[width=0.6\linewidth]{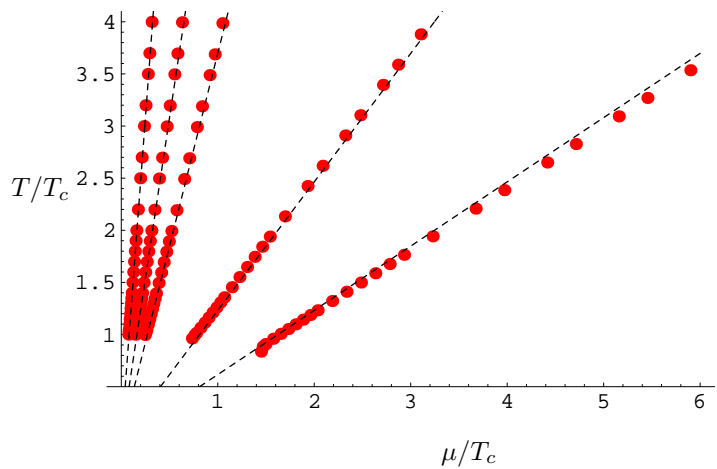}
\setlength{\unitlength}{1cm}
\begin{picture}(6,0)
\put(4,0){\makebox(0,0){\footnotesize $\mu/T_c$}}
\put(-1.7,3.5){\makebox(0,0){\footnotesize $T/T_c$}}
\end{picture}
\caption{%
Curves of constant $s/n=100,50,30,10,5$ 
for the HTL model (dots). Also shown are the 
tree-level results (dashed straight lines) for the corresponding
values of $s/n$.}
\label{fig:snconst}
\end{center}
\end{figure}

Using the quasiparticle model EOS one can calculate the curves of
constant $s/n$ in the $T,\mu$-plane, shown in Fig.~\ref{fig:snconst}.
Surprisingly, the curves are very well approximated by straight lines
having slopes consistent with the tree-level approximation
\beq
\frac{s}{n}\simeq \frac{T^3 \left(\frac{32\pi^2}{45}+\frac{28 \pi^2}{30}%
\right)}{2 \mu T^2}=\frac{T}{\mu} \frac{74 \pi^2}{90},
\eeq
as can be seen in Fig.~\ref{fig:snconst}; only at very large $\mu$
the tree level approximation seemingly starts to break down.

Provided one has knowledge about the value of $T,\mu$ that the system 
reaches at freeze-out (e.g. by analyzing experimental 
data using thermal model descriptions of 
particle production \cite{Braun-Munzinger:2003zd}) as well as an estimate
of the initial energy density created after heavy-ion collision, the
above results allow an estimate of the initial temperature
and chemical potential.
For example, \textsc{Braun-Munzinger} {\em et al.} obtained
$(T,\mu)\simeq(174 {\rm MeV},15 {\rm MeV})$ as freeze-out values 
for the central 
Au-Au collisions at RHIC with
$\sqrt{s}=130 {\rm GeV}$, which corresponds to a value of $s_0/n_0\simeq 100$.
Assuming furthermore an initial energy density of 
$\epsilon_0=20 {\rm GeV/fm^3} \simeq 175 T_c^4$ 
\cite{Braun-Munzinger:2003zd} and
again $T_c=172 {\rm MeV}$ one obtains 
$(T_0,\mu_0)=(344 {\rm MeV},27 {\rm MeV})$ 
as initial temperature and chemical potential. Consequently, one can calculate
the time behavior of the system by e.g. integrating Eq.(\ref{conseq}),
\beq
n=\frac{n_0 \tau_0}{\tau},
\eeq
where $n_0=n(T(\mu_0),\mu_0)$ and $\tau_0$ is expected to be on the order
of $\tau_0\simeq 1 {\rm fm/c}$. In this model, one finds that the system
freeze-out time is given by $\tau_f \simeq 12 {\rm fm/c}$. It is clear 
however, that the above result is only a rough estimate 
of the time the system remains in a quark-gluon plasma phase
since one has to take into account viscous effects of the fluid as well
as the transversal expansion of the system that 
cannot be neglected at late times.

\section{Summary}

In this chapter, I have set up a possible extension of the HTL quasiparticle
model that includes the full plasmon effect, however at the price of
having to introduce another parameter. 
The values of the parameters used to fit the quasiparticle
models to the $N_f=2$ lattice data were given and it was shown 
that the resulting running
coupling turns out to be comparable to the 2-loop perturbative coupling.
I then investigated the quark-number susceptibilities at vanishing chemical
potential and showed that they agree very well with independent lattice 
data; for the HTL model, it was also shown that the higher-order derivatives
represent the trend of another independent lattice study.
Furthermore, it turned out that the quasiparticle pressure can be very
well described by these $\mu=0$ susceptibilities, even up to rather
large values of chemical potential.
Moreover, I calculated the lines of constant pressure starting from 
the critical temperature at $\mu=0$, which were shown to be compatible
with lattice studies in the validity region of the latter. Extrapolating
these results to large chemical potentials and small temperatures
I obtained an equation of state for cold dense matter which should 
represent an improvement over existing perturbative results.
As a first application, I then derived the mass-radius relationship of 
so-called quark-stars, finding results that fit nicely with the 
values inferred for RX J1856.5-3754, a possible candidate for a quark star.
As a second application, I calculated the expansion of a quark-gluon plasma
created after a heavy-ion collision in the Bjorken model, finding
that the system cools along nearly straight lines in the
$T,\mu$-diagram. Finally, I estimated the time that the system remains
in the quark-gluon plasma phase using freeze-out values of the temperature
and chemical potential inferred for Au-Au collisions at RHIC.

\chapter{Results independent of the lattice}
\label{qpmodels2-chap}

In the last chapters I used an effective coupling to describe lattice data
at vanishing chemical potential which eventually gave a 
plausible equation of state for arbitrary chemical potentials. Although
this approach seems to work very well it is somewhat unsatisfactory to
be limited to available lattice data and a phenomenological ansatz for an 
effective coupling. Therefore, I want to investigate in this chapter
how the results of the previous chapters are modified if one renounces
all lattice input safe one number, namely the ratio 
$T_c/\Lambda_{\overline{\hbox{\scriptsize MS}}}$,
which for $N_f=2$ I take to be given by $0.49$, as was used before.
Instead of the effective coupling I will use the standard two-loop running
coupling of Eq.(\ref{alpha2loop}) with $\bar{\mu}$ ranging between
$\pi T$ and $4 \pi T$ at $\mu=0$, therefore testing the theoretical error
of the model predictions. Furthermore, since the quasiparticle models 
represent an implementation of the thermodynamic quantities which is correct
at least to leading perturbative order, one is then able to make predictions
of these quantities ``uncontaminated'' by lattice artifacts using 
the machinery of the previous sections. I will then compare these predictions
to both perturbational approaches and lattice results (whenever available)
as well as those obtained in the previous chapters.

\section{Setup}

\begin{figure}
\begin{minipage}[t]{\linewidth}
\includegraphics[width=0.48\linewidth]{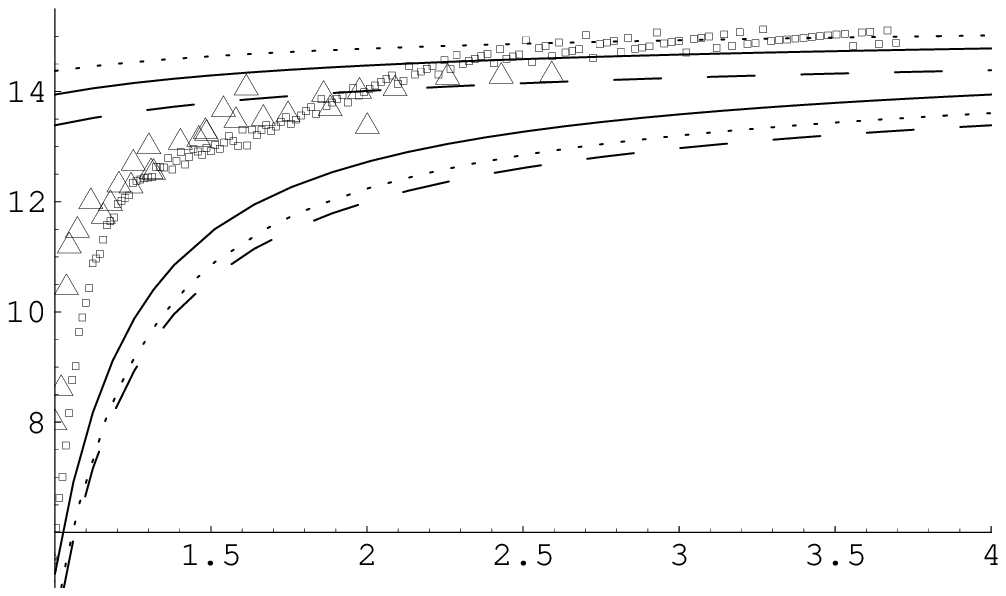}
\hspace*{0.3cm}
\includegraphics[width=0.48\linewidth]{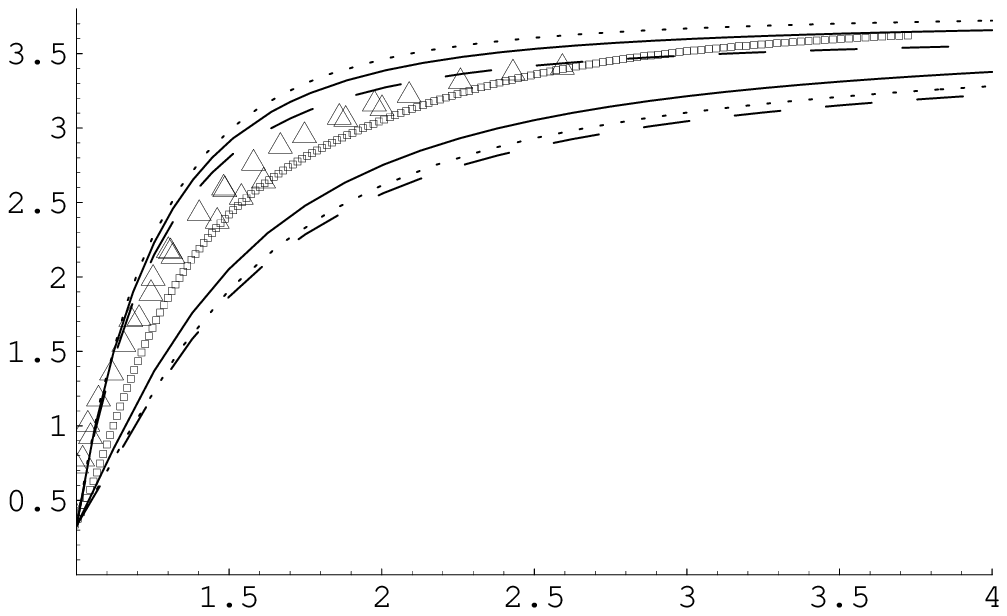}
\setlength{\unitlength}{1cm}
\begin{picture}(6,0)
\put(3.5,5){\makebox(0,0){(a)}}
\put(3.5,0){\makebox(0,0){\footnotesize $T/T_c$}}
\put(-0.5,3){\makebox(0,0){\footnotesize $s/T^3$}}
\end{picture}
\setlength{\unitlength}{1cm}
\begin{picture}(6,0)
\put(5,5){\makebox(0,0){(b)}}
\put(5,0){\makebox(0,0){\footnotesize $T/T_c$}}
\put(1,3){\makebox(0,0){\footnotesize $p/T^4$}}
\end{picture}
\end{minipage}
\caption{%
Two-flavor entropy (a) and pressure (b)
for HTL (long-dashed lines) and NLA quasiparticle models (full lines
for $c_{\Lambda}=1$, dotted lines for $c_{\Lambda}=2,\frac{1}{2}$), where
the renormalization scale has been taken to be $\bar{\mu}=4 \pi T$ and 
$\bar{\mu}=\pi T$, respectively. Also shown are $N_f=2$ lattice 
results from Refs.~\cite{AliKhan:2001ek} (triangles) and \cite{Karsch:2000ps}
(boxes), respectively.}
\label{fig:Smu0allcomp}
\end{figure}

Using the HTL and NLA quasiparticle models for the pressure and the two-loop
perturbative coupling, the expression for the entropy becomes that of
\textsc{Blaizot} {\em et.al} 
\cite{Blaizot:2000fc}. Evaluating the entropy as a function
of $T$ by using $T_c/\Lambda_{\overline{\hbox{\scriptsize MS}}}=0.49$
for $N_f=2$
(which differs from the value used in the original reference) one can 
compare the 
result to lattice data for $N_f=2$ from Ref.~\cite{AliKhan:2001ek} 
(used for the fits in the previous sections) and Ref.~\cite{Karsch:2000ps}
in Fig.~\ref{fig:Smu0allcomp}a. As can be seen, the NLA model results
are in general agreement with the lattice data (especially 
since a rigorous extrapolation of the
lattice data to the continuum limit is still missing) while the HTL model
seems to predict slightly too small values for the entropy. 

The pressure of the respective models is then given by Eq.(\ref{HTLPmodel}),
where $B_0$ is fixed in such a way 
that the pressure at $T=T_c$ is equal to the pressure
of a free gas of 3 massless flavors of pions,
\beq
p(T_c,0)/T_c^4=\frac{\pi^2}{30}\simeq 0.329.
\eeq
This completes the setup of the model. The resulting pressure at $\mu=0$
as a function of the temperature is shown in  
Fig.~\ref{fig:Smu0allcomp}b; although the model pressure is not as close
to the lattice data as in the previous chapters where a phenomenological
fit of the coupling was used, it is still remarkable that the resulting
band turns out to cover the lattice results and not much more while
the lattice input has been minimized to one number.

\section{Results}

\begin{figure}
\begin{center}
\includegraphics[width=0.6\linewidth]{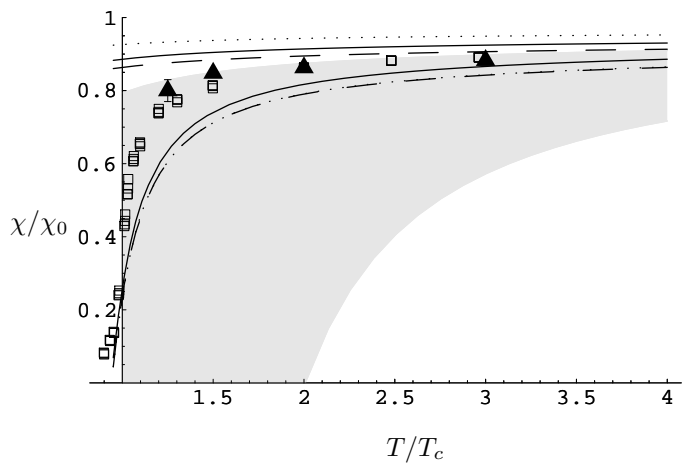}
\setlength{\unitlength}{1cm}
\begin{picture}(6,0)
\put(3.5,0){\makebox(0,0){\footnotesize $T/T_c$}}
\put(-1.5,3){\makebox(0,0){\footnotesize $\chi/\chi_0$}}
\end{picture}
\caption{%
Two-flavor susceptibilities 
for HTL (long-dashed lines) and NLA quasiparticle models (full lines
for $c_{\Lambda}=1$, dotted lines for $c_{\Lambda}=4,\frac{1}{4}$), where
the renormalization scale has been taken to be $\bar{\mu}=4 \pi T$ and 
$\bar{\mu}=\pi T$, respectively. Also shown are $N_f=2$ lattice 
results from Ref.~\cite{Gavai:2001ie} (full triangles) and\cite{Fodor:2002km}
(open boxes), respectively. The gray band corresponds to results from 
strict perturbation theory to order $\alpha_s^{3}\ln{\alpha}$ 
\cite{Vuorinen:2002ue} (see text for details).}
\label{fig:QP2Sus}
\end{center}
\end{figure}

The characteristic curves again take shapes resembling ellipses as was 
the case in the previous chapters. Calculating the quark-number 
susceptibilities one finds the results shown in Fig.~\ref{fig:QP2Sus}.
As can be seen, the agreement with lattice data is still fairly good
at high temperatures, whereas at temperatures comparable to $T_c$ the 
variations from varying $\bar{\mu}$ by a factor of $2$ around $2 \pi T$ 
get rather large. Note however that the results from the NLA models
for the susceptibilities clearly differ from those obtained originally
by \textsc{Blaizot}%
 {\em et al.} \cite{Blaizot:2001vr}. The discrepancy is due to
the fact that here the fermionic contributions were 
divided into soft and hard part similar to the gluonic contributions while
in the original reference the fermions have been treated differently.
As a consequence of treating the fermionic sector as in 
Ref.~\cite{Rebhan:2003wn}, the NLA model susceptibilities for $N_f=2$
lie only slightly higher than the HTL results, as shown
in Fig.~\ref{fig:QP2Sus}. Therefore, the NLA results are in good agreement
with existing lattice data \cite{Gavai:2001ie,Fodor:2002km} while the
original NLO results did not have any overlap with the data.
I also show the results for the quark-number susceptibilities from strict
perturbation theory to order $\alpha_s^{3}\ln{\alpha_s}$ by
\textsc{Vuorinen} \cite{Vuorinen:2002ue,Vuopriv} 
in Fig.~\ref{fig:QP2Sus}; the 
band shown corresponds to varying $\bar{\mu}=\pi T$ to  $\bar{\mu}=4\pi T$ while
the unknown $\alpha_s^{3}$ coefficient is taken close to zero 
\cite{Vuopriv}.

When calculating the slope of the lines of constant pressure one obtains
$T_c \frac{d T}{d \mu^2}\simeq-0.055/-0.065$ for the HTL quasiparticle 
model for $\bar{\mu}=\pi T$ (first value) and $\bar{\mu}=4 \pi T$
(second value). The variations for the NLA model are slightly larger 
since one also has to consider the variation in the coefficient 
$c_{\Lambda}$; one finds $T_c \frac{d T}{d \mu^2}\simeq-0.052/-0.065$
for the NLA model when varying $\bar{\mu}$ as above and $c_{\Lambda}$
by a factor of $2$ around $1$. Therefore, also the slopes of constant
pressure turn out to be very similar to what was found in chapter 
\ref{qpmodels-chap},
indicating that the agreement between these results and lattice calculations
for $N_f=2$ is a quite robust prediction of the quasiparticle models
and in fact nearly independent of the fit data used in the previous chapters.

Extrapolating the $N_f=2$ quasiparticle results to the region of small
temperatures one can calculate an estimate for the critical density $\mu_0$
where $p(0,\mu_0)=0$. Again varying $\bar{\mu}$ 
one finds for the HTL model $\mu_0\simeq3.49 T_c$ and $\mu_0=2.93T_c$ 
for $\bar{\mu}=\pi T$ and $\bar{\mu}=4 \pi T$, respectively. For the NLA model
the results turn out to be very similar, with the only difference being
that the upper prediction of $\mu_0$ 
is slightly increased to $\mu_0\simeq3.52 T_c$.
Comparing these to the results from the fitted quasiparticle models one
finds that for $\bar{\mu}=\pi T$ the estimates for $\mu_0$ are
noticeably higher in the models not fitted to lattice data at $\mu=0$,
while for larger values of $\bar{\mu}$ the results become very similar
(i.e. for $\bar{\mu}=4 \pi T$ the HTL model result is only about $2$\%
lower than in chapter \ref{qpmodels-chap}).

\begin{figure}
\begin{center}
\begin{minipage}[t]{.48\linewidth}
\includegraphics[width=.9\linewidth]{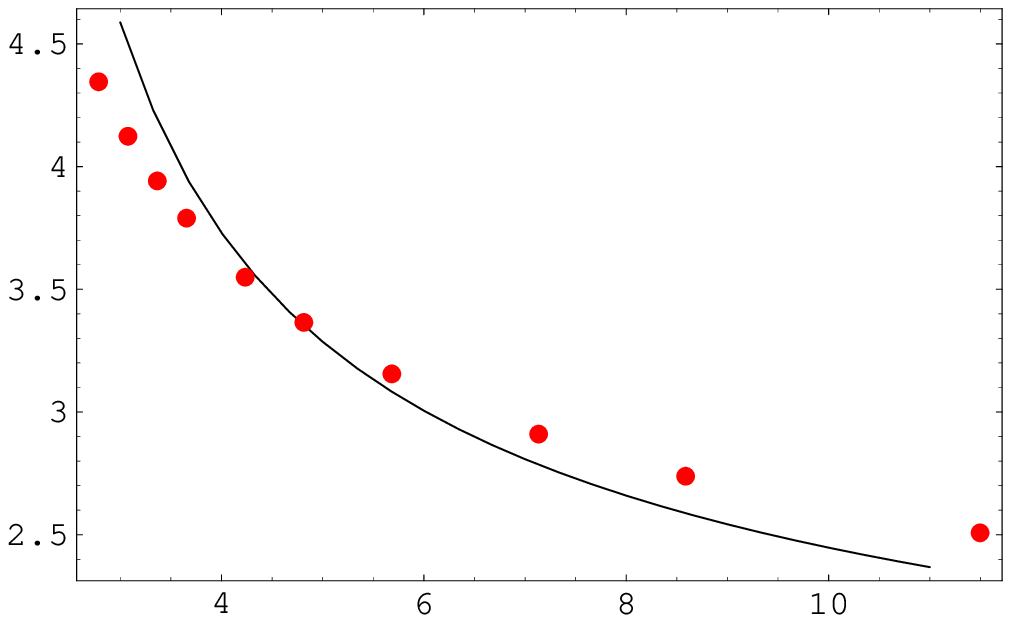}
\setlength{\unitlength}{1cm}
\begin{picture}(5,0)
\put(3.5,0){\makebox(0,0){\footnotesize $\mu/T_c$}}
\put(3.5,4.7){\makebox(0,0){\footnotesize (a)}}
\put(-0.5,2.5){\makebox(0,0){$4 \pi \alpha_s$}}
\end{picture}
\end{minipage}\hfill
\begin{minipage}[t]{.48\linewidth}
\includegraphics[width=.9\linewidth]{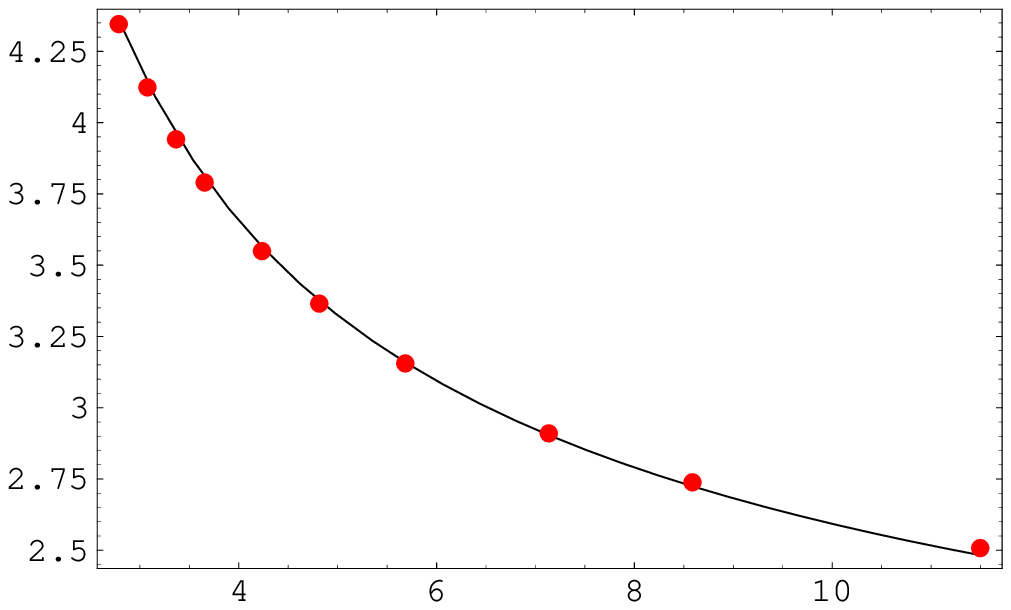}
\setlength{\unitlength}{1cm}
\begin{picture}(5,0)
\put(3.5,0){\makebox(0,0){\footnotesize $\mu/T_c$}}
\put(3.5,4.7){\makebox(0,0){\footnotesize (b)}}
\put(-0.5,2.5){\makebox(0,0){$4 \pi \alpha_s$}}
\end{picture}
\end{minipage}
\end{center}
\caption{Two-loop (a) and one loop (b) fits (full line) to the quasiparticle
model coupling at $T=0$ (points).}
\label{fig:GTOLC}
\end{figure}

Finally, it is also interesting to investigate closer what form the 
coupling constant takes in the region $T\ll \mu$. For example, one can try to
describe $\alpha_s(T=0,\mu)$ by the standard two-loop result with
$\bar{\mu}=y \mu$, where $y$ is to be determined by a least-square
fit. This approach turns out to be not too successful, as can be seen
in the exemplary plot of Fig.~\ref{fig:GTOLC}a, where such a fit has been
tried on the HTL model result. Remarkably, the situation is very different
when using the {\em one-loop} form for the running coupling 
Eq.(\ref{alpha1loop}) instead
of the two-loop result. For the one-loop coupling,
the fits with $\bar{\mu}=y \mu$ typically look like the one shown
in Fig.~\ref{fig:GTOLC}b, both for the HTL model as well as for the NLA 
models. Moreover, it turns out that the fit constant $c$ is always  
proportional to the scale used at $\mu=0$, i.e. when the 
renormalization scale was $\bar{\mu}=4 \pi T$ at $\mu=0$ the scale that
fits the coupling best at $T=0$ is given by 
$\bar{\mu}\simeq 4 \pi \mu \, 0.37$ while for an original scale 
of $\bar{\mu}=\pi T$ one finds $\bar{\mu}\simeq \pi \mu \, 0.33$
at $T=0$. 

\subsection{Notes on changing 
$T_c/\Lambda_{\overline{\hbox{\scriptsize MS}}}$}

As stressed before, the above results depend only on one external parameter,
namely the value of $T_c/\Lambda_{\overline{\hbox{\scriptsize MS}}}$.
It is therefore interesting to investigate how the results derived
above will change when this parameter turns out to be smaller or bigger
than $0.49$. To this effect it is sufficient to see that the running
coupling only depends on the scale through 
$\bar{\mu}/\Lambda_{\overline{\hbox{\scriptsize MS}}}$
and accordingly
\beq
\frac{\bar{\mu}}{\Lambda_{\overline{\hbox{\scriptsize MS}}}}=
\frac{\bar{\mu}}{T_c} \frac{T_c}{\Lambda_{\overline{\hbox{\scriptsize MS}}}}.
\eeq
Increasing the value of $T_c/\Lambda_{\overline{\hbox{\scriptsize MS}}}$
therefore amounts to increasing the ratio renormalization scale
over critical temperature. As an effect,
the entropy and the susceptibilities will rise, whereas the slopes of
constant pressure will become more negative. Finally, the estimate
for the critical density will also take smaller values.

\section{Summary}

In this chapter I investigated the results from various quasiparticle
models when using only one number, namely the value of
$T_c/\Lambda_{\overline{\hbox{\scriptsize MS}}}$ as an input. It turns
out that the pressure and the quark number susceptibilities 
at $\mu=0$ are still described fairly well. The results for the 
slopes of constant pressure are close to what has been found in the
previous chapters as long as $\bar{\mu}$ is not much smaller than
$2 \pi T$, as is the case for the estimates of the critical density
at $T=0$. Furthermore, it turns out that 
the resulting coupling at $T=0$ can be very well described by 
the standard one-loop running coupling. I conclude by pointing out
how the results derived in this chapter have to be modified when
$T_c/\Lambda_{\overline{\hbox{\scriptsize MS}}}$ changes from 
the value adopted here.

\chapter{The anisotropic quark-gluon plasma}
\label{asi-chap}

In the previous chapters I have limited myself to the consideration of 
isotropic systems. However, as pointed out in the introduction,
the only system above the deconfinement transition one can (probably) study
on earth, namely that of a fireball following 
an ultrarelativistic heavy-ion collision, is 
rather anisotropic in nature. 
Therefore, it is interesting to investigate which effects 
the presence of an anisotropy has on the dynamics of the system
and  which differences one can expect when comparing to the usually studied
isotropic case.
Note that 
I will consider only systems which are anisotropic in {\em momentum space},
therefore taking a snapshot of the system at some short time after 
the collision and before the system had time to convert the anisotropy from
momentum space to configuration space. Although eventually one would require
a full treatment including anisotropies in configuration space, 
I will show that one can learn a lot about anisotropic systems by considering
momentum anisotropies alone; moreover, a treatment of configuration space
anisotropies would involve a full non-equilibrium quantum field theoretical
description of the system, which despite recent progress is not available
yet.

In this chapter, I will therefore analyze the collective modes of 
high-temperature QCD in the case when there is an anisotropy in the 
momentum-space distribution function. I will present results (published
in Ref.~\cite{Romatschke:2003ms})
for a class of anisotropic distribution functions which can be obtained
by stretching or squeezing an isotropic distribution function along one
direction, thereby preserving a cylindrical symmetry 
in momentum space.

\section{Gluon self-energy in an anisotropic system}

Generalizing the classification of quasiparticles in
an isotropic quark-gluon plasma in
chapter \ref{peshier-chap} one first derives
the HTL resummed gluon self-energy in an anisotropic system within 
semi-classical transport theory \cite{Mrowczynski:1993qm,Mrowczynski:1994xv,%
Mrowczynski:1997vh}, which has been shown to be equivalent to
the HTL diagrammatic approach \cite{Mrowczynski:2000ed}. Within semi-classical
transport theory partons are described by their phase-space densities and their
time evolution is given by Vlasov-type transport equations 
\cite{Elze:1989un,Blaizot:2001nr}. Concentrating
on the physics at the soft scale, $k \sim \alpha_s^{1/2} T \ll T$, 
the magnitude of the field fluctuations at this scale
is $A \sim \alpha_s^{1/4} T$ and derivatives are of the scale 
$\partial_x \sim \alpha_s^{1/2} T$. With this power-counting 
a systematic truncation of the terms contributing to the transport 
equations for soft momenta can be realized.

At leading order in the coupling constant the color current 
$J^\mu$ induced by a soft gauge field $A^\mu$ with four-momentum 
$K=(\omega,{\bf k})$ can be obtained by performing a covariant 
gradient expansion of the quark and gluon Wigner functions in mean-field 
approximation.
The result is
\beq
J^{\mu,a}_{\rm ind}(X)=(4 \pi \alpha_s)^{1/2} \int \frac{d^3 p}{(2\pi)^3} V^{\mu}%
 (2 N \delta n^a(p
,X)+N_f ( \delta f_{+}^{a}(p,X)-\delta f_{-}^{a}(p,X)) \; ,
\label{current}
\eeq
where $V^{\mu} = (1,{\bf k}/\omega)$ is the velocity of the hard plasma 
constituents,
$\delta n^a(p,X)$ is the fluctuating part of the gluon density,
and $\delta f^a_{+}(p,X)$ and $\delta f^a_{-}(p,X)$ 
are the fluctuating parts of the quark and anti-quark densities, 
respectively. Note that 
$\delta n^a$ transforms as a vector in the adjoint representation 
($\delta n \equiv \delta n^a T^a$) and $\delta f^a_{\pm}$ transforms 
as a vector in the fundamental representation 
($\delta f_{\pm} \equiv \delta f^a_{\pm} t^a$).

Neglecting the collision terms which only enter at subleading order, 
the quark and gluon density matrices above satisfy the following 
transport equations:
\bqa
\label{qvlasov}
\left[ V \cdot D_{X}, \delta f_{\pm}(p,X) \right] &=& 
  \mp (4 \pi \alpha_s)^{1/2} V_{\mu} F^{\mu \nu}(p,X) \partial_{\nu} f_{\pm}({\bf p}) \; ,\\
\left[ V \cdot D_{X}, \delta n(p,X) \right] &=& 
  - (4 \pi \alpha_s)^{1/2} V_{\mu} F^{\mu \nu}(p,X) \partial_{\nu} n({\bf p}) \; ,
\label{gvlasov}
\eqa
where $D_X = \partial_X + i (4 \pi \alpha_s)^{1/2} A(X)$ is the covariant 
derivative and $n({\bf p})$ and $f_{\pm}({\bf p})$ are the initial anisotropic
gluon and quark distribution functions (which in the isotropic limit
would correspond to the Bose-Einstein and Fermi-Dirac distribution functions,
respectively).

Solving the transport equations (\ref{qvlasov}) and (\ref{gvlasov}) 
for the fluctuations 
$\delta n$ and  $\delta f_\pm$ gives the induced current 
via Eq.(\ref{current}),
\bqa
J^{\mu}_{\rm ind}(X)&=&%
4 \pi \alpha_s \int \frac{d^3 p}{(2\pi)^3} V^{\mu} V^{\alpha} 
        \partial^{\beta}_{(p)} h({\bf p}) \nonumber \\
&&\times \int d\tau \, 
        U(X,X-V\tau) F_{\alpha \beta}(X-V\tau) U(X-V\tau,X) \; ,
\eqa
where $U(X,Y)$ is a gauge parallel transporter defined by the 
path-ordered integral
\beq
U(X,Y) = \mathcal{P} \, {\rm exp}\left[ - i (4 \pi \alpha_s)^{1/2}%
 \int_X^Y d Z_\mu A^\mu(Z) \right] \; ,
\eeq
$F_{\alpha \beta} = \partial_\alpha A_\beta - \partial_\beta A_\alpha%
 - i (4 \pi \alpha_s)^{1/2} [A_\mu,A_\nu]$ 
is the gluon field strength tensor, and
\beq
h({\bf p}) = 2 N n({\bf p})+ N_{f} (f_{+}({\bf p})+f_{-}({\bf p})) \; 
\eeq
is a particular combination of the gluon and quark distribution functions.
Neglecting terms of subleading order in $\alpha_s$ (implying $U\rightarrow1$ 
and $F_{\alpha \beta} 
\rightarrow \partial_\alpha A_\beta - \partial_\beta A_\alpha$) 
and performing a Fourier 
transform of the induced current to momentum space one obtains
\beq
J^{\mu}_{\rm ind}(K)=4 \pi \alpha_s \int \frac{d^3 p}{(2\pi)^3} V^{\mu} %
\partial^{\beta}_{(p)} h({\bf p}) \left( g_{\gamma \beta} - %
\frac{V_{\gamma} K_{\beta}}{K\cdot V + i \epsilon}\right) A^{\gamma}(K) \; ,
\eeq
where $\epsilon$ is a small parameter that has to be sent to zero in the end. 

From this expression of the induced current the self-energy is obtained 
through the relation
\beq
\Pi^{\mu \nu}(K)=\frac{\delta J^{\mu}_{\rm ind}(K)}{\delta A_{\nu}(K)} \; ,
\eeq
which gives the gluon self-energy of an anisotropic system,
\beq
\Pi_{\mu \nu}(K)= 4 \pi \alpha_s \int \frac{d^3 p}{(2\pi)^3} V_{\mu} %
\partial^{\beta}_{(p)} h({\bf p}) \left( g_{\nu \beta} - %
\frac{V_{\nu} K_{\beta}}{K\cdot V + i \epsilon}\right) \; .
\label{selfenergy1}
\eeq
Note that this result can also be obtained 
using diagrammatic methods if one assumes that the distribution function 
is symmetric under 
${\bf p} \rightarrow - {\bf p}$ \cite{Mrowczynski:2000ed}. 
Furthermore, in the isotropic limit
$h({\bf p})\rightarrow h_{\rm iso}(p)$, one recovers the well-known HTL gluon
self-energy from Eq.(\ref{selfenergy1}).

After a little bit of algebra one finds that this tensor is symmetric, 
$\Pi^{\mu\nu}(K)=\Pi^{\nu\mu}(K)$, and transverse, 
$K^\mu\Pi^{\mu\nu}(K)=0$, if the distribution function vanishes on 
a two-sphere at infinity,
$\lim_{{\bf p}\rightarrow \infty} h({\bf p})=0$, since e.g.
\beq
K^{\mu} \Pi_{\mu 0} = -4 \pi \alpha_s k_{j} \int \frac{d^3 p}{(2 \pi)^3}
\partial^{j} h({\bf p}). 
\eeq
Therefore, only the spatial components of the self-energy 
are needed for the dispersion relations of the quasi-gluons, which can be
shown as follows: in 
the linear approximation the equations of motion for the gauge 
fields can be obtained
by expressing the induced current in terms of the self-energy
\beq
J^\mu_{\rm ind}(K) = \Pi^{\mu\nu}(K) A^\nu(K) \; ,
\eeq
and plugging this into Maxwell's equations
\beq
-iK_\mu F^{\mu\nu}(K) = J^\nu_{\rm ind}(K) + J_{\rm ext}^\nu(K) \; ,
\eeq
to obtain
\beq
[K^2 g^{\mu\nu} - K^\mu K^\nu + \Pi^{\mu\nu}(K)]A_\nu(K) =%
 - J_{\rm ext}^\nu(K) 
\; ,
\eeq
where $J^\nu_{\rm ext}$ is an external current.
Using the gauge covariance of the self-energy in the HTL-approximation
one can write this 
in terms of a physical electric field by specifying a particular
gauge. Using the temporal axial gauge ($A_0=0$) which has already
been used in the isotropic case in chapter \ref{peshier-chap} one obtains 
\beq
[(k^2-\omega^2)\delta^{ij} - k^i k^j + \Pi^{ij}(K)] E^j(K) = 
  (\Delta^{-1}(K))^{ij} E^j(K) = i \omega \, J_{\rm ext}^i(K) \; .
\eeq
Inverting the propagator allows one to determine the response of
the system to the external source
\beq
E^i(K) = i \omega \, \Delta^{ij}(K) J_{\rm ext}^j(K) \; .
\eeq
The dispersion relations for the collective modes can then be obtained as
in chapter \ref{peshier-chap}, by 
finding the poles of the propagator $\Delta^{ij}(K)$.

\subsection{Tensor decomposition}

Since an anisotropic system possesses a 
preferred spatial direction\footnote{If I were to consider a system where also
the cylindrical 
symmetry is lost there would be two preferred spatial directions
and the tensor basis for 
this system would accordingly be more complicated, but still can be
constructed by the method proposed here.}, 
the gluon propagator cannot simply be decomposed
into a transversal and longitudinal part as was the case for the isotropic
system treated in chapter \ref{peshier-chap};
therefore one needs to construct a new tensor basis.
 
As mentioned above, the gluon 
self-energy is symmetric and transverse; as a result 
not all components of $\Pi^{\mu \nu}$ are 
independent and one can restrict the considerations
to the spatial part of $\Pi^{\mu \nu}$, denoted $\Pi^{i j}$.
One therefore needs to construct a basis for a symmetric 3-tensor that
-- apart from the momentum $k^{i}$ -- also depends on a 
fixed anisotropy three-vector $\hat{n}^{i}$, with $\hat{{\bf n}}^2=1$.
Following Ref.~\cite{Kobes:1991dc} one first defines
the projection operator
\begin{equation}
A^{ij}=\delta^{ij}-k^{i}k^{j}/k^2,
\end{equation}
and uses it to construct $\tilde{n}^{i}=A^{ij} \hat{n}^{j}$ which obeys 
$\tilde{n} \cdot k =0$. With this one can construct the remaining
three symmetric 3-tensors
\begin{equation}
B^{ij}=k^{i}k^{j}/k^2
\end{equation}
\begin{equation}
C^{ij}=\tilde{n}^{i} \tilde{n}^{j} / \tilde{n}^2
\end{equation}
\begin{equation}
D^{ij}=k^{i}\tilde{n}^{j}+k^{j}\tilde{n}^{i}.
\end{equation}
The basis spanned by the four tensors ${\bf A},{\bf B},{\bf C},$ and ${\bf D}$
therefore allows one to decompose 
any symmetric 3-tensor ${\bf T}$ into
\beq
{\bf T}=a\,{\bf A}+b\,{\bf B}+c\,{\bf C}+d\,{\bf D} \; ;
\eeq
furthermore, the inverse of any such tensor is then given as
\beq
{\bf T}^{-1}=a^{-1}{\bf A}+\frac{(a+c){\bf B}-a^{-1}(bc-\tilde n^2 k^2 d^2 ){\bf
 C}-d{\bf D}}{
b(a+c)- \tilde n^2 k^2 d^2 } \; ,
\label{inversion}
\eeq
as a short amount of algebra reveals.
With the above tensor decomposition one is now able
to decompose the propagator into modes similar to what is done in 
the isotropic case.
For this, one first needs to decompose the self-energy into 
``self-energy structure functions'', which are the equivalents
of the transversal and longitudinal parts of the self-energy in the
isotropic case.

\subsection{Self-energy structure functions}

\begin{figure}
\begin{center}
\includegraphics[width=.4\linewidth]{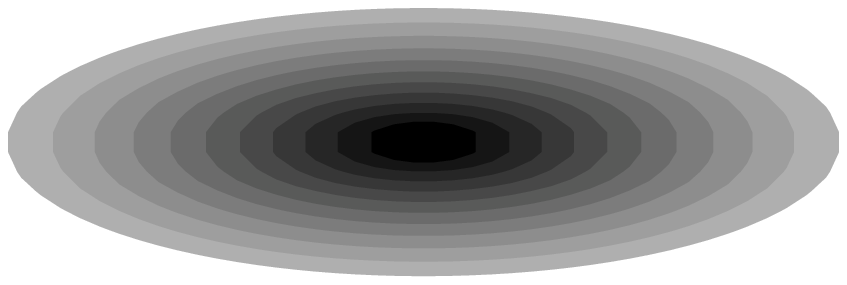}
\setlength{\unitlength}{1cm}
\begin{picture}(0,0)
\put(-3.2,1){\line(0,4){3.8}}
\put(-3.06,4.8){\makebox(0,0){\begin{rotate}{180} \textsf{V} \end{rotate}}}
\put(0,2.79){\makebox(0,0){\begin{rotate}{90} \textsf{V} \end{rotate}}}
\put(-3.2,5){\makebox(0,0){\footnotesize $p_z$}}
\put(-6,2.92){\line(4,0){6}}
\put(0.5,2.92){\makebox(0,0){\footnotesize $p_T$}}
\end{picture}
\caption{Contour plot of an anisotropic version Eq.(\ref{squashing}) 
of the Fermi-Dirac distribution function with positive anisotropy parameter.
The anisotropy vector is taken to be along the $p_z$-direction.}
\label{fig:squashing}
\end{center}
\end{figure}

From Eq.(\ref{selfenergy1}) 
the spatial part of the general gluon self-energy tensor
is found to be 
\begin{equation}
\Pi^{i j}(K) = - 4 \pi \alpha_s%
 \int \frac{d^3 p}{(2\pi)^3} v^{i} \partial^{l} h({\bf p})
\left( \delta^{j l}+\frac{v^{j} k^{l}}{K\cdot V + i \epsilon}\right) \; .
\label{selfenergy2}
\end{equation}
Since the collision term was set to zero, 
the distribution function $h({\bf p})$ is completely arbitrary at this point,
so in order to proceed one needs to assume a specific form for the
distribution function.  In what follows I will require that
$h({\bf p})$ can be obtained from an (arbitrary) isotropic distribution 
function by the rescaling of one direction in momentum space, 
\begin{equation}
h({\bf p})=h_{\xi}({\bf p}) = N(\xi) \ %
h_{\rm iso}\left(\sqrt{{\bf p}^2+\xi({\bf p}\cdot{\bf \hat n})^2}\right)%
\; ,
\label{squashing}
\end{equation}
where ${\bf \hat n}$ is the direction of the anisotropy, $\xi>-1$ is a
parameter reflecting the strength of the anisotropy and $N(\xi)$ is a 
normalization constant\footnote{%
In the original reference \cite{Romatschke:2003ms} this normalization
constant has been set to one since it does not affect the collective
modes qualitatively. However, the correct value has to be used
in order to make quantitative 
predictions of physical observables (e.g. applying the results of this 
chapter to the calculation of the 
heavy quark energy loss in chapter \ref{eloss-chap}). 
}.  Note that $\xi>0$ corresponds to a
contraction of the distribution in the ${\bf \hat n}$ direction 
(shown in Fig.~\ref{fig:squashing})
whereas $-1<\xi<0$ 
corresponds to a stretching of the distribution in the 
${\bf \hat n}$ direction. 
 
The normalization constant $N(\xi)$ has been introduced to preserve the 
relation 
\beq
\frac{d^3 n}{d\ k^3} = h({\bf p})
\eeq
between number density $n$ and the momentum-space distribution function
also in the case where the latter is anisotropic. $N(\xi)$
is then simply determined by requiring $n$
to be the same both for isotropic and anisotropic systems,
\beq
\int \frac{d^3 p}{(2\pi)^3} h_{\rm iso}(p)=%
\int \frac{d^3 p}{(2\pi)^3} h_{\xi}({\bf p})=N(\xi)%
\int \frac{d^3 p}{(2\pi)^3} %
h_{\rm iso}\left(\sqrt{{\bf p}^2+\xi({\bf p}\cdot{\bf \hat n})^2}\right)
\eeq
and can be evaluated to be 
\beq
N(\xi)=\sqrt{1+\xi}
\eeq
by performing a change of
variables to $\tilde p$
\begin{equation}
\tilde{p}^2=p^2\left(1+\xi ({\bf v}\cdot{\bf \hat{n}})^2\right) \; .
\end{equation}
The same change of variables then also 
allows one to simplify (\ref{selfenergy2}) since then it is possible 
to integrate out the $|\tilde p|$-dependence, giving
\begin{equation}
\Pi^{i j}(K) = m_{D}^2 \sqrt{1+\xi}\int \frac{d \Omega}{4 \pi} v^{i}%
\frac{v^{l}+\xi({\bf v}.{\bf \hat{n}}) n^{l}}{%
(1+\xi({\bf v}.{\bf \hat{n}})^2)^2}
\left( \delta^{j l}+\frac{v^{j} k^{l}}{K\cdot V + i \epsilon}\right) \; ,
\end{equation}
where 
\beq
m_D^2 = -{2 \alpha_s\over \pi} \int_0^\infty d p \,  
  p^2 {d h_{\rm iso}(p^2) \over dp} \; ,
\eeq
which corresponds to Eq.(\ref{mD}).
Using the tensor basis from above, one can then decompose 
the self-energy into four structure functions
\begin{equation}
\Pi^{ij}=\alpha A^{i j}+\beta B^{ij} + \gamma C^{ij} + \delta D^{ij} \; ,
\end{equation}
which are determined by taking the contractions
\begin{eqnarray}
k^{i} \Pi^{ij} k^{j} & = & k^2 \beta \; , \nonumber \\
\tilde{n}^{i} \Pi^{ij} k^{j} & = & \tilde{n}^2 k^2 \delta \; , \nonumber \\
\tilde{n}^{i} \Pi^{ij} \tilde{n}^{j} & = & \tilde{n}^2 (\alpha+\gamma) \; , %
\nonumber \\
{\rm Tr}\,{\Pi^{ij}} & = & 2\alpha +\beta +\gamma \; .
\label{contractions}
\end{eqnarray}
Since the form of the structure functions is somewhat unwieldy,
the integral expressions following from 
Eq.(\ref{contractions}) have been relegated to appendix \ref{exp-chap}.

\subsection{Evaluation of the structure functions}

The four structure functions $\alpha,\beta,\gamma,\delta$ 
depend on the Debye mass $m_D$, the frequency and spatial momentum $\omega$
and $k$, the strength of the anisotropy $\xi$ and the angle between the 
spatial momentum and the anisotropy direction 
${\bf \hat{k}}\cdot{\bf \hat n}=\cos\theta_n$. 
For symmetry reasons one can restrict
$0<\theta_n<\frac{\pi}{2}$ in the following.
In the isotropic limit ($\xi \rightarrow 0$) 
the structure functions $\alpha$ and $\beta$ 
reduce to the isotropic HTL self-energies and $\gamma$ 
and $\delta$ vanish,
\begin{figure}
\begin{center}
\begin{minipage}[t]{.48\linewidth}
\includegraphics[width=.9\linewidth]{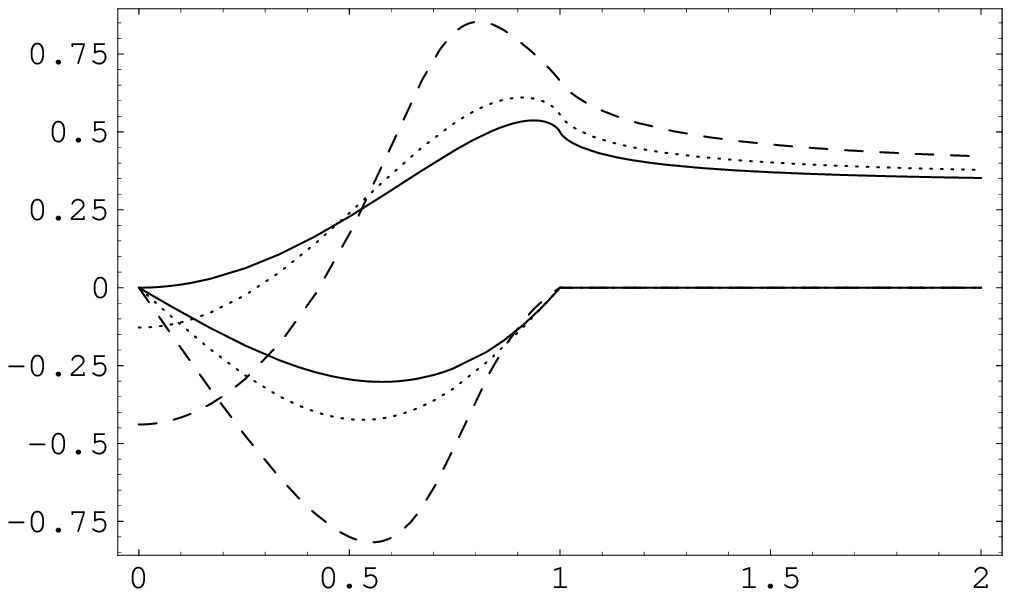}
\setlength{\unitlength}{1cm}
\begin{picture}(5,0)
\put(3.5,4.7){\makebox(0,0){\footnotesize (a)}}
\put(3.5,0){\makebox(0,0){\footnotesize $\omega/k$}}
\put(-0,1.5){\makebox(0,0){\begin{rotate}{90}%
\footnotesize $({\rm Re}\alpha, {\rm Im}\alpha)/m_D^2$
\end{rotate}
}}
\end{picture}
\end{minipage}\hfill
\begin{minipage}[t]{.48\linewidth}
\includegraphics[width=.9\linewidth]{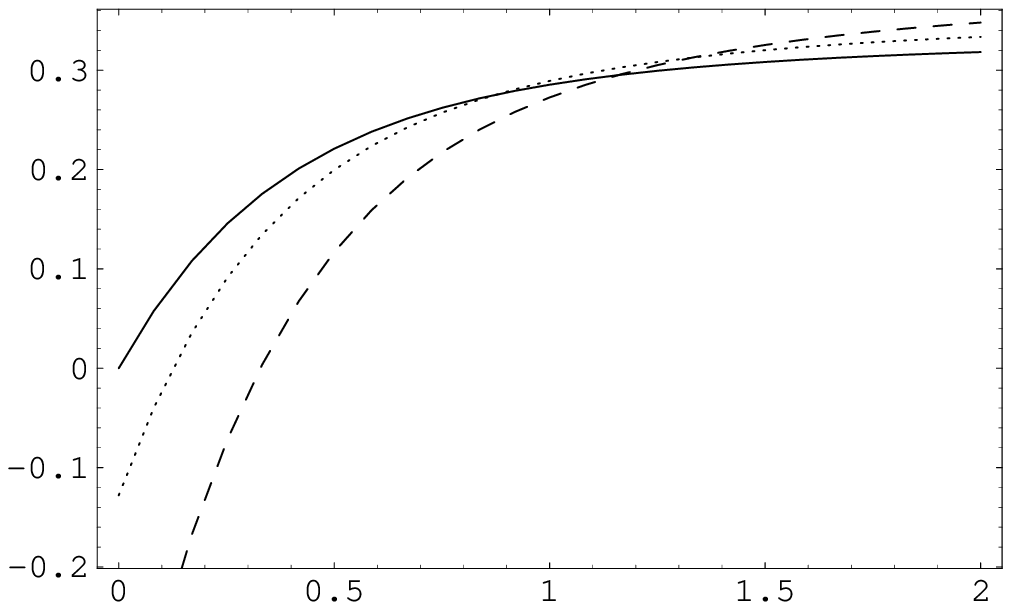}
\setlength{\unitlength}{1cm}
\begin{picture}(5,0)
\put(3.5,4.7){\makebox(0,0){\footnotesize (b)}}
\put(3.5,0){\makebox(0,0){\footnotesize $\Gamma/k$}}
\put(-0,2){\makebox(0,0){\begin{rotate}{90}%
\footnotesize ${\rm Re}\alpha/m_D^2$
\end{rotate}
}}
\end{picture}
\end{minipage}
\end{center}
\caption[a]{Real and imaginary parts of $\alpha/m_D^2$ as a function 
of real $\omega/k$ 
(a) and (b) the real part of $\alpha/m_D^2$ 
for $\omega/k = i \Gamma/k$ with $\theta_n = \pi/4$ and 
$\xi = \{0,1,10\}$ (full line, dotted line and dashed line, respectively)
in both cases.}
\label{struct-fig}
\end{figure}
\bqa
\alpha(K,0) &=& \Pi_T(K) \; , \nonumber \\
\beta(K,0) &=& - {\omega^2\over k^2} \Pi_L(K) \; , \nonumber \\
\gamma(K,0) &=& 0 \; , \nonumber \\
\delta(K,0) &=& 0 \; ,
\label{isolimit}
\eqa
where $\Pi_T(K)$ and $\Pi_L(k)$ are 
given by Eq.(\ref{HTLPiL}).
Note that 
for finite $\xi$ the analytic structure of $\alpha,\beta,\gamma,\delta$  
is the same as for $\Pi_L,\Pi_T$ in the isotropic case, namely 
there is a cut in the 
complex $\omega$ plane which can be chosen
to run along the real $\omega$ axis from $-k<\omega<k$.  
For real-valued $\omega$ the 
structure functions are complex for all $\omega < k$ (corresponding
to the Landau damping regime of chapter \ref{peshier-chap}) and real 
for $\omega >k$ while for 
imaginary-valued $\omega$ all four structure functions are real-valued.
As an example, a plot of the structure function $\alpha$ for real and 
imaginary
values of $\omega$, $\xi=\{0,1,10\}$, and $\theta_n =\pi/4$ is shown
in Fig.~\ref{struct-fig}.

Using these structure functions it is then possible to 
write the inverse propagator ${\bf \Delta}^{-1}(K)$ in terms
of the above tensor basis 
\beq
{\bf \Delta}^{-1}(K) = (k^2 - \omega^2 + \alpha){\bf A} + 
(\beta - \omega^2){\bf B} 
        + \gamma {\bf C} + \delta {\bf D} \; \ .
\eeq
Applying the inversion formula (\ref{inversion}) one obtains 
a decomposition for the gluon propagator into modes
\beq
{\bf \Delta}(K) =  \Delta_A {\bf A} + (k^2 - \omega^2 + \alpha +%
 \gamma)\Delta_G {\bf B}
              + [(\beta-\omega^2)\Delta_G - \Delta_A] {\bf C} -%
 \delta \Delta_G {\bf D} \; ,
\label{gluonpropmodes}
\eeq
where
\bqa
\Delta_A^{-1}(K) &=& k^2 - \omega^2 + \alpha \; , \label{propfnc1} \\
\Delta_G^{-1}(K) &=& (k^2 - \omega^2 + \alpha + \gamma)(\beta-\omega^2)-k^2 %
\tilde n^2 \delta^2 \; .
\label{propfnc2}
\eqa
In the following it is convenient to use the form
\beq
\Delta(K) =  \Delta_A \, [{\bf A}-{\bf C}] 
        + \Delta_G \, [(k^2 - \omega^2 + \alpha + \gamma) {\bf B} + %
(\beta-\omega^2) {\bf C}  - \delta {\bf D}] \; ,
\eeq
so that the poles of the propagator correspond to the poles of 
$\Delta_A$ and $\Delta_G$, respectively.

\subsection{The quark propagator in an anisotropic system}

Using the equivalence of semi-classical kinetic theory and the HTL diagrammatic
approach one can derive the quark propagator in an anisotropic system 
similar to the gluon propagator above. However,
unlike the gluon propagator the quark propagator in an anisotropic system
can be shown to behave very similar to its isotropic counterpart
\cite{MrowHeidel:2002}. Consequently, the quark collective 
modes will look very similar to those derived in chapter \ref{peshier-chap}
and therefore the quark propagator in an anisotropic system
will not be treated here explicitly.

\section{Collective modes of an anisotropic quark-gluon plasma}

\sectionmark{Collective modes of an anisotropic plasma}

The dispersion relations for the gluonic modes in an anisotropic
quark-gluon plasma are determined by setting $\Delta_A^{-1}=0$ and
$\Delta_G^{-1}=0$ and then solving these equations using 
Eqs.(\ref{propfnc1},\ref{propfnc2}). A subsequent comparison to the
results of chapter \ref{peshier-chap} will show the qualitative
differences with respect to the isotropic case.

\subsection{Static limit}

As a first test of how the momentum-space anisotropy in the 
distribution functions affects the properties of the system one can
consider
the response of the system to static electric and magnetic fluctuations. 
For this test, one needs to examine the 
limit $\omega\rightarrow0$ of the propagators (\ref{propfnc1}) and 
(\ref{propfnc2}):
approaching along the real $\omega$ axis one finds that to leading order 
$\alpha \sim \gamma \sim O(\omega^0)$, $\beta \sim O(\omega^2)$, and 
$\delta \sim O(i \omega)$;
therefore, one can define the four mass scales
\bqa
m_\alpha^2 &=& \lim_{\omega\rightarrow0} \alpha \; , \nonumber \\
m_\beta^2 &=& \lim_{\omega\rightarrow0} - {k^2\over\omega^2} \beta \; , %
\nonumber \\
m_\gamma^2 &=& \lim_{\omega\rightarrow0} \gamma \; , \nonumber \\
m_\delta^2 &=& \lim_{\omega\rightarrow0}  {\tilde n k^2\over\omega} %
{\rm Im}\,\delta \; ,
\label{massdef}
\eqa
which will be used in the following. The static limit of the 
propagators $\Delta_A$ Eq.(\ref{propfnc1}) and $\Delta_G$
Eq.(\ref{propfnc2}) in terms of these masses is then given by the 
expressions
\bqa
\Delta_A^{-1} &=& k^2 + m_\alpha^2 \; \, \\
\Delta_G^{-1} &=& -{\omega^2\over k^2}\left[(k^2+m_\alpha^2+m_\gamma^2)%
(k^2+m_\beta^2)  - m_\delta^4 \right] \; .
\eqa
Furthermore, $\Delta_G^{-1}$ can be factorized into 
\beq
\Delta_G^{-1} = -{\omega^2\over k^2}(k^2 + m_+^2)(k^2+m_-^2) \; ,
\label{deltaGfac}
\eeq
where
\beq
2 m_{\pm}^2 = {\mathcal M}^2 \pm \sqrt{{\mathcal M}^4-4(m_\beta^2(m_\alpha^2+%
m_\gamma^2)-m_\delta^4)} \; ,
\label{mpm}
\eeq
and
\beq
{\mathcal M}^2 = m_\alpha^2+m_\beta^2+m_\gamma^2 \; .
\eeq
In the isotropic limit 
$\xi\rightarrow0$, most of the mass parameters vanish, 
$m_\alpha^2=m_\gamma^2=m_\delta^2=m_-^2\rightarrow0$, while $m_+^2$ becomes
equal to the Debye mass squared, 
$m_+^2 \rightarrow m_D^2$, so
that one recovers the results obtained in chapter \ref{peshier-chap}.
\begin{figure}
\begin{center}
\begin{minipage}[t]{.48\linewidth}
\includegraphics[width=.9\linewidth]{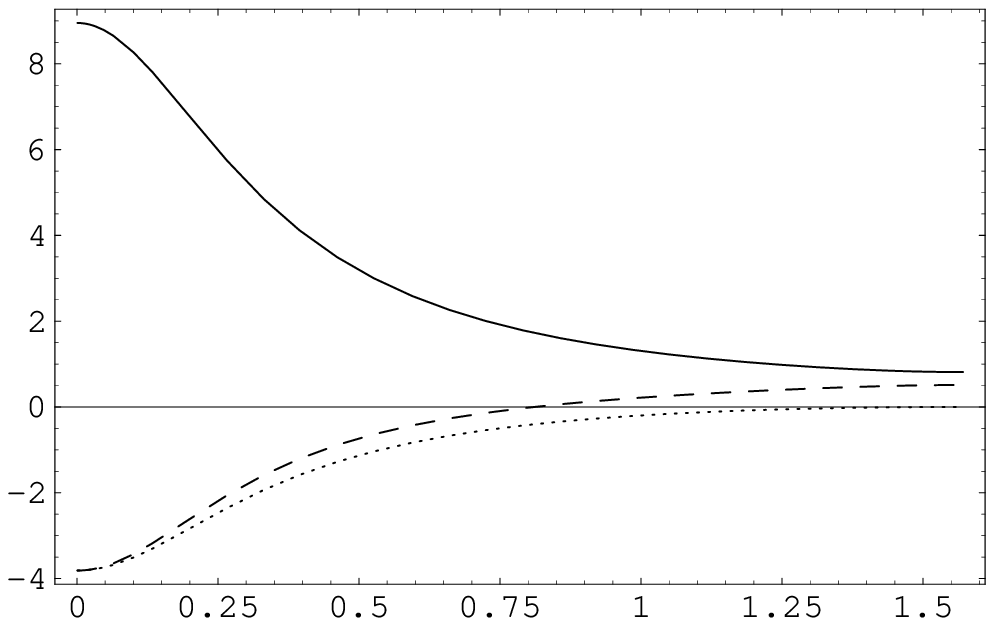}
\setlength{\unitlength}{1cm}
\begin{picture}(5,0)
\put(3.5,4.7){\makebox(0,0){(a)}}
\put(3.5,0){\makebox(0,0){\footnotesize $\theta_n$}}
\put(0,1){\makebox(0,0){\begin{rotate}{90}%
\footnotesize $(m_+^2,m_-^2,m^2_\alpha)/m_D^2$
\end{rotate}
}}
\end{picture}
\end{minipage}
\hfill
\begin{minipage}[t]{.48\linewidth}
\includegraphics[width=.9\linewidth]{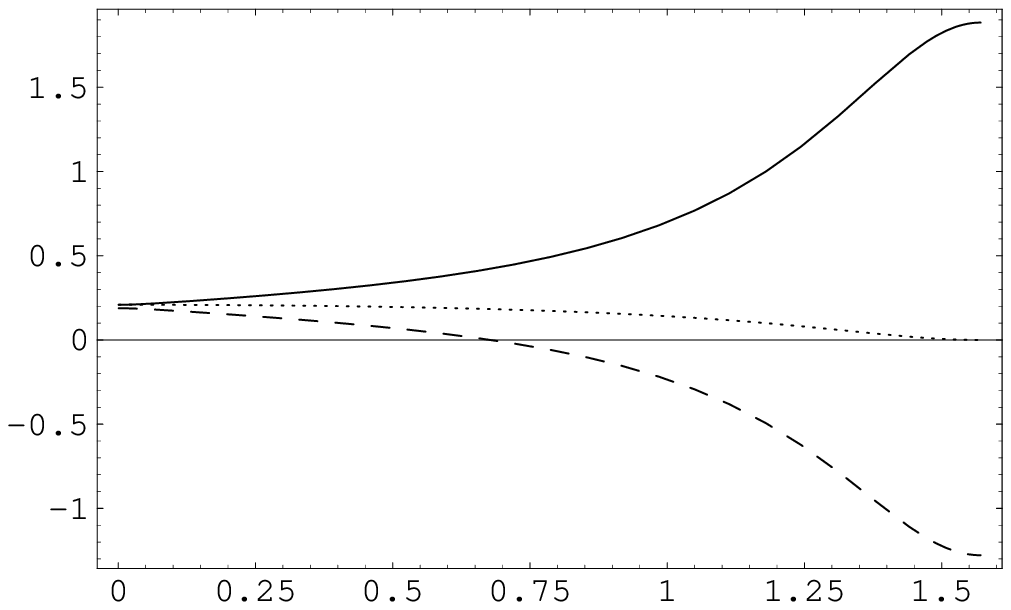}
\setlength{\unitlength}{1cm}
\begin{picture}(5,0)
\put(3.5,4.7){\makebox(0,0){(b)}}
\put(3.5,0){\makebox(0,0){\footnotesize $\theta_n$}}
\put(0,1){\makebox(0,0){\begin{rotate}{90}%
\footnotesize $(m_+^2,m_-^2,m^2_\alpha)/m_D^2$
\end{rotate}
}}
\end{picture}
\end{minipage}
\end{center}
\caption[a]{Angular dependence of $m_\alpha^2$, $m_+^2$, and $m_-^2$
(dotted line, full line and dashed line, respectively)
at fixed (a) $\xi=10$ and (b) $\xi=-0.9$.}
\label{static-fig}
\end{figure}
For finite $\xi$ it is possible to evaluate all four masses defined above
analytically (the results for $m_\alpha$ and $m_\beta$ are listed in
Appendix \ref{exp-chap}), but since the resulting expressions are quite
unwieldy it is more straightforward to use a numerical evaluation. 
As an example, the angular dependence
 of $m_\alpha^2$, $m_+^2$, and $m_-^2$ at fixed $\xi=10$ and $\xi=-0.9$
is plotted in Fig.~\ref{static-fig}. 
In the case $\xi>0$ (Fig.~\ref{static-fig}a), one can see that
the scale $m_+^2$ is bigger than $m_D^2$  
for small $\theta_n$ while being smaller than $m_D^2$ 
for $\theta_n$ near $\pi/2$; indeed one finds that this 
corresponds to the screening of electrostatic modes in the isotropic limit, 
the only difference being an angular dependence of the screening length 
for nonzero $\xi$.

Non-vanishing 
scales $m_\alpha^2$ and $m_-^2$, however, should naively correspond
to a screening of magnetostatic modes, which is absent in isotropic systems.
Indeed, the fact that anisotropic systems allow
for non-vanishing magnetostatic screening masses
was already found by \textsc{Cooper} {\em et al.} \cite{Cooper:2002ff};
however, \textsc{Cooper} {\em et al.} neglected to consider the fact that the 
resulting mass squares might be negative (as is the case for small 
$\theta_n$ in Fig.~\ref{static-fig}a) and
would therefore not correspond to a screening of the magnetic interaction.
 
More precisely, the fact that these quantities are negative 
indicates that for $\xi>0$ the 
system possesses an instability to transverse and ``mixed'' external 
perturbations associated with $m_\alpha^2$ and $m_-^2$, respectively,
as has now been verified by \textsc{Birse} {\rm et al.} \cite{Birse:2003qp} and
discussed in a more general context by \textsc{Arnold}, \textsc{Lenaghan} and 
\textsc{Moore} \cite{Arnold:2003rq}.
As can be seen in Fig.~\ref{static-fig}a the transverse instability 
is present for any $\theta_n\neq\pi/2$ while the mixed
instability is only present for $\theta_n < \theta^{\rm mixed}_c$ with
$\theta^{\rm mixed}_c$ depending on the value of $\xi$.  

For the case $\xi<0$, which is shown in Fig.~\ref{static-fig}b, 
the $\theta_n$-dependence of the scales $m_+^2$ and $m_{-}^2$
is exactly reversed with respect to the case $\xi>0$, 
as one would expect naively from Eq.(\ref{squashing}). The scale 
$m_{\alpha}^2$, however, which was strictly negative for $\xi>0$, now
turns out to be strictly positive in the case of $\xi<0$, so the
transverse instability vanishes in the case of $\xi<0$.
For $\theta_n \gtrsim \pi/4$ the mixed instability is still 
present, since the scale $m_-^2$ is again negative.

The role of these unstable modes will be treated later on in more detail.

\subsection{Stable modes}

\begin{figure}
\begin{center}
\begin{minipage}[t]{.325\linewidth}
\includegraphics[width=.98\linewidth]{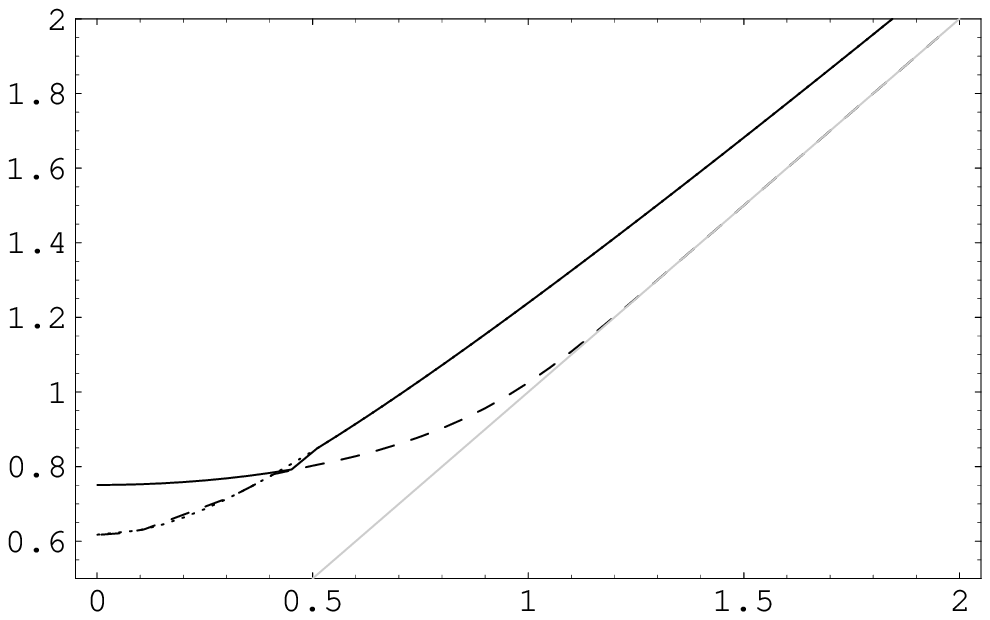}
\setlength{\unitlength}{1cm}
\begin{picture}(5,0)
\put(2.5,0){\makebox(0,0){\footnotesize $k/m_D$}}
\put(2.5,3.5){\makebox(0,0){\footnotesize (a)}}
\put(0,0.5){\makebox(0,0){\begin{rotate}{90}%
\footnotesize $(E_\alpha,E_+,E_-)/m_D$
\end{rotate}
}}
\end{picture}
\end{minipage}
\hfill
\begin{minipage}[t]{.32\linewidth}
\includegraphics[width=.98\linewidth]{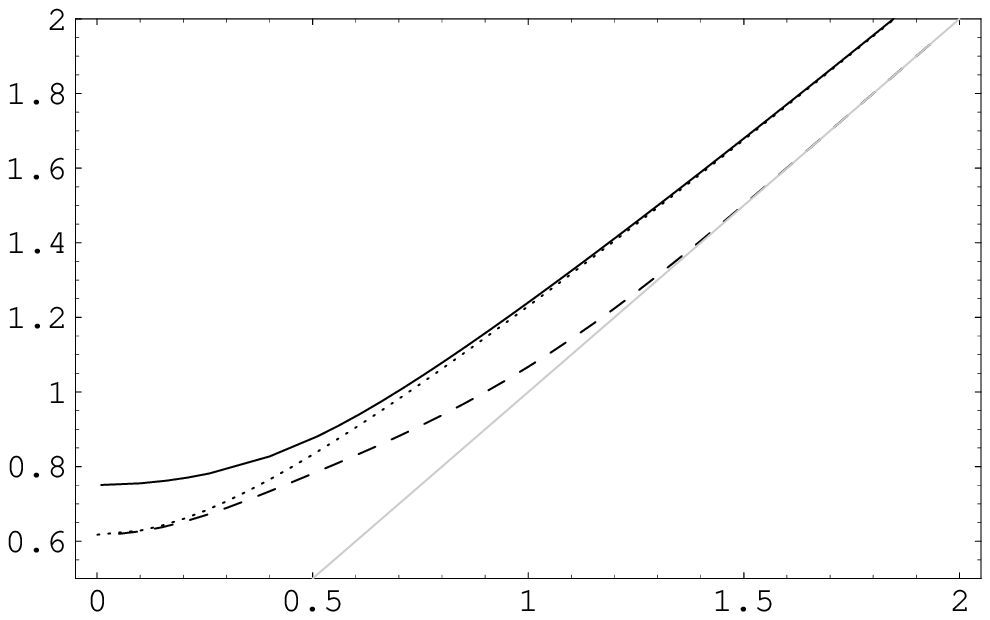}
\setlength{\unitlength}{1cm}
\begin{picture}(5,0)
\put(2.5,0){\makebox(0,0){\footnotesize $k/m_D$}}
\put(2.5,3.5){\makebox(0,0){\footnotesize (b)}}
\end{picture}
\end{minipage}
\hfill
\begin{minipage}[t]{.325\linewidth}
\includegraphics[width=.98\linewidth]{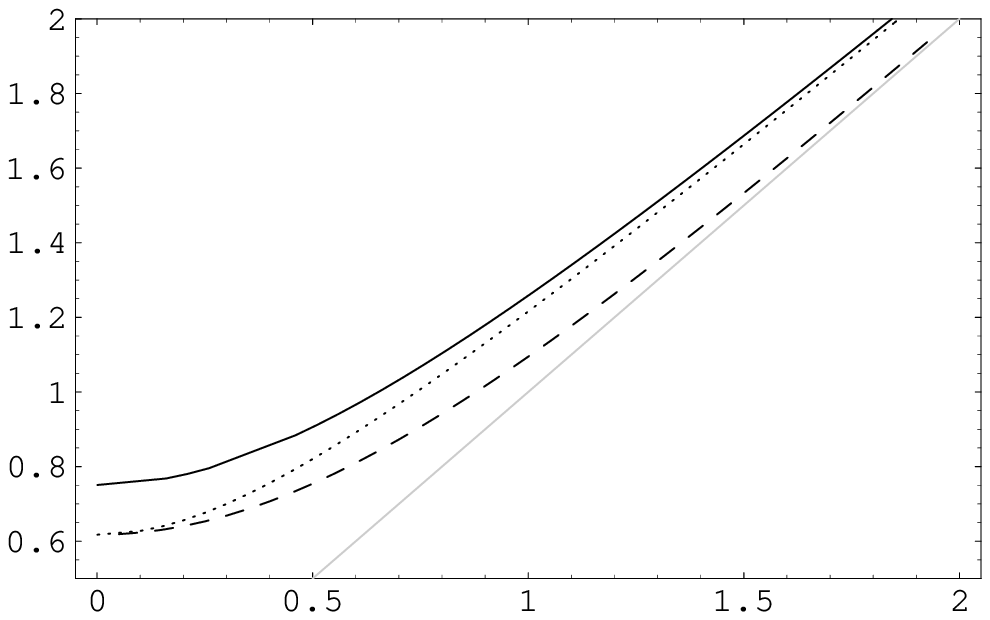}
\setlength{\unitlength}{1cm}
\begin{picture}(5,0)
\put(2.5,0){\makebox(0,0){\footnotesize $k/m_D$}}
\put(2.5,3.5){\makebox(0,0){\footnotesize (c)}}
\end{picture}
\end{minipage}
\end{center}
\caption[a]{Angular dependence of $E_\alpha$ (dotted line), 
$E_+$ (full line), and $E_-$ (dashed line) for
$\xi=10$, and $\theta_n=\{0,\pi/4,\pi/2\}$ (plots (a), (b) and (c), 
respectively).}
\label{smodes-fig}
\end{figure}

In the non-static case,
a factorization of $\Delta_G^{-1}$ similar to Eq.(\ref{deltaGfac})
can be achieved, which allows
the determination of the dispersion relations for all of the 
collective modes in the system.

Considering first the stable collective modes which correspond to poles 
of the propagator at
real-valued $\omega>k$ one factorizes $\Delta_G^{-1}$ as
\beq
\Delta_G^{-1} = (\omega^2 - \Omega_+^2)(\omega^2-\Omega_-^2) \; ,
\eeq
where
\beq
2 \Omega_{\pm}^2 = \bar\Omega^2 \pm \sqrt{\bar\Omega^4- 4 ((\alpha+\gamma+%
k^2)\beta-k^2\tilde n^2\delta^2) } \; ,
\label{omegapm}
\eeq
and
\beq
\bar\Omega^2 = \alpha+\beta+\gamma+k^2 \; .
\eeq
Since the quantity under the square root in (\ref{omegapm}) 
can be written as
$(\alpha-\beta+\gamma+k^2)^2+4k^2\tilde n^2\delta^2$ (which 
is always positive for
real $\omega>k$), there are at most two stable modes coming 
from $\Delta_G$, while 
the remaining stable collective mode comes from the zero of $\Delta_A^{-1}$.

The dispersion relations for all of the collective modes are then given by
the solutions to 
\bqa
E^2_\pm &=& \Omega_\pm^2(E_\pm) \; , \\
E^2_\alpha &=& k^2 + \alpha(E_\alpha) \; .
\eqa
In the isotropic limit (\ref{isolimit}) one finds the correspondence
 $E_\alpha = E_+ = E_T$ and $E_- = E_L$.
Accordingly, for finite $\xi$, there are 
three stable quasiparticle modes with dispersion
relations that depend on the angle of propagation with respect to the 
anisotropy
vector, $\theta_n$.  The resulting dispersion relations for all three modes
for the case $\xi=10$, and $\theta_n=\{0,\pi/4,\pi/2\}$ are plotted 
in Fig.~\ref{smodes-fig}.

\subsection{Unstable modes}

For non-zero $\xi$ the propagator also has poles along the imaginary $\omega$ 
axis\footnote{Checking for poles at complex $\omega$ can be done
numerically but no poles on the physical sheet have been found; indeed,
for certain special cases one can prove that there are no other modes
than the ones listed above \cite{Romatschke:2004jh}.
However, there exist solutions on the unphysical sheet which come very
close to the border of the physical sheet 
for sufficiently large values of the anisotropy
parameter $\xi$ (see Ref.\cite{Romatschke:2004jh}).}, that correspond
to exponentially growing or decaying field amplitudes, depending on the
sign of the imaginary part. The exponentially growing solutions correspond
to the unstable modes of the system with 
$\omega \rightarrow i \Gamma$ and $\Gamma$ the real-valued solution 
to the equations
\bqa
\Delta_G^{-1} &=& (\Gamma^2 + \Omega_+^2)(\Gamma^2 + \Omega_-^2)=0 \;%
,\nonumber\\
\Delta_A^{-1} &=& (\Gamma^2+k^2+\alpha)=0.
\eqa
It turns out that 
in contrast to the stable modes there is at most one solution for 
$\Delta_G^{-1}(\omega=i \Gamma)=0$ 
since numerically one finds that $\Omega_+^2>0$ for all $\Gamma>0$, while
$\Delta_A^{-1}(\omega=i \Gamma)=0$ has only solutions for $\xi>0$.
Therefore, the system possesses one or two unstable modes depending on
the sign of the anisotropy parameter.

\begin{figure}
\begin{center}
\begin{minipage}[t]{.48\linewidth}
\includegraphics[width=.9\linewidth]{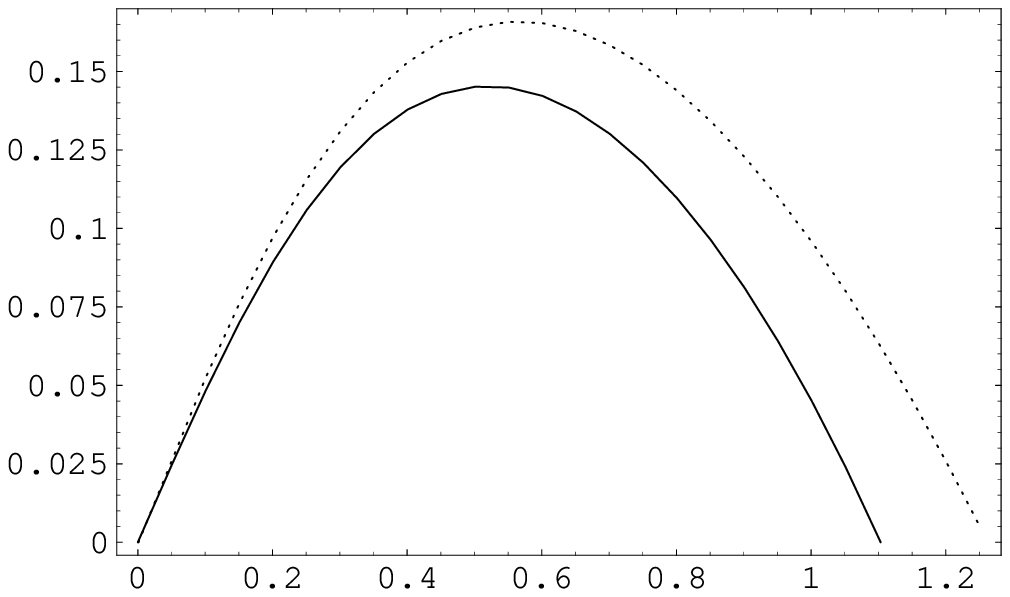}
\setlength{\unitlength}{1cm}
\begin{picture}(5,0)
\put(3.5,0){\makebox(0,0){\footnotesize $k/m_D$}}
\put(3.5,4.7){\makebox(0,0){\footnotesize (a)}}
\put(0,1.5){\makebox(0,0){\begin{rotate}{90}%
\footnotesize $(\Gamma_\alpha, \Gamma_-)/m_D$
\end{rotate}
}}
\end{picture}
\end{minipage}
\hfill
\begin{minipage}[t]{.48\linewidth}
\includegraphics[width=.9\linewidth]{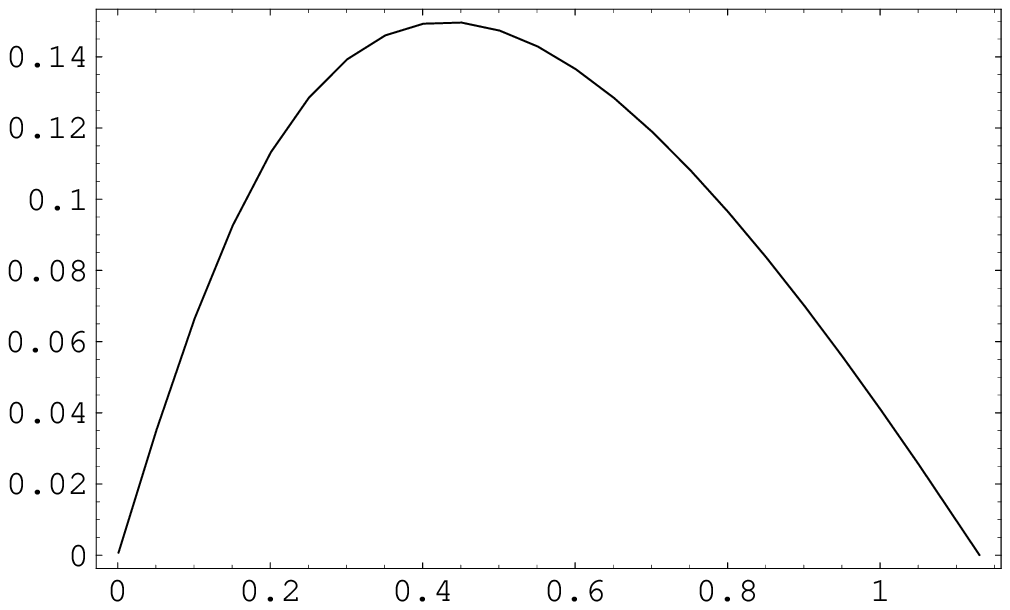}
\setlength{\unitlength}{1cm}
\begin{picture}(5,0)
\put(3.5,0){\makebox(0,0){\footnotesize $k/m_D$}}
\put(3.5,4.7){\makebox(0,0){\footnotesize (b)}}
\put(0,2){\makebox(0,0){\begin{rotate}{90}%
\footnotesize $\Gamma_-/m_D$
\end{rotate}
}}
\end{picture}
\end{minipage}
\end{center}
\caption[a]{$\Gamma_\alpha(k)$ (dotted line) and $\Gamma_-(k)$ 
as a function of $k$ with
(a) $\xi=10$ and $\theta_n=\pi/8$ and (b) $\xi=-0.9$ and $\theta_n=\pi/2$.}
\label{umodes-fig}
\end{figure}
In Fig.~\ref{umodes-fig}a a plot of the growth rates $\Gamma_\alpha(k)$ 
and $\Gamma_-(k)$ with $\xi=10$ and $\theta_n=\pi/8$ is shown, reflecting
the presence of two unstable modes; in 
Fig.~\ref{umodes-fig}b $\Gamma_-(k)$ is shown for $\xi=-0.9$ 
and $\theta_n=\pi/2$.
 
\section{Small $\xi$ limit}

In the small-$\xi$ limit it is possible to obtain analytic expressions 
for all of the
structure functions order-by-order in $\xi$, which provide a well-defined
test case to the otherwise numerically conducted analysis.  
To linear order in $\xi$ one finds
\bqa
\alpha &=& \Pi_T(z) + \xi\Bigg[
          {z^2\over12}(3+5\cos2\theta_n)m_D^2 
          -{1\over6}(1+\cos2\theta_n)m_D^2 \nonumber \\
          && \hspace{4cm} 
        + {1\over4}\Pi_T(z)\left( (3+3\cos2\theta_n) -%
 z^2(3+5\cos2\theta_n) \right)\Bigg] \; , \nonumber \\
z^{-2} \beta &=& -\Pi_L(z) + \xi\left[{1\over6}(1+3\cos2\theta_n) m_D^2
        - \frac{\Pi_L(z)}{2} %
\left(\cos2\theta_n(2-3 z^2)+1-z^2\right)%
\right] \; , \nonumber \\
\gamma &=& {\xi\over3}(3 \Pi_T(z) - m_D^2)(z^2-1)\sin^2\theta_n \; ,%
 \nonumber \\
\delta &=& {\xi\over3k}(4 z^2 m_D^2+3 \Pi_T(z)(1-4z^2))\cos\theta_n \; ,
\label{smallxi}
\eqa
where $z=\omega/k$ and again ${\bf \hat{k}}\cdot {\bf \hat{n}}=\cos{\theta_n}$.

\subsection{Static Limit}

Using the linear expansions from above and the well-known static 
limits $\Pi_L \rightarrow m_D^2$
and $\Pi_T \rightarrow - i \pi \omega/(4 k)$ one obtains analytic
expressions for the mass scales (\ref{massdef}), 
\bqa
\hat m_\alpha^2 &=& - {\xi\over6}(1 + \cos 2\theta_n) \; , \nonumber \\
\hat m_\beta^2 &=& 1 + {\xi\over6}(3 \cos 2\theta_n+2) \; , \nonumber \\
\hat m_\gamma^2 &=& {\xi\over3}\sin^2\theta_n \; ,  \nonumber \\
\hat m_\delta^2 &=& -\xi{\pi\over4}\sin\theta_n\cos\theta_n \; ,
\eqa
where $\hat m^2 = m^2/m_D^2$.  Using these one obtains expansions
to linear order in $\xi$ for $m_\pm^2$ defined in (\ref{mpm})
\bqa
\hat m_+^2 &=& 1 + {\xi\over6}(3 \cos 2\theta_n+2) \; , \nonumber \\
\hat m_-^2 &=& -{\xi\over3}\cos2\theta_n \; .
\eqa

Clearly, the mass scales in the small-$\xi$ limit have the same 
qualitative behavior as their exact counterparts, as one can see by comparing
the above expressions to Fig.~\ref{static-fig}. Therefore, one gains
confidence that the small-$\xi$ limit conserves the basic features of the full
anisotropic problem while having the advantage of being analytically tractable.

\subsection{Collective modes}

The collective modes in the small-$\xi$ limit are most easily found 
by an expansion of Eq.(\ref{gluonpropmodes}),
\bqa
\Delta_A^{-1} &=& k^2-\omega^2+\alpha = 0 \nonumber \\
\Delta_G^{-1} &=& (k^2-\omega^2+\alpha+\gamma)(\beta-\omega^2) = 0 \; ,
\label{SXdec}
\eqa
where $\alpha$, $\beta$, and $\gamma$ are given by (\ref{smallxi})
and $\delta^2$ can be ignored in the expansion since it is of order 
$O(\xi^2)$.
Solving the above equations one finds that again both the stable and 
unstable modes in the small-$\xi$ limit correspond qualitatively
to the full problem.
Note that there is again only one unstable mode coming from $\Delta_G^{-1}$ 
since $\beta(i\Gamma)>0$ for all $\Gamma>0$.

\section{Large $\xi$ limit}

\begin{figure}
\begin{center}
\includegraphics[width=.48\linewidth]{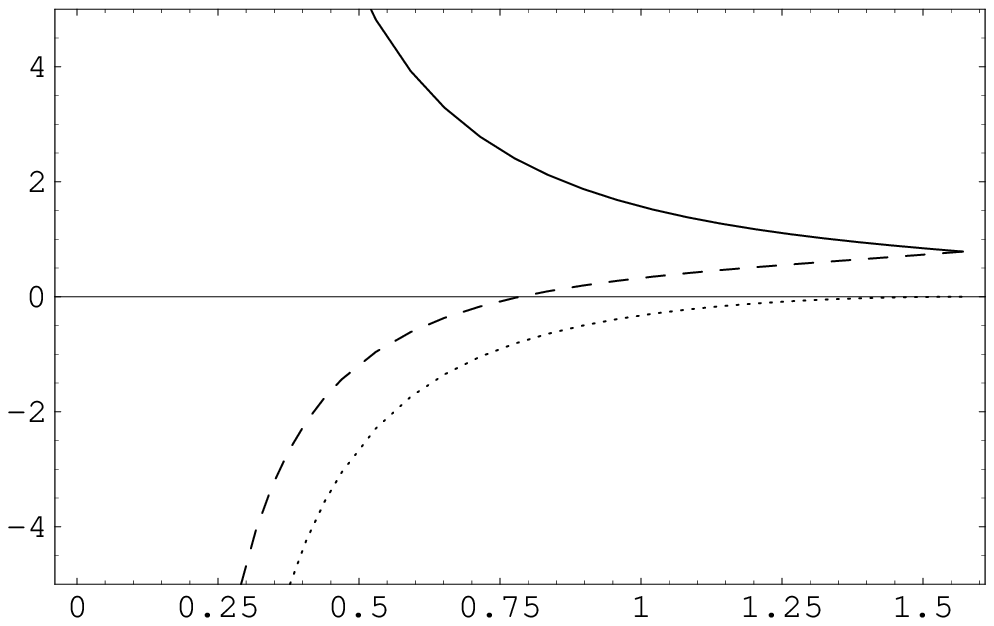}
\end{center}
\setlength{\unitlength}{1cm}
\begin{picture}(5,0)
\put(7,0.5){\makebox(0,0){\footnotesize $\theta_n$}}
\put(3.5,1.5){\makebox(0,0){\begin{rotate}{90}%
\footnotesize $(m_+^2,m_-^2,m^2_\alpha)/m_D^2$
\end{rotate}}}
\end{picture}
\caption{%
Angular dependence of the mass-scales $m^2_\alpha,m_+^2,m_-^2$ (dotted line,
full line and dashed line, respectively) in $\xi\rightarrow \infty$ limit.
}
\label{fig:LXmasses}
\end{figure}

Another case where one is able to calculate analytic results for the structure
functions is the limit $\xi\rightarrow \infty$. By considering the action
of the distribution Eq.(\ref{squashing}) on some smooth test function
$\Phi({\bf p})$,
\beq
\digamma=\lim_{\xi\rightarrow \infty}\int \frac{d^3 p}{(2\pi)^3}%
h_{\xi}({\bf p})%
\Phi({\bf p})
\label{LXlimit1}
\eeq
and choosing ${\bf p}\cdot{\bf \hat n}$ to be parallel to the $p_z$-axis
one can rescale %
$\cos{\theta}\rightarrow \frac{\cos{\theta}}{\sqrt{\xi}}$
so that the
limit $\xi \rightarrow \infty$ can be performed, 
\beq
\digamma= \int \frac{p^2 dp d\phi }{(2\pi)^3} \int_{-\infty}^{\infty} %
d\cos{\theta} \ %
h_{\rm iso}\left(p \sqrt{1+\cos^2 {\theta}}\right)%
\Phi\left(\begin{array}{c} p \cos{\phi}\\p \sin{\phi} \\ 0 \end{array}\right).
\eeq
This means that in the limit $\xi\rightarrow \infty$ Eq.(\ref{squashing}) 
takes the form
\beq
\lim_{\xi\rightarrow \infty} 
h_{\xi}({\bf p})
\rightarrow \delta({\bf \hat{p}}\cdot{\bf \hat n}) \int_{-\infty}^{\infty} dx\ 
h_{\rm iso}\left(p \sqrt{1+x^2}\right),
\eeq
which corresponds to the extreme anisotropic case considered by 
\textsc{Arnold, Lenaghan} and \textsc{Moore} 
\cite{Arnold:2003rq}. As a consequence,
one can make use of 
this form by partially integrating Eq.(\ref{selfenergy2}) 
\beq
\Pi^{i j}(K)=4 \pi \alpha_s \int \frac{d^3 p}{(2\pi)^3} \frac{h_{\xi}%
({\bf p})}{p}%
\left[\delta^{i j}-\frac{k^i v^j+k^j v^i}{-K\cdot V-i\epsilon}+%
\frac{(-\omega^2+k^2)v^i v^j}{(-K\cdot V-i \epsilon)^2}%
\right]
\label{selfenergy3}
\eeq
and applying the techniques from Ref.~\cite{Arnold:2003rq} to obtain 
analytic expressions for the structure functions in the large $\xi$ limit.
Using 
\beq
\lim_{\xi\rightarrow \infty}%
4 \pi \alpha_s \int \frac{d^3 p}{(2\pi)^3} \frac{h_{\xi}%
({\bf p})}{p}= m_D^2\frac{\pi}{4}
\eeq
one immediately finds
\beq
\Pi^{i j}(K)=\frac{m_D^2 \pi }{4} \int_{0}^{2 \pi} \frac{d \phi}{2 \pi}%
\left[\delta^{i j}-\frac{k^i v^j+k^j v^i}{-K\cdot V-i\epsilon}+%
\frac{(-\omega^2+k^2)v^i v^j}{(-K\cdot V-i \epsilon)^2}%
\right].
\eeq
The remaining average over angles can easily be done, obtaining
the structure functions through the contractions Eq.(\ref{contractions}),
giving
\bqa
\alpha&=&\frac{m_D^2 \pi}{4} \left[-{\rm cot}^2{\theta_n} +%
\frac{\hat{\omega}}{\sin^2{\theta_n}}%
\left(\hat{\omega}+\frac{1-\hat{\omega}^2}{%
\sqrt{\hat{\omega}+\sin{\theta_n}}\sqrt{\hat{\omega}-\sin{\theta_n}}%
}\right)\right]\nonumber\\
\beta&=&\frac{m_D^2 \pi}{4} \hat{\omega}^2\left[-1+\hat{\omega} \frac{\hat{\omega}^2+\cos{2\theta_n}}{%
(\hat{\omega}+\sin{\theta_n})^{3/2}(\hat{\omega}-\sin{\theta_n})^{3/2}%
}\right] \nonumber \\
\gamma&=&\frac{m_D^2 \pi}{4} \frac{1-\hat{\omega}^2}{4\sin^2{\theta_n}}\left[%
6+2\cos{2 \theta_n}+\hat{\omega} \frac{3-6 \hat{\omega}^2-2(1+\hat{\omega}^2) \cos{2\theta_n}-\cos{%
4 \theta_n}}{%
(\hat{\omega}+\sin{\theta_n})^{3/2}(\hat{\omega}-\sin{\theta_n})^{3/2}%
}\right]\nonumber\\
\delta&=&\frac{m_D^2 \pi}{4 k} \frac{\cos{\theta_n}}{\sin^2{\theta_n}}%
\hat{\omega}\left[\hat{\omega}+\frac{-(1-\hat{\omega}^2)^2+(1-2 \hat{\omega}^2) \cos^2{\theta_n}}{%
(\hat{\omega}+\sin{\theta_n})^{3/2}(\hat{\omega}-\sin{\theta_n})^{3/2}%
}\right].
\label{LXstruf}
\eqa
It is then straightforward to extract the static limit of these structure
functions, giving the mass scales $m_{\alpha}^2,m_+^2,m_-^2$ in the large-$\xi$
limit. The result is plotted in Fig.~\ref{fig:LXmasses} as a function
of the angle $\theta_n$. As can be seen, the mass scales diverge 
at $\theta_n=0$, whereas $m_-^2$ and $m_+^2$ coincidence at $\theta_n=\pi/2$
and $m^2_\alpha$ vanishes.
Also, one recovers the results from \textsc{Arnold, Lenaghan} and 
\textsc{Moore} 
\cite{Arnold:2003rq} that the scale $m_-^2$ is only negative for
$\sin\theta_n<1/\sqrt{2}$ (corresponding to the presence of the 
electric instability in
this reference) and the various growth rates of the unstable modes.
Finally, one can show that the expressions of the structure functions for
general $\xi$ converge towards the above analytic results when
$\xi$ becomes large.

\section{$\xi\rightarrow -1$ limit}

To get some insight in the case of $\xi\rightarrow -1$ where the 
 distribution function Eq.(\ref{squashing}) becomes
extremely elongated one can again consider its action on a test function
$\Phi({\bf p})$,
\beq
\digamma=\lim_{\xi\rightarrow -1} \int \frac{d^3 p}{(2\pi)^3}%
h_{\xi}({\bf p})
\Phi({\bf p}).
\label{LXlimit2}
\eeq
Performing a simple scaling $p_z\rightarrow\frac{p_z}{%
\sqrt{1+\xi}}$ one obtains the interesting result
\beq
\digamma=\lim_{\xi\rightarrow -1} \int \frac{d^2 p_{\perp} d\phi}{(2\pi)^3}%
\int_{-\infty}^{\infty}dp_z \ 
h_{\rm iso}\left(\sqrt{{\bf p}_{\perp}^2+p_z^2}\right)
\Phi\left(\begin{array}{c}{\bf p_{\perp}} \\ \frac{p_z}{\sqrt{1+\xi}}%
 \end{array}\right)
\eeq
indicating 
that the argument of the test function $\Phi$ is pushed to very large
values, where $\Phi$ is usually taken to be vanishing. However, one can
get a bit more insight by compactifying momentum space to a 
sphere and considering the coordinate map 
\beq
{\bf p}\rightarrow \frac{{\bf p}^{\prime}}{{\bf p}^{\prime \ 2}},
\label{northpole}
\eeq
with the volume element transforming to
\beq
d^3 p \rightarrow \frac{d^3 p^{\prime}}{p^{\prime \  6}}. 
\eeq
One then considers the action of the distribution function on a test
function $\Phi^{\prime}$ in the primed coordinate system,
\beq
\digamma=\lim_{\xi\rightarrow -1} N(\xi) \int \frac{d^3 p^{\prime}}{(2\pi)^3}%
\frac{1}{p^{\prime\ 6}}
h_{\rm iso}\left(\frac{\sqrt{{\bf p^{\prime}}^2+%
\xi({\bf p^{\prime}}\cdot{\bf \hat n})^2}}{{\bf p^{\prime}}^2}\right)%
\Phi^{\prime}({\bf p}^{\prime}).
\eeq
Taking again ${\bf \hat{p}^{\prime}}\cdot{\bf \hat n}=\cos{\theta^{\prime}}$ 
one scales 
\beq
p^{\prime}\rightarrow p^{\prime} \sqrt{1+ \xi \cos^2{\theta^{\prime}}} 
\eeq
to obtain
\beq
\digamma=\lim_{\xi\rightarrow -1} \int \frac{d^3 p^{\prime}}{(2\pi)^3}%
\frac{1}{p^{\prime\ 6}} \frac{N(\xi)}{(1+\xi \cos^{2}\theta^{\prime})^{3/2}}
h_{\rm iso}\left(\frac{1}{p^{\prime}}\right)
\Phi^{\prime}({\bf p^{\prime}}\sqrt{1+ \xi \cos^2\theta^{\prime}}).
\eeq
Since the isotropic distribution function consisting of Bose-Einstein
and Fermi-Dirac statistical factors lifts the possible singularity at
$p^{\prime}=0$, the only ``dangerous'' term in the integrand is
$1+\xi \cos^2\theta$ which becomes zero for $\xi\rightarrow -1$ and 
small angles $\theta^{\prime}$. 
Substituting therefore $\cos^2 \theta=1-(1+\xi) \chi$
one can perform the limit $\xi\rightarrow -1$, obtaining
\beq
\digamma=\int \frac{d p^{\prime} d\phi d\chi}{(2\pi)^3}%
\frac{1}{p^{\prime\ 4}} \frac{1}{(1+\chi)^{3/2}}
h_{\rm iso}\left(\frac{1}{p^{\prime}}\right)
\Phi^{\prime}(0)=4\pi \Phi^{\prime}(0)\int_{0}^{\infty}dp \ p^2 h_{\rm iso}(p).
\eeq
Therefore, in the limit $\xi\rightarrow -1$, Eq.(\ref{squashing}) simply 
is given by
\beq
\lim_{\xi\rightarrow -1} 
h_{\xi}({\bf p})
d^3p%
\rightarrow \delta^3({\bf p^{\prime}}) d^3 p^{\prime} %
4\pi \int_0^\infty dp \ p^2 h_{\rm iso}(p),
\eeq
which upon insertion into Eq.(\ref{selfenergy3}) shows that in this
limit the self-energy and consequently all structure functions vanish. 
Therefore, the limit $\xi\rightarrow -1$ unfortunately does not correspond
to the interesting physical situation of a line momentum distribution
but rather to the free limit.

\section{Discussion of instabilities}

The existence of unstable modes in an anisotropic quark-gluon plasma 
(first studied by \textsc{Mr\'owczy\'nski} in a series of papers 
\cite{Mrowczynski:1993qm,Mrowczynski:1994xv,Mrowczynski:1997vh})
may play an important role in the dynamical evolution of the quark-gluon 
plasma, since their exponential growth can lead to a more
rapid thermalization and isotropization of the plasma. 
Recently, \textsc{Mr\'owczy\'nski} and \textsc{Randrup} \cite{Randrup:2003cw}
have performed a phenomenological
estimate of the growth rate of the instabilities for scenarios relevant
to the quark-gluon plasma produced at RHIC or LHC, 
using however a different implementation of the anisotropy than
Eq.(\ref{squashing}). They found that the degree of amplification of the
instabilities is not expected to dominate the dynamics of the quark-gluon
plasma, but instead their effect would be comparable to the contribution
from elastic Boltzmann collisions. However, they also pointed out that
if a large number of unstable modes would be excited then their combined
effect on the overall dynamics could well be significant.
More recently, \textsc{Arnold, Lenaghan} and \textsc{Moore}
\cite{Arnold:2003rq}
investigated the case corresponding to the $\xi\rightarrow \infty$ limit
from above, arguing that it drastically modifies the 
scenario of ``bottom-up thermalization'' advocated by 
\textsc{Baier, Mueller, Schiff}
and \textsc{Son} \cite{Baier:2000sb}, which would then have to be replaced by a
different scheme. 

Furthermore, the presence of instabilities will in general prohibit
the calculation of physical quantities in a perturbative framework, since
they correspond to unregularized singularities of the propagator.
This problem as well as special exception to it will be discussed 
in more detail in chapter \ref{eloss-chap}.

Because of these findings, it is probably fair to say that 
a more intensive study of the instabilities of anisotropic systems
and especially their saturation and effect on the thermalization will 
be of great interest in the near future.

\section{Summary}

In this chapter I calculated the gluon self-energy in a system that has an
anisotropy in momentum space. I presented a tensor basis that can be used
to write the gluon self-energy as the sum of four 
``structure functions'' and showed that in the isotropic limit two of
these structure functions vanish while the others correspond to the
longitudinal and transversal part of the self-energy of chapter 
\ref{peshier-chap}. Considering the static limit of the self-energy it was 
found
that for anisotropic systems there is in general 
a non-vanishing magnetostatic mass-scale, but since the associated mass
squared is negative, this corresponds to the existence of unstable modes.

Furthermore, solving the dispersion relations three stable and 
either one or two unstable modes were found, 
depending on whether the distribution function
is stretched or squeezed. I also considered the limit of weak anisotropies
and two cases where the anisotropy gets extremely strong, deriving
analytic expressions for the structure functions which serve as verification
of the general result and provide the basis for further analytic studies.
I conclude by reviewing
the current knowledge about the effects of the unstable modes on the 
thermalization of the quark-gluon plasma produced in heavy-ion collisions.

 \chapter{Energy loss of a heavy parton}
\label{eloss-chap}

In this chapter I will make use of the resummed self-energies derived
for isotropic and anisotropic systems obtained in chapters 
\ref{peshier-chap} and \ref{asi-chap} to calculate the collisional energy loss 
of a heavy parton in a deconfined QCD plasma. 
In the context of heavy-ion collisions, a theoretical understanding
of the heavy parton energy loss in a quark-gluon plasma
is important for a proper comparison
with experimental results, e.g. for the inclusive electron spectrum (measured
recently at RHIC \cite{Adcox:2002cg}) and for jet suppression rates.
In the following I will focus on the collisional energy loss (for a review
on the radiative energy loss see e.g. \cite{Baier:2000mf,Accardi:2003gp}) 
using the technique
of \textsc{Braaten} and \textsc{Thoma}
\cite{Braaten:1991jj,Braaten:1991we} which gives
the complete leading-order result for this quantity.

The main idea behind this technique is to consider independently the 
contributions from soft (involving momenta $q\sim m_D$) and 
hard ($q\sim T$) gluon exchange, which are separated by some arbitrary
momentum scale $q^{*}$ that cuts off the UV and IR divergences of the 
soft/hard contributions, respectively. 
Moreover, it was found that in the weak-coupling limit $\alpha_s\ll 1$ the
condition $m_D\ll q^* \ll T$ could be used to expand the resulting
integral expressions for the soft and hard contributions further, giving
an analytic result for the energy loss independent of the separation scale
$q^*$. However, in the case of QCD where the coupling becomes quite large,
this approach may give unphysical results for energy loss, as will be shown
later on.

It turns out that when not expanding the integral expressions with respect
to the condition $m_D\ll q^* \ll T$, in the isotropic case
both the hard and soft contributions to the energy loss 
{\em always} stay positive, even for very large coupling, as has been
shown in QED \cite{Romatschke:2003vc}. However, this comes at the
expense of giving up independence of the complete result on the scale $q^*$,
which then has to be fixed somehow. Fortunately, it turns out that in the
weak coupling limit the scale dependence becomes very small, too, so that 
unless $q^*$ is taken to be very large ($q^*\gg T$) 
or very small ($q^*\ll m_D$) one recovers the 
result found by \textsc{Braaten} and \textsc{Thoma}. 
When the coupling is increased one
can then still fix $q^*$ by the ``principle of minimal sensitivity'',
which means that $q^*$ is chosen such that the energy loss (and therefore
also the variation with respect to $q^*$) is minimized.
As will be shown, this procedure eliminates the unphysical result of negative
energy loss in the isotropic case
and also provides an estimate on the theoretical uncertainty
of the result by varying $q^*$ by a fixed amount around the minimizing value.

Moreover, it will be shown that the technique by \textsc{Braaten} and 
\textsc{Thoma} can also
be applied to anisotropic systems. This is in fact non-trivial since
the presence of plasma instabilities in general could render the calculation
of the soft part divergent; however, the collisional energy loss turns 
out to be protected by an interesting mechanism dubbed ``dynamical shielding''
\cite{Romatschke:2003vc}, rendering the calculation safe at least to 
leading order, as will be shown in the following.

\section{Soft contribution and effect of instabilities}

The soft contribution to the collisional energy loss can be calculated 
using classical field theory methods. Starting from the 
classical expression for the parton energy loss per unit of time given by
\beq
\left({{\rm d} W\over {\rm d} t}\right)_{\rm soft} = {\rm Re} \, %
\int d^3{\bf x}
 \; {\bf J}^{a}_{\rm ext}(X)\cdot{\bf E}^{a}_{{\rm ind}}(X) \; ,
\label{eloss1}
\eeq
where $a$ is a color index, %
$X=(t,{\bf x})$, and ${\bf J}_{\rm ext}$ is the current induced by 
a test parton propagating with velocity ${\bf v}$:
\bqa
{\bf J}^{a}_{\rm ext}(X) &=& {\mathcal Q}^{a} {\bf v} \delta^{(3)}({\bf x}-{\bf v}t) \; , \nonumber
 \\
{\bf J}^{a}_{\rm ext}(Q) &=& (2 \pi)  {\mathcal Q}^{a}%
 {\bf v} \delta(\omega - {\bf q}\cdot{\bf v})
 \; ,
\eqa
with $Q=(\omega,{\bf q})$ and ${\mathcal Q}^a$ being the color charge.  Using
\beq
E^{i,a}_{{\rm ind}}(Q) = i \omega \, \left(\Delta^{ij}(Q)-\Delta_{0}^{ij}(Q)\right)%
 J^{j,a}_{\rm ext}(Q) \; ,
\eeq
where $\Delta^{ij}_0$ denotes the free propagator, 
and Fourier transforming to coordinate space one obtains
\beq
E^{i,a}_{{\rm ind}}(X) = i {\mathcal Q}^a %
 \int {d^3 {\bf q} \over (2 \pi)^3} \, ({\bf q}\cdot{\bf v}) (\Delta^{ij}(Q)-\Delta^{ij}_0(Q)) v^j%
  e^{i({\bf q}\cdot{\bf x}-({\bf q}\cdot{\bf v}) t) } \; ,
\eeq
where $Q=({\bf q}\cdot{\bf v},{\bf q})$. 
Using the above equation one finds for the soft contribution 
Eq.(\ref{eloss1}) to 
the heavy parton energy-loss 
\beq
-\left({{\rm d} W\over {\rm d} t}\right)_{\rm soft} = {\mathcal Q}^2%
  \, {\rm Im} \, \int {d^3 {\bf q} \over (2 \pi)^3} \, ({\bf q}\cdot{\bf v}) v^i %
\left[\Delta^{ij}(Q)-\Delta^{ij}_0(Q)\right] v^j  \; .
\label{eloss2}
\eeq

\subsection{Dynamical shielding of instabilities}

In the case of an anisotropic system unstable modes are present
that manifest themselves by poles of the propagator in the static limit
$\lim_{\omega\rightarrow 0} \Delta_{i j}^{-1}=0$, as has been discussed
in chapter \ref{asi-chap}. To simplify the argument, one can use the
tensor decomposition of the propagator in chapter \ref{asi-chap} and
concentrate on the poles of $\Delta_{A}$ given in Eq.(\ref{gluonpropmodes}) 
only, while the poles of
$\Delta_{G}$ can be treated in a similar way. Schematically the soft 
contribution to the energy loss Eq.(\ref{eloss2}) is then given as
\bqa
\left({{\rm d} W\over {\rm d} x} \right)_{\rm A,\,soft} &\sim& 
{\rm Im} \, \int d\Omega \int q\,dq \, %
 \, \frac{\hat{\omega}}{(q^2-q^2\hat{\omega}^2 +\alpha)} \; .
\label{elossSchematic}
\eqa
Since the structure function $\alpha$ for $\xi>0$ is negative-valued, 
\beq
\lim_{\omega\rightarrow0} \alpha(\omega,q) = M^2(-1 + i D \hat\omega) +%
 {\cal O}(\omega^2) \; ,
\label{alphastatl}
\eeq
in the static limit the integrand in Eq.(\ref{elossSchematic})
seems to contain
an unregulated singularity at $q=M$, which therefore renders the result
divergent. Thus the presence of instabilities in a system 
is believed to generically prohibit the calculation of quantities in 
a perturbative framework since those quantities will be plagued by unregulated
divergencies \cite{MooreBiel}. 

The schematic integrand for the energy loss in Eq.(\ref{elossSchematic}),
however, turns out to be ``shielded'' from this singularity, as can be
seen by taking the limit 
\bqa
\lim_{\hat\omega\rightarrow 0} \, \lim_{q\rightarrow M} 
\frac{\hat\omega}{q^2-q^2\hat{\omega}^2 + M^2(-1+iD\hat\omega)} 
\;\rightarrow\; \frac{1}{M^2(\hat{\omega}+ i D)} 
\;\rightarrow\; -\frac{i}{M^2 D} 
\; .
\label{elossSchematic2}
\eqa
This limit exists as 
long as the coefficient $D$ in Eq.(\ref{alphastatl}) is non-vanishing,
being somewhat similar to dynamical screening of the magnetic sector
of finite temperature QCD.
Therefore, the integral Eq.(\ref{elossSchematic}) may safely be performed,
as long as $D\neq 0$. 

\subsubsection{Proof of dynamical shielding I: small $\xi$}

Since dynamical shielding relies crucially on the fact that the imaginary
part of the structure function is of $O(\omega)$ when the real part is 
negative-valued, a simple proof of this assumption
valid for the weak-anisotropy or small-$\xi$ regime will be given here. 
For $\Delta_A$
the proof is straightforward since ${\rm Im}\ \Delta_A^{-1}={\rm Im}\ \alpha$
and from Eqs.(\ref{smallxi}) one finds
\beq
\lim_{\hat\omega\rightarrow0} 
\alpha = m_D^2 \left[ - {\xi\over6} (1+\cos2\theta_n) 
- {i \pi \over 4} \hat\omega \left( 1
 + {3 \xi \over 4} (1+\cos2\theta_n) 
 \right) \right] \, .
\eeq
From this expression it is clear that the 
imaginary value cannot change sign for any positive value of $\xi$, whereas
for negative value of $\xi$, the static limit of $\alpha$ is positive-valued,
so no instability exists in $\Delta_A$.
To prove that $\Delta_G$ is also safe, one first uses its small-$\xi$
decomposition Eq.(\ref{SXdec}) to see that only the $\alpha+\gamma$ 
term can become negative-valued. Using once again the expressions in 
Eq.(\ref{smallxi}) one finds after a little bit of algebra
\beq
\lim_{\hat\omega\rightarrow0} \alpha+\gamma=%
m_D^2 \left[ - {\xi\over6} (-2+4\cos^2\theta_n) 
- {i \pi \over 4} \hat\omega \left( 1
 + {\xi \over 4} (-4+10\cos^2\theta_n) 
 \right) \right].
\eeq
For $\xi>0$ the instability is present for 
$\cos^2 \theta_n>\frac{1}{2}$, while the imaginary part may change its sign
only for $\cos^2 \theta_n<\frac{4}{10}$; for $\xi<0$, the instability
occurs for $\cos^2 \theta_n<\frac{1}{2}$ while even for the ``worst
case'' $\xi=-1$ the imaginary part can only change its sign for 
$\cos^2 \theta_n>\frac{8}{10}$, so also the $\Delta_G$ contribution to the
energy loss is protected by dynamical
shielding. This completes the proof for the case of \hbox{small $\xi$}.

\subsubsection{Proof of dynamical shielding II: large $\xi$}

Also for very large $\xi$ it can be shown analytically 
that the energy loss calculation
is protected by dynamical shielding. Considering the $\Delta_A$ contribution
first one can make use of the analytic expressions in Eq.(\ref{LXstruf})
to find
\beq
\lim_{\hat\omega\rightarrow0} 
\alpha = \frac{m_D^2 \pi}{4} \left[ -{\rm cot}^2\theta_n-\frac{%
i\hat{\omega}}{\sin^3{\theta_n}}\right] \, ,
\eeq
where again it can immediately be seen that the $O(\omega)$ contribution to
the imaginary part never vanishes.
For $\Delta_G$ it turns out that a factorization similar to the small-$\xi$
case is possible,
\beq
\Delta_G(\hat{\omega},q)=\hat{\omega}^2(q^2+q_+^2)(q^2+q_-^2),
\eeq
where
\bqa
q_+^2&=&\frac{m_D^2 \pi}{16 \sin^2 \theta_n}\left[2(1+\cos^2\theta_n)+2%
\sqrt{\cos^4 \theta_n+4 \cos^2\theta_n}\right]+O(\hat{\omega}) \nonumber\\
q_-^2&=&\frac{m_D^2 \pi}{16 \sin^2 \theta_n}\left[4-2\cos^2\theta_n-2%
\sqrt{\cos^4 \theta_n+4 \cos^2\theta_n}\right.+\nonumber\\
&&\left.+\frac{i \hat{\omega}}{(4+\cos^2 \theta_n)\sin{\theta_n}}%
\left(8+2 \cos^2\theta_n-3 \sqrt{2} \sqrt{%
(8+2 \cos^2 \theta_n) \cos^2\theta_n}\right)\right]. \qquad \;
\eqa
As can be seen, $q_+^2>0$ in the whole region $0<\theta_n<\frac{\pi}{2}$,
so one can limit the considerations to $q_-^2$. There a little bit of
algebra shows
that $q_-^2<0$ for \hbox{$\cos^2\theta_n>\frac{1}{2}$}, while on the other hand
the imaginary part changes sign at \hbox{$\cos^2\theta_n=\frac{1}{2}$}, so all
angles $\theta_n \neq \frac{\pi}{4}$ are protected through dynamical
shielding, while for the angle $\theta_n=\frac{\pi}{4}$ the singularity
at $q=0$ of $\Delta_G$ is rendered harmless by the volume element of 
the integration. This completes the proof for the case of large $\xi$.

\subsection{Application of dynamical shielding}

Unfortunately a proof that shows that dynamical shielding protects the 
soft contribution to the energy loss of naked singularities is rather 
difficult for general $\xi$. Numerical evidence however suggests that 
dynamical shielding also works for general $\xi$,
which is not covered by the above proofs. Therefore, one can
be reasonably confident that the singularities coming from the unstable
modes are rendered safe everywhere and one may proceed to evaluate 
the soft contribution to the energy loss.

Using the tensor basis derived in chapter \ref{asi-chap} to invert the
propagator in Eq.(\ref{eloss2}) and taking the contractions 
\bqa
v^i A^{ij} v^j &=& v^2 - (\hat{\bf q}\cdot{\bf v})^2 \; , \nonumber \\
v^i B^{ij} v^j &=& (\hat{\bf q}\cdot{\bf v})^2 \, , \nonumber \\
v^i C^{ij} v^j &=& (\tilde{\bf n}\cdot{\bf v})^2/\tilde{n}^2  \, ,\nonumber \\
v^i D^{ij} v^j &=& 2 ({\bf q}\cdot{\bf v}) ( \tilde{\bf n}\cdot{\bf v})  \; ,
\eqa
one obtains
\bqa
-\left({{\rm d} W\over {\rm d} x}\right)_{\rm soft} &=& \frac{{\mathcal Q}%
^2}{v}  \, {\rm Im} \, \int {d^3 {\bf q} \over (2 \pi)^3} \, \omega \,
         \left(\Delta_A(Q)-{1\over q^2-\omega^2}\right) \, \left[ v^2 - {\omega^2 \over q^2} - {(\tilde{\bf n}\cdot{\bf v})^2\over\tilde{n}^2} \right] \nonumber
 \\
       && \hspace{-7mm} + \,\omega \, \Delta_G(Q) \, \left[ {\omega^2\over q^2}(q^2 - \omega^2 + \alpha + \gamma) + (\beta-\omega^2) {(\tilde{\bf n}\cdot{\bf v}
)^2 \over \tilde{n}^2}  - 2 \delta \omega ( \tilde{\bf n}\cdot{\bf v}) \right]
        \; \nonumber \\
       && \hspace{-7mm} + \, {1\over\omega(q^2-\omega^2)} \, \left[ {\omega^2\over q^2}(q^2 - \omega^2) - \omega^2 {(\tilde{\bf n}\cdot{\bf v})^2 \over \tilde{n}^2}  \right]
        \; ,
\label{elossReg}
\eqa
where $dW/dx=v^{-1} dW/dt$ and $\omega = {\bf q}\cdot{\bf v}$.  
After some algebraic transformations involving scaling out the momentum 
one obtains
\bqa
-\left({{\rm d} W\over {\rm d} x} \right)_{\rm soft} &=& \frac{{\mathcal Q}%
^2}{v}
\, {\rm Im} \, \int {d^3 {\bf q} \over (2 \pi)^3} \, \frac{\hat{\omega}}{q (1-\hat{\omega}^2)} \, 
\Biggl[\frac{- \alpha}{(q^2-q^2\hat{\omega}^2 +\alpha)} (v^2-
\hat{\omega}^2-{(\tilde{\bf n}\cdot{\bf v})^2\over\tilde{n}^2}) \nonumber \\
&& \hspace{5.5cm} + \frac{q^2 \mathcal{A}+\mathcal{B}}{q^4 \mathcal{C}+q^2 \mathcal{D}+ \mathcal{E}}\Biggr],
\label{elossReg2}
\eqa
where 
\bqa
\mathcal{A}&=&(1-\hat{\omega}^2)^2 \beta+\hat{\omega}^2%
{(\tilde{\bf n}\cdot{\bf v})^2\over\tilde{n}^2} (\alpha+\gamma)-2 \hat{%
\omega}(1-\hat{\omega}^2)(\tilde{\bf n}\cdot{\bf v}) \hat{\delta} \nonumber \; ,\\
\mathcal{B}&=&\left((\alpha+\gamma)\beta-\tilde{\bf{n}}^2\hat{\delta}^2\right)%
(1-\hat{\omega}^2-{(\tilde{\bf n}\cdot{\bf v})^2\over\tilde{n}^2})\nonumber \; ,\\
\mathcal{C}&=&-\hat{\omega}^2(1-\hat{\omega}^2)\nonumber \; ,\\
\mathcal{D}&=&-\hat{\omega}^2(\alpha+\gamma)+(1-\hat{\omega}^2)\beta \nonumber%
\\
\mathcal{E}&=&(\alpha+\gamma)\beta-\tilde{\bf{n}}^2\hat{\delta}^2 \, ,
\eqa
with $\hat{\omega}=\hat{\bf q}\cdot{\bf v}$ and $\hat{\delta}=q \delta$. 
Since all the
momentum dependence has been made explicit and possible singularities coming
from the instabilities 
are known to be shielded one can perform the $q$ integration, obtaining
\bqa
-\left({{\rm d} W\over {\rm d} x}\right)_{\rm soft}\!\!\!\!\!\!\! &=& %
\!\!\! \frac{{\mathcal Q}%
^2}{v}  \, {\rm Im} \, %
\int {d \Omega_{q}\over (2 \pi)^3} \, %
\frac{\hat{\omega}}{(1-\hat{\omega}^2)} \, %
\Biggl[-\alpha %
\frac{(v^2-\hat{\omega}^2-{(\tilde{\bf n}\cdot{\bf v})^2\over\tilde{n}^2})}
{2(1-\hat{\omega}^2)}%
\ln{\frac{q^{* 2}(1-\hat{\omega}^2)+\alpha}{\alpha}} \nonumber \\
&& \hspace{5.5cm} + F(q^{\star})-F(0)\Biggr] \,,
\label{Elosssoftfinal}
\eqa
where
\beq
F(q)=\frac{\mathcal{A}}{4 \mathcal{C}} \ln{\left(%
-4\mathcal{C}\left(\mathcal{C} q^4+\mathcal{D}q^2+\mathcal{E}\right)%
\right)}
+\frac{\mathcal{AD}-2\mathcal{BC}}{4 %
\mathcal{C}\sqrt{\mathcal{D}^2-4 \mathcal{CE}}}\ln{\frac{%
\sqrt{\mathcal{D}^2-4 \mathcal{CE}}+\mathcal{D}+2\mathcal{C}q^2}{
\sqrt{\mathcal{D}^2-4 \mathcal{CE}}-\mathcal{D}-2\mathcal{C}q^2}} \; ,
\eeq
and a UV momentum cutoff $q^{*}$ has been introduced on the $q$ integration. 

\subsection{Behavior of the soft part}

\begin{figure}
\begin{center}
\includegraphics[width=.4\linewidth]{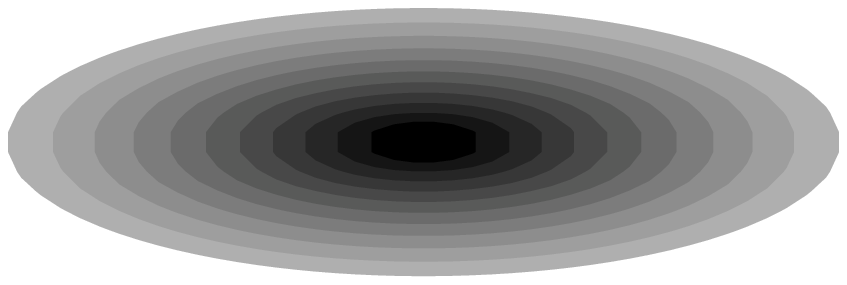}
\setlength{\unitlength}{1cm}
\begin{picture}(0,0)
\put(-3.2,1){\line(0,4){3.8}}
\put(-3.06,4.8){\makebox(0,0){\begin{rotate}{180} \textsf{V} \end{rotate}}}
\put(-3.2,5){\makebox(0,0){\footnotesize ${\bf \hat{n}}$}}
\put(-2.4,4.8){\makebox(0,0){\footnotesize ${\bf v}$}}
\put(-3,4){\makebox(0,0){\footnotesize $\theta$}}
\put(-6,2.92){\line(4,0){6}}
\put(-3.2,2.92){\line(2,5){0.69}}
\put(-2.39,4.6){\makebox(0,0){\begin{rotate}{159} \textsf{V} \end{rotate}}}
\end{picture}
\caption{Sketch of the directional dependence of the energy loss: 
$\bf \hat{n}$ is the direction of the anisotropy, ${\bf v}$ the velocity 
of the parton and $\theta$ the angle between these.}
\label{fig:ELsquas}
\end{center}
\end{figure}

\begin{figure}
\hfill
\begin{minipage}[t]{.45\linewidth}
\includegraphics[width=0.9\linewidth]{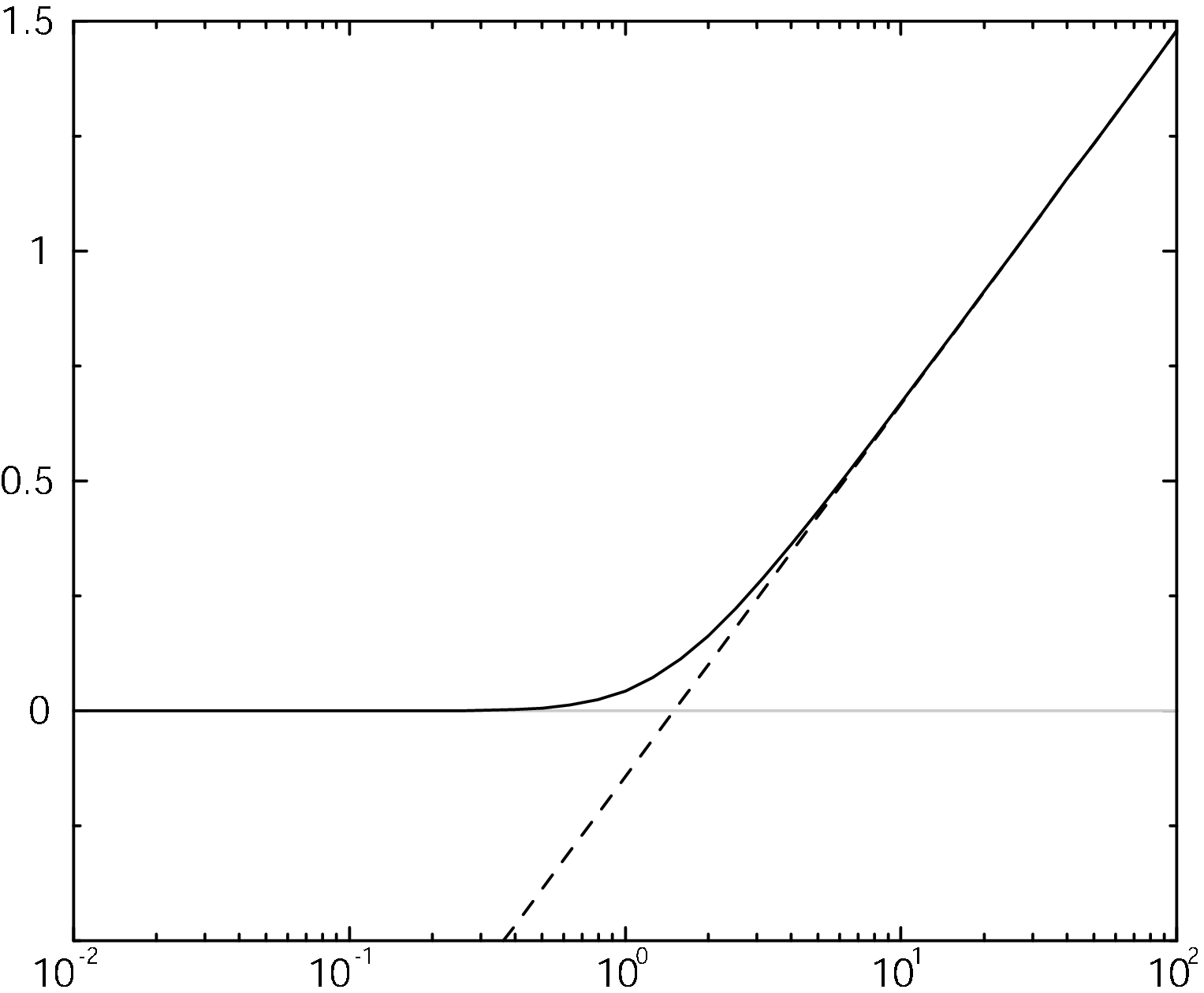}
\setlength{\unitlength}{1cm}
\begin{picture}(9,0)
\put(-0.3,1.2){\makebox(0,0){\begin{rotate}{90}%
\footnotesize
$-\left(\frac{dW}{dx}\right)_{\rm soft,iso}%
/\left(\frac{{\mathcal Q}^2 m_D^2}{8 \pi}\right)$ 
\end{rotate}}}
\put(3,0.25){\makebox(0,0){\footnotesize $q^{*}/m_D$}}
\end{picture}
\vspace{-1cm}
\caption{Soft contribution to the energy loss (full line) as a function
of $q^{*}/m_D$ for $v=0.5$ compared to the Braaten-Thoma result
(dashed line).}
\label{fig:softisoBT}
\end{minipage} \hfill
\hfill
\begin{minipage}[t]{.45\linewidth}
\includegraphics[width=0.9\linewidth]{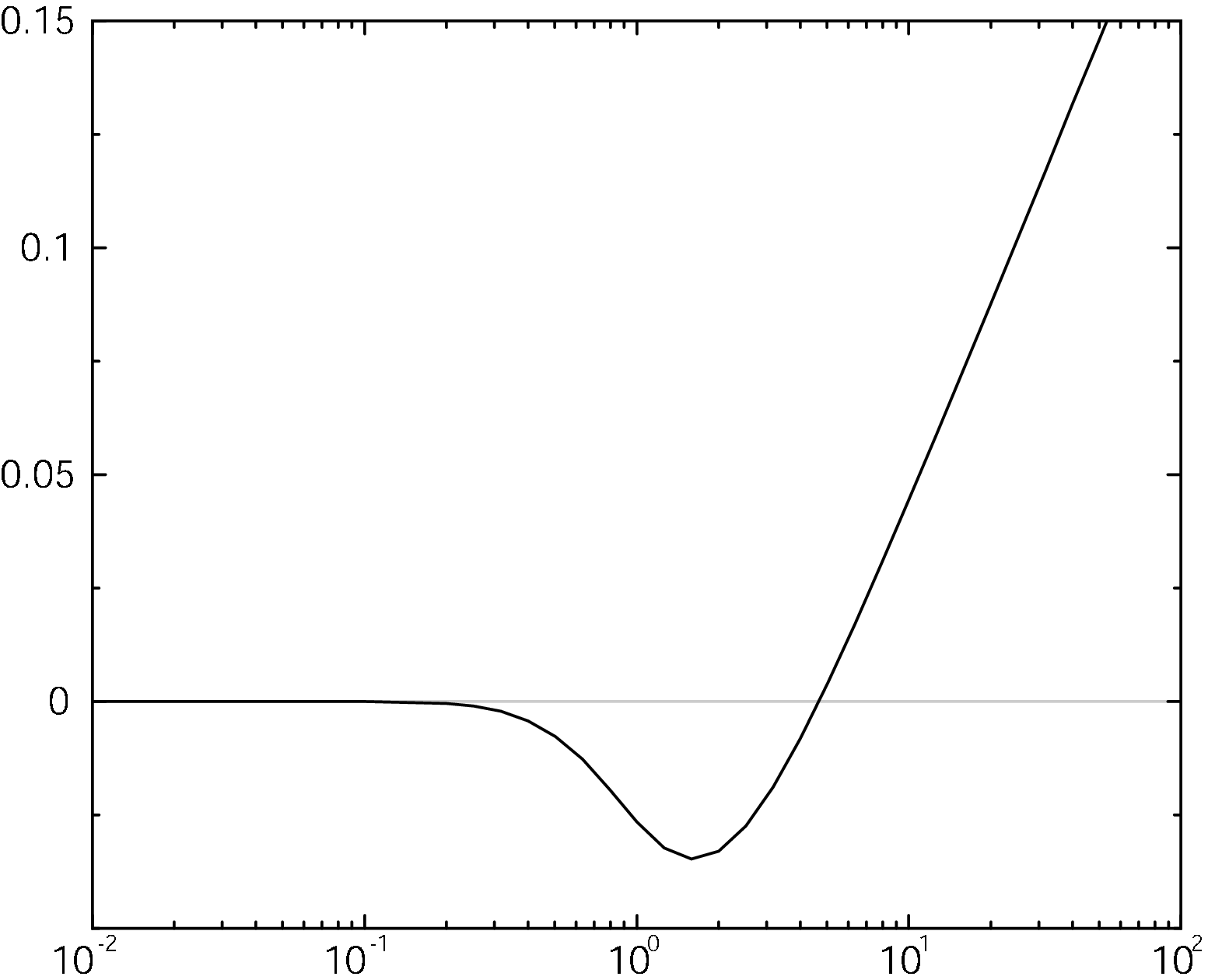}
\setlength{\unitlength}{1cm}
\begin{picture}(9,0)
\put(-0.3,1.2){\makebox(0,0){\begin{rotate}{90}%
\footnotesize $-\left(\frac{dW}{dx}\right)_{\rm soft,\xi_2}%
/\left(\frac{{\mathcal Q}^2 m_D^2}{8 \pi}\right)$ 
\end{rotate}}}
\put(3,0.25){\makebox(0,0){\footnotesize $q^{*}/m_D$}}
\end{picture}
\vspace{-1cm}
\caption{The contribution $-\left({dW/dx}\right)_{\rm soft,\xi_2}$
as a function of $q^{*}/m_D$ for $v=0.5$.}
\label{fig:softxi2}
\end{minipage} \hfill
\end{figure}

To get some idea on how the soft part behaves as a function of the parameters
it is useful to consider once again the limit of small $\xi$ in 
Eq.(\ref{Elosssoftfinal}). It turns out that in this limit one obtains
\beq
-\left({{\rm d} W\over {\rm d} x}\right)_{\rm soft,\,small-\xi}\!\!\!\! =\!\! 
-\left[\left({{\rm d} W\over {\rm d} x}\right)_{\rm soft,iso}\!\!\!
+\xi \left(
\left({{\rm d} W\over {\rm d} x}\right)_{\rm soft,\xi_{1}}+\!\!%
\frac{({\bf v} \cdot {\bf \hat{n}})^2}{v^2}%
\left({{\rm d} W\over {\rm d} x}\right)_{\rm soft,\xi_{2}}\right)\right],
\label{softsmallxi}
\eeq
with 
\pagebreak
\bqa
-\left({{\rm d} W\over {\rm d} x}\right)_{\rm soft,iso} &=& %
\frac{{\mathcal Q}^2}{v^2}  \, {\rm Im} \, %
\int_{-v}^{v} {d \hat{\omega} \over (2 \pi)^2} \, %
\hat{\omega}\left[- \frac{v^2-\hat{\omega}^2}{2 (1-\hat{\omega}^2)^2} %
\Pi_T \ln{\frac{(1-\hat{\omega}^2)q^{*2}+\Pi_T}{\Pi_T}}\right.\nonumber\\
&& \hspace*{3cm}\left.-\frac{\Pi_L}{2}%
\ln{\frac{q^{*2}-\Pi_L}{-\Pi_L}}
\right]
\label{isosoft}
\eqa
being the isotropic result for general $q^*$ (calculated originally 
by Thoma and Gyulassy \cite{Thoma:1991fm}). Note that Eq.(\ref{isosoft}) 
corresponds to the Braaten-Thoma result when expanding the 
logarithms under the assumption $q^*\gg m_D$. The respective results
are compared in Fig.~\ref{fig:softisoBT}: while the results are identical
for large $q^*/m_D$ the Braaten-Thoma result becomes negative for
small $q^*/m_D$
whereas the unexpanded result obtained through numerical integration
is positive for all values of $q^*/m_D$.

The effects of the anisotropy are encoded in the functions
$\left({{\rm d} W/{\rm d} x}\right)_{\rm soft,\xi_{1}}$
and $\left({{\rm d} W / {\rm d} x}\right)_{\rm soft,\xi_{2}}$ which resemble
that of $\left({{\rm d} W / {\rm d} x}\right)_{\rm soft,iso}$, but
since they consist of many more terms than the isotropic contribution 
they will not be listed here explicitly. The direction of the
heavy parton enters the small-$\xi$ result only through the explicit term
$({\bf v} \cdot {\bf \hat{n}})^2/v^2$. 
The function $-\left({{\rm d} W / {\rm d} 
x}\right)_{\rm soft,\xi_{2}}$  which multiplies this term and therefore
controls the directional dependence of the soft part is plotted in
Fig.~\ref{fig:softxi2}. As can be seen in this figure, the 
function is negative for small $q^*/m_D$ but becomes positive for
large $q^*/m_D$. Since the function will eventually be evaluated at some
$q^*/m_D$ that minimizes the overall energy loss depending
on the velocity and the coupling constant, the directional trend
of the soft part reverses once this value of $q^*/m_D$ crosses the
point where $\left({{\rm d} W / {\rm d} x}\right)_{\rm soft,\xi_{2}}=0$.
Therefore, one expects the energy loss to be peaked along ${\bf \hat{n}}$
for some couplings and velocities, while being peaked transverse to 
${\bf \hat{n}}$ for others (see sketch in Fig.~\ref{fig:ELsquas}).

For general $\xi$ it turns out that the results obtained in the small 
$\xi$ limit still hold qualitatively although quantitatively the 
predictions differ considerably once $\xi$ becomes large. Fortunately,
in the limit $\xi\rightarrow \infty$ one can again use the
analytic results for the structure functions of chapter \ref{asi-chap}
as a verification of quantitative results.

\section{Hard Contribution}

\begin{figure}[t]
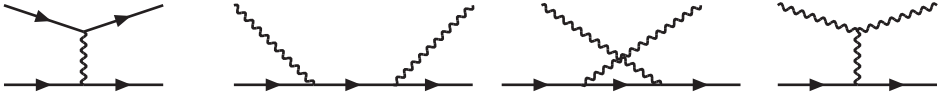

\bqa \nonumber
\picb{\Lqu(0,0)(30,0) \Lqu(30,0)(60,0) \Lgl(30,20)(30,0) \Lqu(0,30)(30,20)%
 \Lqu(30,20)(60,30)}~\hspace{1cm}
\picb{\Lqu(0,0)(30,0) \Lqu(30,0)(60,0) \Lqu(60,0)(90,0) \Lgl(0,30)(30,0)%
 \Lgl(90,30)(60,0)}~
\hspace{1.5cm}
\picb{\Lqu(0,0)(30,0) \Lqu(30,0)(60,0) \Lqu(60,0)(90,0) \Lgl(75,30)(30,0)%
 \Lgl(15,30)(60,0)}
\hspace{1.75cm}
\picb{\Lqu(0,0)(30,0) \Lqu(30,0)(60,0) \Lgl(30,20)(30,0) \Lgl(0,30)(30,20)%
 \Lgl(30,20)(60,30)}
~\hspace{1cm}
\eqa
\caption[a]{%
Tree-level Feynman diagrams for the scattering processes 
$Q q\rightarrow Q q$ (first diagram) and 
$Q g \rightarrow Q g$ (remaining diagrams).}
\label{hard-diagrams}
\end{figure}

The hard contribution can be separated into two parts: one contribution
coming from the scattering of the heavy parton on quarks in the plasma
and another one that takes into account the scattering on gluons (corresponding
to the tree-level diagrams shown in Fig.\ref{hard-diagrams}). Assuming
the velocity of the parton to be much higher then the ratio of the plasma
temperature to the energy of the parton, $v\gg T/E$, the contribution
coming from quark-quark scattering can be reduced to \cite{Braaten:1991we}
\bqa
-\left({\rm d} W \over {\rm d} x\right)_{\rm hard}^{Qq} \!\! &=&%
 {2 (4 \pi)^3 N_f \alpha_s^2 \over 3v}
        \int {d^3{\bf k} \over (2 \pi)^3} {f_{\xi}({\bf k})\over k} 
        \int {d^3{\bf k^\prime} \over (2 \pi)^3} {1 - f_{\xi}({\bf k}^\prime) %
\over k^\prime}
        \delta(\omega-{\bf v}\cdot{\bf q}) \nonumber \\
&&       \hspace*{-7mm}\times \; \Theta(q-q^*) {\omega \over (\omega^2-q^2)^2} 
        \left[ 2 (k-{\bf v}\cdot{\bf k})^2%
                + {1 - v^2 \over 2} (\omega^2 - q^2) \right], \;
\label{hQq}
\eqa
after performing the Dirac traces and evaluating the sum over spins while
the contribution coming from quark-gluon scattering gives
\bqa
-\left({\rm d} W \over {\rm d} x\right)_{\rm hard}^{Qg} \!\!&=&%
 {(4 \pi)^3 \alpha_s^2 \over 2v}
        \int {d^3{\bf k} \over (2 \pi)^3} {n_{\xi}({\bf k})\over k} 
        \int {d^3{\bf k^\prime} \over (2 \pi)^3} {1 + n_{\xi}({\bf k}^\prime) %
\over k^\prime}
        \delta(\omega-{\bf v}\cdot{\bf q}) \Theta(q-q^*) \nonumber \\
 &&       \times \; \omega 
        \left[ \frac{(1-v^2)^2}{(k-{\bf v}\cdot{\bf k})^2}+ 8 \frac{%
(k-{\bf v}\cdot{\bf k})^2+\frac{1-v^2}{2}(\omega^2-q^2)}{(\omega^2-q^2)^2}
\right].\quad
\eqa
Here $f_{\xi}({\bf k})$ and $n_{\xi}({\bf k})$ 
are the anisotropic versions of the tree-level Fermi-Dirac and 
Bose-Einstein distribution functions at zero chemical potential and 
$\omega = k^\prime - k$ while ${\bf q} = {\bf k}^\prime - {\bf k}$. Note
also that $q^*$ acts as IR cutoff for the $q$ integration. Since 
the integrand is odd under the interchange 
${\bf k} \leftrightarrow {\bf k}^\prime$ the terms involving the products 
$f_{\xi}({\bf k})f_{\xi}({\bf k}^\prime)$ and 
$n_{\xi}({\bf k})n_{\xi}({\bf k}^\prime)$ vanish
since they are symmetric. Redefining the origin of the ${\bf k}^\prime$
integration so that it becomes an integration over ${\bf q}$ one obtains
\bqa
-\left({\rm d} W \over {\rm d} x\right)^{Qq}_{\rm hard}%
 &=&{ 16 N_f \alpha_s^2 \over 3v}
        \int {d^3{\bf k} \over (2 \pi)^3} {f_{\xi}({\bf k})\over k} 
        \int_{q^*}^\infty q^2 dq \, \int d\Omega_{q} \;
        \frac{\delta(\omega-{\bf v}\cdot{\bf q})}{|\bf{q}+\bf{k}|} \nonumber \\
        \hspace{1.5cm} && \times \; {\omega \over (\omega^2-q^2)^2} 
        \left[ 2 (k-{\bf v}\cdot{\bf k})^2 + {1 - v^2 \over 2} (\omega^2 - q^2) \right]\; , \nonumber \\
-\left({\rm d} W \over {\rm d} x\right)_{\rm hard}^{Qg} &=&%
 {4\alpha_s^2 \over v}
        \int {d^3{\bf k} \over (2 \pi)^3} {n_{\xi}({\bf k})\over k} %
        \int_{q^*}^\infty q^2 dq \, \int d\Omega_{q} \;
        \frac{\delta(\omega-{\bf v}\cdot{\bf q})}{|\bf{q}+\bf{k}|} \nonumber \\
        &&\times \; \omega 
        \left[ \frac{(1-v^2)^2}{(k-{\bf v}\cdot{\bf k})^2}+ 8 \frac{%
(k-{\bf v}\cdot{\bf k})^2+\frac{1-v^2}{2}(\omega^2-q^2)}{(\omega^2-q^2)^2}
\right], \; \quad \quad
\label{elosshard1}
\eqa
where now $\omega=|{\bf q}+{\bf k}|-k.$ 

Choosing $\bf{v}$ to be the z-axis for the 
$\bf{q}$ and $\bf{k}$ integration
one can rewrite the delta-function as
\bqa
\delta(|{\bf q}+{\bf k}|-k-{\bf v}\cdot{\bf q})&=&\nonumber \\
&& \hspace*{-4.5cm}\frac{\delta(\phi_q-\phi_{0})%
\Theta(k+{\bf v}\cdot{\bf q}) 2 |{\bf q}+{\bf k}|}
{q \sqrt{4 k^2 \sin^2 \theta_k \sin^2 \theta_q-(q(1-v^2 \cos^2 \theta_q)+%
2 \cos \theta_q (k \cos \theta_k-k v))^2}},
\eqa
where $\phi_{0}$ is the solution of the equation 
\beq
\cos (\phi_0-\phi_k)=-\frac{q(1-v^2 \cos^2 \theta_q)+2 \cos \theta_q (k \cos %
\theta_k-k v)}{2 k \sin \theta_k \sin \theta_q}.
\eeq
Including a factor of 2 because of the symmetry 
$\phi_0 \leftrightarrow 2 \pi-\phi_0$ the $\phi_q$ integration is 
straightforward. Moreover, the $q$ integration can be done by scaling 
$k=q z$ as well as $\omega=q v \cos{\theta_q}$, since then 
\beq
n_{\xi}({\bf k})\rightarrow\sqrt{1+\xi}n(q z \sqrt{1+\xi ({\bf \hat{k}}\cdot%
{\bf \hat{n}})^2})
\eeq
and one only needs to consider the integrals over Fermi-Dirac and 
Bose-Einstein distributions,
\bqa
\frac{1}{x^2 T^2}\int_{xT}^{\infty} dq\, q \ f(q)\!\!\!&=&\!\!\!%
\frac{x \ln(1+\exp{(-x)})-{\rm Li}_2(-\exp{(-x)})}{x^2} = F_1(x)\qquad 
\label{F11} \\
\frac{1}{x^2 T^2}\int_{xT}^{\infty} dq\, q \ n(q)\!\!\!&=&\!\!\!%
\frac{-x \ln(1-\exp{(-x)})+{\rm Li}_2(\exp{(-x)})}{x^2} = F_2(x),\qquad
\eqa
where 
$x=\frac{q^{*} z}{T}\sqrt{1+\xi(n_x \sin\theta_k \cos\phi_k+n_z%
 \cos\theta_k)^2}$,
with  $n_z=\bf{\hat{n}}\cdot\hat{\bf v}$ and \hbox{$1=n_x^2+n_z^2$}.
The hard contributions to the energy loss then take the form
\bqa
-\left({\rm d} W \over {\rm d} x\right)_{\rm hard}^{Qq} &\!\!\!=\!\!& %
{8 \alpha_s^2 N_f (\hat{q}^{*})^2 T^2 \sqrt{1+\xi} \over 3 \pi^3 v}
        \int_{0}^{\infty} z dz \int_{-1}^{1} d\cos \theta_k 
        \left[\int_{0}^{2 \pi} \! d \phi_k F_1 (x)\right] %
       \nonumber \\
        && \int_{-1}^{1} d\cos{\theta_{q}} %
     \times \; {\mathcal T}
        {v \cos \theta_q \over (v^2 \cos^2 \theta_q-1)^2} 
        \left[ 2 z^2(1-v \cos \theta_k)^2\right.\nonumber\\
&&\hspace{5cm}\left.  + {1 - v^2 \over 2} (v^2 \cos^2 \theta_q - 1) \right], %
\nonumber\\
-\left({\rm d} W \over {\rm d} x\right)_{\rm hard}^{Qg} &\!\!\!=\!\!& %
{2 \alpha_s^2 (\hat{q}^{*})^2 T^2 \sqrt{1+\xi} \over \pi^3 v}
        \int_{0}^{\infty} z dz \int_{-1}^{1} d\cos \theta_k 
        \left[\int_{0}^{2 \pi} \! d \phi_k F_2 (x)\right] %
       \nonumber \\
        && \int_{-1}^{1} d\cos{\theta_{q}} %
     \times \; {\mathcal T} v \cos \theta_q
 \left[\frac{(1-v^2)^2}{z^2(1-v \cos \theta_k)^2}\right.\nonumber\\
&&\hspace{2cm}\left.
+8 \frac{z^2(1-v \cos \theta_k)^2+\frac{1-v^2}{2}(v^2 \cos^2 \theta_q-1)}
{(v^2 \cos^2 \theta_q-1)^2}\right],\nonumber\\
\label{myelosshard3}
\eqa
where ${\mathcal T}$ denotes the unwieldy expression
\beq
{\mathcal T}=        \Theta(z+v \cos{\theta_q})
        \frac{\Theta(4 z^2 \sin^2 \theta_k \sin^2 \theta_q-(1-v^2 \cos^2 \theta_
q+%
        2 \cos \theta_q z (\cos \theta_k- v))^2)}%
        {\sqrt{4 z^2 \sin^2 \theta_k \sin^2 \theta_q-(1-v^2 \cos^2 \theta_q+%
        2 \cos \theta_q z (\cos \theta_k- v))^2}}.
\eeq
The remaining integrations have to be performed numerically.

\begin{figure}
\begin{center}
\includegraphics[width=0.5\linewidth]{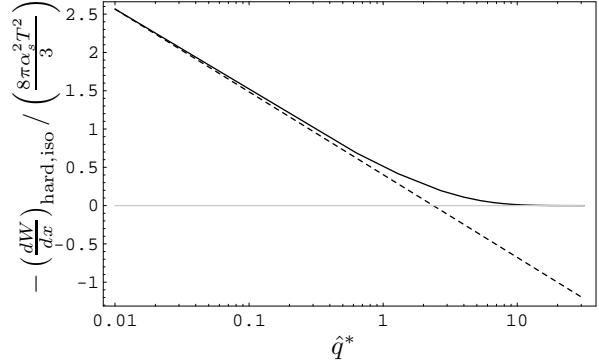}
\end{center}
\setlength{\unitlength}{1cm}
\begin{picture}(5,0)
\put(3.5,1.5){\makebox(0,0){\begin{rotate}{90}%
\footnotesize
$-\left(\frac{dW}{dx}\right)_{\rm hard,iso}%
/\left(\frac{8 \pi \alpha_s^2 T^2}{3}\right)$ 
\end{rotate}}}
\put(7.5,0.8){\makebox(0,0){\footnotesize $\hat{q}^{*}$}}
\end{picture}
\vspace{-1cm}
\caption{Hard contribution to the isotropic energy loss 
for $N_f=2$ (full line) as a function of $\hat{q}^{*}$ for $v=0.5$ 
compared to the Braaten-Thoma result (dashed line).}
\label{fig:hardisoBT}
\end{figure}

\subsection{Behavior of the hard part}

Similar to what has been done for the soft part it is possible to gain
some insight on the parametric dependence of the hard contribution
by considering the anisotropy to be weak ($\xi\ll 1$) \cite{Romatschke:2003vc}.
One can then compare the unexpanded isotropic ($\xi=0$) result to the 
Braaten-Thoma result, shown in Fig. \ref{fig:hardisoBT}.
As one can see,
the two results are identical for small $\hat{q}^*=q^*/T$ while 
for larger $\hat{q}^*$ the Braaten-Thoma result becomes negative
and the unexpanded result Eq.(\ref{myelosshard3}) is positive for all
values of $\hat{q}^*$.

\section{Isotropic results}

\begin{figure}
\hfill
\begin{minipage}[t]{.45\linewidth}
\includegraphics[width=0.9\linewidth]{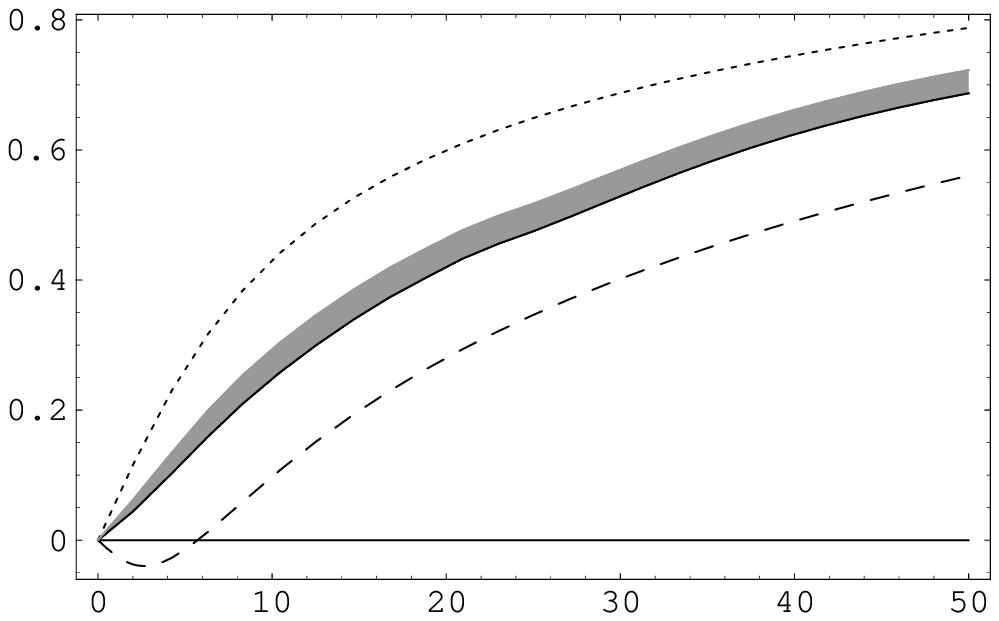}
\setlength{\unitlength}{1cm}
\begin{picture}(9,0)
\put(-0.3,1){\makebox(0,0){\begin{rotate}{90}%
\footnotesize
$-\left(\frac{dW}{dx}\right)_{\rm iso}%
[{\rm GeV/fm}]$ 
\end{rotate}}}
\put(3,0.25){\makebox(0,0){\footnotesize $p [{\rm GeV}]$}}
\put(3,4.5){\makebox(0,0){a}}
\end{picture}
\end{minipage} \hfill
\hfill
\begin{minipage}[t]{.45\linewidth}
\includegraphics[width=0.9\linewidth]{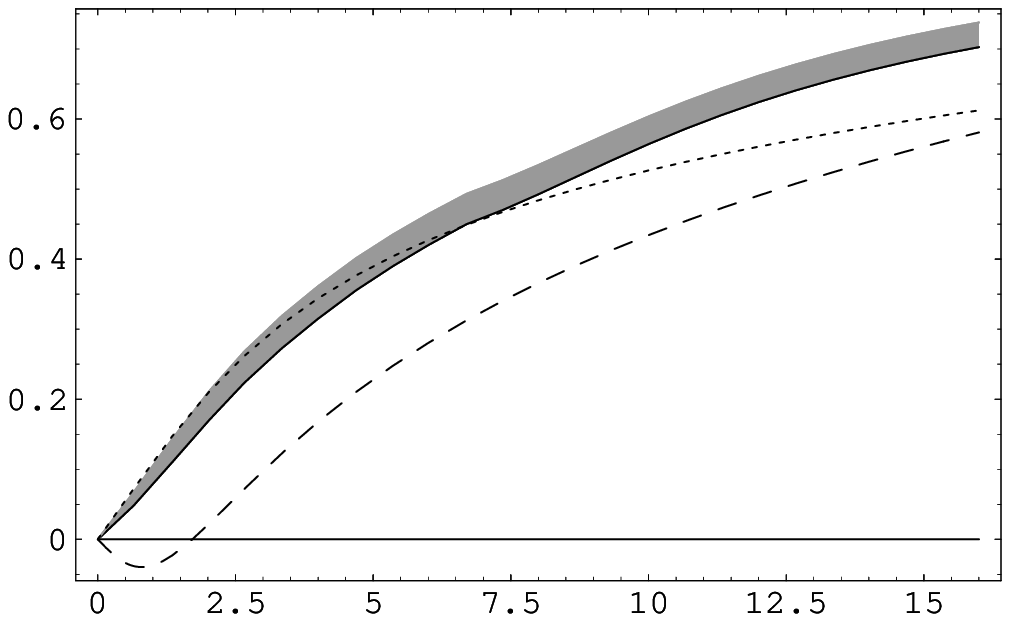}
\setlength{\unitlength}{1cm}
\begin{picture}(9,0)
\put(-0.3,1){\makebox(0,0){\begin{rotate}{90}%
\footnotesize $-\left(\frac{dW}{dx}\right)_{\rm iso}%
[{\rm GeV/fm}]$ 
\end{rotate}}}
\put(3,0.25){\makebox(0,0){\footnotesize $p [{\rm GeV}]$}}
\put(3,4.5){\makebox(0,0){b}}
\end{picture}
\vspace{-1cm}
\end{minipage} \hfill
\caption{Isotropic 
energy loss of a beauty (a) and charm quark (b) as a function
of momentum $p$ for $T=250$MeV and $\alpha_s=0.3$. Shown are the respective
results from Ref.~\cite{Bjorken:1982tu} 
(dotted line) and Ref.~\cite{Braaten:1991we} (dashed line)
as well as the evaluation
of Eq.(\ref{Elossiso}) (full line) with a variation of $q^*$ (gray band).}
\label{fig:QCDELq}
\end{figure}

In the isotropic limit, the total collisional energy loss 
of a heavy parton is obtained by adding
Eq.(\ref{isosoft}) and Eq.(\ref{myelosshard3}) with $\xi=0$,
\beq
-\left(\frac{dW}{dx}\right)_{\rm iso}=-\left[%
\left(\frac{dW}{dx}\right)_{\rm soft,iso}+%
\left(\frac{dW}{dx}\right)_{\rm hard,iso}\right].
\label{Elossiso}
\eeq
It is a function of the strong coupling
$\alpha_s$, the particle velocity $v$, the temperature $T$ and the
momentum separation scale $q^*$.
As has already been discussed above, the $q^*$ dependence of the result
is found to become weak for small values of the
coupling limit $\alpha_s$ (similar to what has been found in QED 
\cite{Romatschke:2003vc}), corresponding to the original result
by \textsc{Braaten} and \textsc{Thoma} \cite{Braaten:1991we}; 
for larger values of the
coupling one fixes $q^*=q^{\rm pms}$ 
using the principle of minimum sensitivity,
\beq
\frac{d }{d q^*} \left. \left(\frac{dW}{dx}\right)_{\rm iso} %
\right|_{q^*=q^*_{\rm pms}}=0.
\label{PMScond}
\eeq 
Note that $-\left. \left(\frac{dW}{dx}\right)_{\rm iso} %
\right|_{q^*=q^*_{\rm pms}}$ always serves as a lower bound on the result
for the energy loss.
To get an estimate of how strongly the result depends on this special
value of $q^{*}$, one can e.g. vary $q^*_{\rm pms}$ by a certain factor 
$c_{q^*}$ and
evaluate the energy loss at $q^*_{\rm pms} c_{q^*}$ and 
$q^*_{\rm pms}/c_{q^*}$. This factor 
$c_{q^*}$ is in
principle arbitrary, but should be such that the resulting $q^*$ is neither
much smaller than $m_D$, nor much bigger than $2 \pi T$; in the following,
I have chosen $c_{q^*}=2$ to be consistent with what has been done in
QED \cite{Romatschke:2003vc}.


In Fig.~\ref{fig:QCDELq}a,b various results for the energy loss of a 
heavy parton in an isotropic quark-gluon plasma are compared: shown
are the results from \textsc{Bjorken} \cite{Bjorken:1982tu}, 
\textsc{Braaten} and 
\textsc{Thoma} \cite{Braaten:1991we}, as well as 
Eq.(\ref{Elossiso}) at $q^*=q^*_{\rm pms}$ 
together with its variation using $c_{q^*}=2$.
For the compilations a temperature of $T=250$MeV and a coupling
constant of $\alpha_s=0.3$ was adopted;
Fig.~\ref{fig:QCDELq}a shows the energy loss of a beauty quark with 
mass $M_{Q}\simeq 5$GeV while in Fig.~\ref{fig:QCDELq}b the energy loss 
of a charm quark with mass $M_{Q}\simeq 1.5$GeV is plotted, both 
as a function of their momenta
\beq
p=\frac{v M_Q}{\sqrt{1-v^2}}.
\eeq

\subsection{Limitations}

Since Eq.(\ref{Elossiso}) has been derived for an infinitely heavy parton,
it breaks down for thermal velocities $v\sim \sqrt{T/M_Q}$ because 
a quark with $v=0$ can only gain energy from collisions with
other particles in the heat bath. A semi-quantitative estimate of 
the velocity where the energy loss becomes negative has been done
in Ref.~\cite{Braaten:1991we} by repeating the above calculation
in the limit $v\rightarrow 0$ and for weak couplings, 
finding $v\sim \sqrt{3 T/M_Q}$ (which corresponds to $p=1.5$GeV and
$p=2.1$GeV for charm and beauty quarks, respectively).
Similarly, Eq.(\ref{Elossiso}) also breaks down for ultrarelativistic
energies $E\gg M_Q^2/T$, with the cross-over energy having been determined
to be $E_{\rm cross}\simeq 1.8 M_Q^2/T$ (corresponding to $v>0.995$ for
both charm and beauty quarks).
However, note that the reason that the Braaten-Thoma result turns negative
for a momentum of $p\simeq 5.7$GeV for the beauty and $p\simeq 1.71$GeV
for the charm quark (as is indicated in Fig.~\ref{fig:QCDELq}a,b)
is not due to this physical reason but rather due to a failure of the
extrapolation from the weak coupling limit to realistic couplings
\cite{Braaten:1991we}.
For the unexpanded result Eq.(\ref{Elossiso}), this unphysical behavior
does not occur and one can therefore expect it to be valid for velocities
down to the original estimate $v\sim \sqrt{3 T/M_Q}$.


Finally, it should be noted that
no estimate on the next-to-leading order (NLO) corrections to the energy loss
has been made here. Therefore, it should be kept in mind that for QCD with
large realistic coupling the inclusion of these NLO corrections might
give energy loss results that are not covered by the variations of $q^*$
in the leading-order result, so these variations should be interpreted
with care.

\section{Anisotropic results}

\begin{figure}
\hfill
\begin{minipage}[t]{.45\linewidth}
\includegraphics[width=0.9\linewidth]{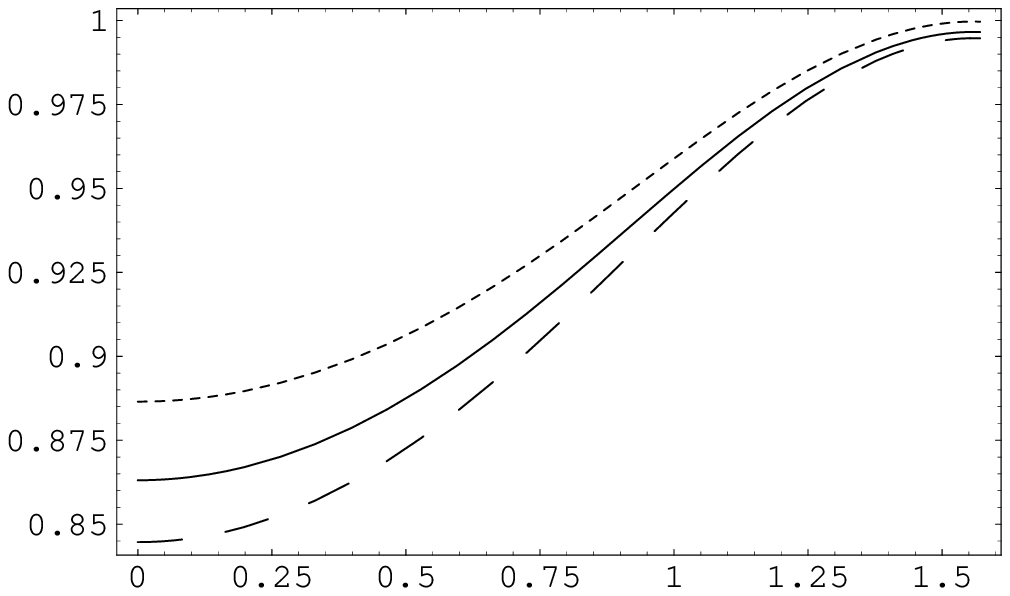}
\setlength{\unitlength}{1cm}
\begin{picture}(9,0)
\put(-0.5,1.5){\makebox(0,0){\begin{rotate}{90}%
\footnotesize
$\left(\frac{dW}{dx}\right)/\left(\frac{dW}{dx}\right)_{\rm iso}$
\end{rotate}}}
\put(3,0.25){\makebox(0,0){\footnotesize $\theta$}}
\end{picture}
\vspace{-1cm}
\caption{Anisotropic energy loss at \hbox{$v=0.5$}, 
$\xi=1.0$ for $\alpha_s=0.2,0.3,0.4$
(dotted, full and dashed line, respectively) as a function of
the \hbox{angle $\theta$} (normalized to the isotropic result).}
\label{fig:ELang1}
\end{minipage} \hfill
\hfill
\begin{minipage}[t]{.45\linewidth}
\includegraphics[width=0.9\linewidth]{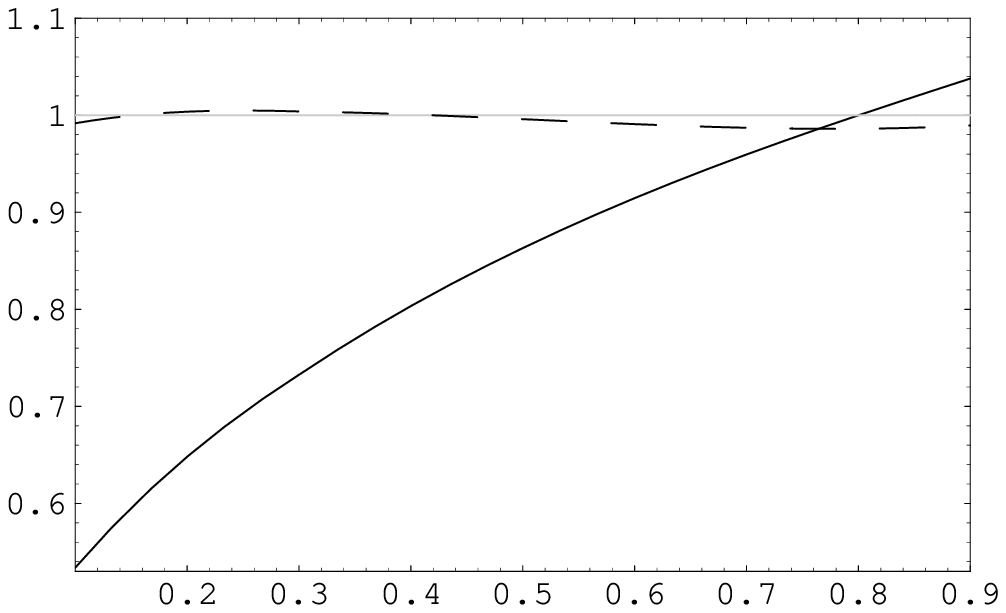}
\setlength{\unitlength}{1cm}
\begin{picture}(9,0)
\put(-0.5,1.5){\makebox(0,0){\begin{rotate}{90}%
\footnotesize $\left(\frac{dW}{dx}\right)/\left(\frac{dW}{dx}\right)%
_{\rm iso}$
\end{rotate}}}
\put(3,0.25){\makebox(0,0){\footnotesize $v$}}
\end{picture}
\vspace{-1cm}
\caption{Anisotropic energy loss for $\xi=1.0$, 
$\alpha_s=0.3$ and $\theta=0$ (full line)
compared to $\theta=\pi/2$ (dashed line) as a function of velocity 
(normalized to the isotropic result).}
\label{fig:ELang2}
\end{minipage} \hfill
\end{figure}

The full result for the heavy parton energy loss 
in an anisotropic system with strength $\xi$
is given by adding the soft contribution from Eq.(\ref{Elosssoftfinal}) 
and the hard contributions from Eq.(\ref{myelosshard3}),
\beq
-\left(\frac{dW}{dx}\right)=-\left[%
\left(\frac{dW}{dx}\right)_{\rm soft}+%
\left(\frac{dW}{dx}\right)_{\rm hard}\right].
\label{Elossgen}
\eeq
It depends on $q^*,T$ and $v$ as was the case for the isotropic energy loss,
but in general also on the angle of the particle direction with respect to
the anisotropy vector $\cos\theta={\bf \hat{v}\cdot \hat{n}}$ (sketched
in Fig.~\ref{fig:ELsquas}).
The angular dependence of the energy loss at $\xi=1$ normalized
to the isotropic energy loss Eq.(\ref{Elossiso}) 
is shown in Fig.~\ref{fig:ELang1} for $v=0.5$
and different couplings $\alpha_s=0.2,0.3,0.4$; as can be seen, for 
all couplings considered the energy loss turns out to be smaller along
${\bf \hat{n}}$ than transverse to it. However, as already noted
in the earlier discussion on the soft contribution to the energy loss,
this dependence may change as a function of the velocity $v$, as can be seen
in Fig.~\ref{fig:ELang2}, where the energy loss Eq.(\ref{Elossgen})
evaluated at $\theta=0$ and $\theta=\pi/2$ has been plotted 
for $\alpha_s=0.3$. From this figure 
it becomes clear that the energy loss is peaked
at $\theta=\pi/2$ for velocities smaller than $v_0$ while it is peaked 
at $\theta=0$ for $v>v_0$, where $v_0\simeq 0.7$ for $\xi=1$ and 
$\alpha_s=0.3$ (see also \cite{RS:2004el2}).
Note that in Figs.~\ref{fig:ELang1}, \ref{fig:ELang2} the variational bands
correspond largely to those of the isotropic results 
(about $40$\% higher).

For simplicity here it was assumed that quark and gluon distribution 
functions have momentum-space anisotropies of the same strength. In general,
this assumption is quite probably not fulfilled, so a more realistic
result for the energy loss will depend on the quark and gluon
anisotropies separately.

%

\section{Effects of finite chemical potential}

Taking finite chemical potential into account, the soft contribution
Eq.(\ref{Elosssoftfinal}) remains unchanged in form with the only
difference being the change of the Debye mass
\beq
m_D^2\rightarrow 4 \pi \alpha_s \left(\frac{6+N_f}{6}T^2+%
\frac{N_f \mu^2}{2 \pi^2}\right),
\eeq
in the structure functions according to Eq.(\ref{mD}). For the hard 
contribution, only the quark-quark scattering term changes, so 
one has to replace
\beq
f_{\xi}({\bf k})(1-f_{\xi}({\bf k^{\prime}}))\rightarrow
\frac{1}{4} (f_{+, \xi}({\bf k})+f_{-, \xi}({\bf k}))\left(1-
(f_{+, \xi}({\bf k})+f_{-, \xi}({\bf k})\right)
\eeq
in Eq.(\ref{hQq}). Again, all terms involving products of distribution
functions can be shown not to contribute to the energy loss, so
one recovers Eq.(\ref{myelosshard3}) with the function $F_1(x)$ being
replaced by
\beq
F_1(x)\rightarrow \frac{1}{2 T^2} \int_{x T}^{\infty} dq\ q 
\left(f_+(q)+f_-(q)\right),
\eeq
which can be calculated analytically similar to Eq.(\ref{F11}).
A subsequent evaluation of the collisional energy loss at finite chemical
potential in the weak coupling limit has been done in Ref.~\cite{Vija:1995is}.
There it was found that the effects played by the quark chemical potential
are rather small unless $\mu/T$ gets large. 

\section{Summary}

In this chapter I have calculated the complete leading-order 
collisional energy loss of a heavy parton propagating through an 
anisotropic quark-gluon plasma. The calculation was based on
results for the collective modes in an anisotropic plasma of chapter 
\ref{asi-chap} and it was demonstrated that the unstable modes found for such 
a system would in principle cause the result to be divergent. However,
it has been shown 
that the collisional energy loss is protected by the mechanism
of dynamical shielding, for which a proof in two analytically tractable
regimes (weak and extremely strong anisotropy) was given.
As a side result, in the isotropic limit the original result for
the energy loss was shown to have a correction which cured the problem
of unphysical energy loss results for heavy partons.

The results were applied for beauty and charm quarks and the applicability
of the calculation was discussed. Also, it was shown that for anisotropic
systems the collisional energy loss has an angular dependence which 
is expected to increase for larger couplings and stronger anisotropies,
possibly having phenomenological implications.
Finally, a refinement of the calculation involving different momentum-space
anisotropies for the quark and gluon distribution functions was suggested
and the effects of finite chemical potential were discussed.

\chapter{Conclusions and outlook}
\label{conc-chap}

In this work, I have investigated the quasiparticle description of 
certain observables in the hot and dense quark-gluon plasma. 
More precisely, the quasiparticle excitations and in general
the collective modes of systems with isotropic as well as anisotropic
momentum-space distribution functions were analyzed in the HTL approximation,
with a special emphasis on the gluonic excitations. 

For isotropic systems, an HTL quasiparticle model for the thermodynamic
pressure based on the two-loop $\Phi$-derivable approximation for the entropy
was proposed, which improves on an existing simpler model by taking into
account the full momentum dependence of the HTL self-energies. 
It was shown that these quasiparticle models can be used to accurately 
describe two-flavor lattice results for the pressure, entropy and quark-number 
susceptibilities at zero chemical potential by adopting an effective
running coupling including two fit parameters. An extension of
these quasiparticle models to finite chemical potential (naturally given by
a flow equation for the strong coupling as a consequence of the stationarity
condition for the quark and gluon propagators) was also shown to
be consistent with independent lattice studies in the region of applicability
of the latter. Therefore, these results demonstrate that 
quasiparticle models can qualitatively reproduce lattice results
for the major bulk thermodynamic
quantities above the phase transition, at least when supplied by 
a phenomenologically inspired effective strong coupling as an input.
Quantitatively the differences with respect to lattice studies were found
to be small but non-vanishing, although it is probably fair to say that
currently the variations in the latter due to different implementations
of fermions and incomplete continuum-extrapolations are comparable in 
magnitude.
Encouragingly, 
the inclusion of the plasmon effect which has devastating effects
on a strictly perturbative calculation for the pressure turns out to lead
to small corrections when incorporated through a quasiparticle model,
as has been shown here.

Therefore, quasiparticle models in general and the HTL model in particular
provide a simple, physically intuitive and rather accurate 
description of the quark-gluon plasma,
both at zero and also at non-vanishing chemical potential.
Moreover, motivated by this success at small chemical potential,
estimates of quantities at large 
chemical potential, such as the critical chemical potential
at very small temperature, 
can and have been calculated.
Although these results represent only estimates
based on the physics at small $\mu$ and are therefore likely to change
somewhat once techniques which are more accurate for the high density regime 
are developed, the  
quasiparticle model calculations give precise and robust
predictions in contrast to lattice studies which (as yet) cannot
reach this domain of high density at all or strictly perturbative results which
suffer from large renormalization scheme dependences. 

However, further refinements are possible: as a first step, 
a complete determination of the 
next-to-leading order corrections to the self-energies would eliminate
the dependence of the NLA model on the free parameter $c_{\Lambda}$,
allowing a precise inclusion of the full plasmon term in the model,
thereby greatly increasing its accuracy.
The next (ambitious) step would be to 
aim for a fully self-consistent calculation
of the one-loop self-energies, which in turn would make the 
quasiparticle model completely self-contained, since both the 
form of the gap-equations as well as the running coupling would result
from such a calculation. Unfortunately, while the first step is
fairly straightforward, the second step probably still has to await progress
concerning gauge-dependencies and the renormalization procedure of 
e.g. the two-loop $\Phi$-derivable approximation for QCD.
Nevertheless, the results of such calculations would be highly interesting
since it is possible that knowledge about such fully non-perturbative
quasiparticle-like excitations might be sufficient to offer 
something that is very close to a full description of the isotropic quark-gluon
plasma.

For anisotropic systems, however, the situation is fairly different: 
the reason for this is that 
systems with an anisotropy in momentum
space 
possess unstable modes that correspond to exponentially growing 
gauge field amplitudes. The presence of these 
unstable modes is believed to have important consequences both on the dynamical
evolution of the quark-gluon plasma as well as on more practical issues
such as the de-facto breakdown of calculations using a perturbative framework,
as has been demonstrated.
However, owing to the fact that in QCD 
the (re-)discovery of these instabilities has only occurred very recently,
many important questions remain
largely unanswered and indeed are subject to active ongoing research.
For example, there are indications that certain observables exist that
are protected from the breakdown of perturbative calculations, as has
been shown explicitly here for the collisional energy loss of a heavy 
parton. However, it is currently unknown if there are any other observables
that are either dynamically shielded or otherwise protected from the 
singularities coming from the unstable modes, so further studies in this
direction are mandatory to
have a better understanding of the effects of the QCD plasma
instabilities.
Also the saturation mechanism of the instabilities deserves
closer investigation: while for QED non-perturbative effects on
the hard particles are responsible for halting the growth of the unstable
modes, for QCD the non-Abelian interaction between the soft fields
might saturate this growth already earlier.
Moreover, a detailed understanding of the role that the instabilities play 
in modifying the bottom-up thermalization scheme is still missing and 
would probably be
very important for the early system evolution following a heavy-ion 
collision. 
Summarizing, the full implications of unstable modes in the anisotropic 
quark-gluon plasma have not yet been worked out, but current knowledge
suggests that these might have (measurable?) consequences on the physics
tested at RHIC and especially the LHC once it becomes operable, since
there the momentum-space anisotropies are probably very strong.

In conclusion, calculations based on the analysis of collective modes offer
qualitative as well as quantitative predictions relevant 
for QCD at high temperature
(and non-vanishing chemical potential) both for isotropic
and anisotropic systems. However, whereas for isotropic systems quasiparticle
models for thermodynamic quantities are well-developed and have been tested
in many occasions, anisotropic systems are less well understood and their
theoretical treatment has only been tested in the case of QED.
Moreover, current calculations (although non-perturbative in character
because of implicit resummations) are essentially limited to including only
leading-order correction terms completely, so the quantitative 
(as well as qualitative?) effects of higher
order corrections might still be substantial for QCD where the coupling
gets large.
Nevertheless -- pending the development of new methods -- quasiparticle
descriptions of hot and dense QCD allow predictions
in a range of situations where other approaches from first principles
become either ambiguous, very complicated or break down altogether, thus 
making the former an invaluable tool in the study of the new form of matter
called the quark-gluon plasma.

\appendix

\chapter{Table of symbols}
\label{appsymb-chap}

For convenience, I include here a 
short explanation of some of the symbols used in the main text
(roughly in order of appearance):

\begin{center}
\begin{tabular}[c]{c|c}
\textbf{Symbol} & \textbf{Meaning}\\
\hline
$T$ & Temperature\\
$\mu$ & Quark chemical potential\\
$T_c$ & Critical temperature at $\mu=0$\\
$p$ & Thermodynamic pressure\\
$p_0$ & Stefan-Boltzmann pressure\\
$\mu_0$ & Chemical potential at $T\simeq0$ where $p=0$\\
$\mu_c$ & Chemical potential at $T\simeq0$ where $p=p(T=T_c,\mu=0)$\\
$\bar{\mu}$ & Renormalization point in $\overline{\hbox{\footnotesize MS}}$
scheme\\
$\Lambda_{\overline{\hbox{\footnotesize MS}}}$& Renormalization
scale in $\overline{\hbox{\footnotesize MS}}$
scheme\\
$s,n,\epsilon$ & Entropy, quark number and energy density\\
$\chi, \bar{\chi}$ & Quark number susceptibilities\\
$\alpha_s$ & Strong coupling constant\\
$N_f$ & Number of quark flavors\\
$m_D$ & Debye mass\\
$T_s,\lambda,B_0$ & Quasiparticle model fit parameters\\
$\xi$ & Anisotropy parameter (see chapter \ref{asi-chap})\\
$\alpha,\beta,\gamma,\delta$ & Anisotropic self-energy structure functions\\
$q^*$ & Momentum separation scale\\
$dW/dx$ & Collisional energy loss\\
\end{tabular}\\
\end{center}

\chapter{Coefficients for the HTL coupling flow equation}
\label{app1-chap}

In order to rewrite Eq.(\ref{Maxepl}) into the form of Eq.(\ref{floweq}),
quite some algebra is necessary. For convenience, I therefore reproduce
here the final coefficients $a_{T},a_{\mu}$ and $b$ which may be directly
used to solve the flow equation.
I find
\bqa
a_{T}&=&\frac{2 N N_f}{N^2-1} \frac{{\mathcal C}}{T}, \nonumber \\
a_{\mu}&=&-\frac{{\mathcal A}}{T^2}-\frac{2 N N_f}{N^2-1} \frac{%
{\mathcal B}}{T^2},\nonumber\\
\frac{b}{4 \pi \alpha_{s,{\rm eff}}}%
&=&\frac{2 N N_f}{N^2-1} \frac{{\mathcal C}}{T} \frac{4 \pi \alpha_{s,{\rm eff%
}}}{%
M^2} \frac{N^2-1}{8 N} T%
-\frac{{\mathcal A}}{T^2} \frac{4 \pi \alpha_{s,{\rm eff}}}{m_D^2}%
\frac{N_f}{\pi^2} \mu\nonumber\\
&&-\frac{2 N N_f}{N^2-1} \frac{{\mathcal B}}{T^2} \frac{4 \pi \alpha_{s,%
{\rm eff}}}%
{M^2} \frac{N^2-1}{8 N} \frac{\mu}{\pi^2},
\eqa
where 
\bqa
{\mathcal A}&=&\int \frac{d^3 k}{(2\pi)^3} \left\{%
\left[\frac{2}{\pi} \int_0^k %
d\omega \ %
dn(\omega) (-\omega) \frac{2 (k^2-\omega^2) ({\rm Im} \Pi_T)^3}%
{\left[(k^2-\omega^2+{\rm Re}\Pi_T)^2+({\rm Im} \Pi_T)^2\right]^2}%
\right.\right.\nonumber\\
&&\left.\left.+\frac{2}{\pi} \int_0^k %
d\omega \ %
dn(\omega) (-\omega) \frac{k^2 ({\rm Im} \Pi_L)^3}%
{\left[(k^2+{\rm Re}\Pi_L)^2+({\rm Im} \Pi_L)^2\right]^2}%
\right]\right.\nonumber\\
&&\hspace*{-2mm}\left.%
-2 E_T^2 \ dn(E_T) \frac{E_T^2-k^2}{|m_D^2E_T^2-%
(E_T^2-k^2)^2|}-E_L^2\ dn(E_L) \frac{E_L^2-k^2}{|k^2-%
E_L^2+m_D^2|}
\right\}\!,\;\;\;\;\; \; \;
\eqa
\bqa
{\mathcal B}&=&\int \frac{d^3 k}{(2\pi)^3} \left\{%
\left[\frac{2}{\pi} \int_0^k %
d\omega \ %
\left[df_+(\omega) (-\omega+\mu)-df_-(\omega)(\omega+\mu)%
\right] \right.\right.\nonumber\\%
&&\left.\left.\qquad \qquad \qquad \qquad \qquad \qquad \qquad \times%
 \frac{(k-\omega) ({\rm Im} \Sigma_+)^3}{\left[(k-\omega+{\rm Re}%
\Sigma_+)^2+({\rm Im} \Sigma_+)^2\right]^2}\right.\right.%
\nonumber\\
&&\left.\left.+\frac{2}{\pi} \int_0^k %
d\omega \ %
\left[df_+(\omega) (-\omega+\mu)-df_-(\omega)(\omega+\mu)%
\right] \frac{(k+\omega) ({\rm Im} \Sigma_-)^3}{\left[(k+\omega+{\rm Re}%
\Sigma_-)^2+({\rm Im} \Sigma_-)^2\right]^2}\right]\right.\nonumber\\
&&\left. +(E_+-k)\left[df_+(E_+) (-E_++\mu)%
-df_-(E_+)(E_++\mu)\right] \frac{E_+^2-k^2}{2 M^2}%
\right.\nonumber \\
&&\left.+(E_-+k)\left[df_+(E_-) (-E_-+\mu)%
-df_-(E_-)(E_-+\mu)\right] \frac{E_-^2-k^2}{2 M^2}%
\right\},
\eqa
\bqa
{\mathcal C}&=&\int \frac{d^3 k}{(2\pi)^3} \left\{%
\left[\frac{2}{\pi} \int_0^k %
d\omega \ %
\left(-df_+(\omega)+df_-(\omega)%
\right) \frac{(k-\omega) ({\rm Im} \Sigma_+)^3}{\left[(k-\omega+{\rm Re}%
\Sigma_+)^2+({\rm Im} \Sigma_+)^2\right]^2}\right.\right.\nonumber\\
&&\left.\left.+\frac{2}{\pi} \int_0^k %
d\omega \ %
\left[-df_+(\omega)+df_-(\omega)%
\right] \frac{(k+\omega) ({\rm Im} \Sigma_-)^3}{\left[(k+\omega+{\rm Re}%
\Sigma_-)^2+({\rm Im} \Sigma_-)^2\right]^2}\right]\right.\nonumber\\
&&\left. +(E_+-k)\left[-df_+(E_+)%
+df_-(E_+)\right]\right.\nonumber \\
&&\left.+(E_-+k)\left[-df_+(E_-)%
+df_-(E_-)\right]\right\}.
\eqa

In the above equations I have used the abbreviations
\bqa
dn(\omega)&=&\frac{- \exp{\omega/T}}{\left(\exp{\omega/T}-1\right)^2}%
\nonumber\\
df_{\pm}(\omega)&=&\frac{- \exp{(\omega\pm \mu)/T}}{%
\left(\exp{(\omega\pm \mu)/T}+1\right)^2}.
\eqa
Furthermore, $E_T, E_L$ and $E_+,E_-$ are the dispersion relations 
$E_T(k),E_L(k)$ and $E_+(k),E_-(k)$ for 
the bosonic and fermion quasiparticles shown in chapter \ref{peshier-chap},
and $\Pi_T,\Pi_L,\Sigma_+,\Sigma_-$ are the bosonic and fermionic
HTL self-energies also presented in this chapter.

\chapter{Analytic expressions for structure functions}
\label{exp-chap}

In this appendix I collect the integral and analytic expressions for the 
structure functions 
$\alpha$, $\beta$, $\gamma$, and $\delta$ defined in Eq.(\ref{contractions}).
Choosing 
${\bf \hat{n}} = \hat{\bf z}$ and ${\bf k}$ to lie 
in the $x\!\!-\!\!z$ plane ($k_x/k_z = \tan\theta_n$) one has 
${\bf \hat{v}}\cdot{\bf \hat{n}}=\cos{\theta}$ 
and ${\bf v}\cdot{\bf k}=v k_{x} \cos{\phi} \sin{\theta}+v k_{z} 
\cos{\theta}$.  Using this parameterization the $\phi$ integration 
for all four structure
functions defined by the contractions in Eq.(\ref{contractions}) 
can be performed analytically.  
\begin{eqnarray}
\alpha(K,\xi)\!\!\!&=&\!\!\!%
\frac{m_{D}^2}{k^2 \tilde{n}^2} \int \frac{d (\cos{\theta})}{2}%
\frac{\omega+ \xi k_{z} \cos{\theta}}{(1+\xi (\cos{\theta})^2)^2}%
\Bigg[\omega-k_{z} \cos{\theta} \nonumber\\
&& \hspace{2cm} +k^2 (s^2-(\cos{\theta}%
-\frac{\omega k_{z}}{k^2})^2)R(\omega-k_{z} \cos{\theta} %
,k_{x} \sin{\theta})\Bigg] ,\\
\beta(K,\xi)\!\!\! &=&\!\!\!%
 -\frac{m_{D}^2 \omega^2}{k^2} \!\! \int \frac{d (\cos{\theta})}
{2} %
\frac{1}{(1\!+\!\xi(\cos{\theta})^2)^2} \nonumber\\
&&\hspace*{2cm}\times\left[1-%
(\omega+ \xi k_{z} \cos{\theta}) R(\omega\!-\!k_{z} \cos{\theta}, %
k_{x}\sin{\theta})\right] ,\\
\gamma(K,\xi)\!\!\!&=&\!\!\! m_{D}^2 \int \frac{d (\cos{\theta})}{2k^2}%
\frac{1}{(1+\xi \cos^2{\theta})^2}\left[\omega^2+\xi k^2 \cos^2{\theta}
-2\frac{k^2}{k_{x}^2}(\omega^2%
-\xi k_{z}^2 \cos^2{\theta})\right.\nonumber\\
&& \hspace{-12mm} 
\left. + \frac{(\omega+\xi k_{z}\cos{\theta} )k^4}{k_{x}^2}%
\left(2(\cos{\theta}-\frac{\omega k_{z}}{k^2})^2-s^2\right)
R(\omega-k_z \cos{\theta},k_{x}\sin{\theta}) \right], \\
\delta(K,\xi)\!\!\!&=&\!\!\!%
\frac{m_{D}^2 \omega}{k^4 \tilde{n}^2} \!\! \int \frac{%
d(\cos{\theta})}{2}%
 \frac{\omega+\xi k_{z} \cos{\theta}}{(1+\xi \cos^2{\theta})^2}%
\left(k_{z}\!+\!(k^2 \cos{\theta}\!-\!\omega k_{z})\right.\nonumber\\
&&\hspace*{5cm}\left.\times R(\omega\!-\!k_{z} \cos{\theta},k_{x}%
\sin{\theta})\right) ,
\end{eqnarray}
where $s^2=(1-\omega^2/k^2)(k_x^2/k^2)$ and
\begin{equation}
R(a,b) = \int_{0}^{2\pi} \frac{d\phi}{2\pi} \frac{1}{a
-b \cos{\phi}+i\epsilon}= \frac{1}{\sqrt{a+b+i\epsilon}\sqrt{a-b+i\epsilon}}%
 \; .
\end{equation}
When $a$ and $b$ are real-valued $R$ can be simplified to
\begin{equation}
R(a,b) = \frac{\rm{sgn}(a) \Theta(a^2-b^2)}{\sqrt{a^2-b^2}}-
\frac{i \Theta(b^2-a^2)}{\sqrt{b^2-a^2}} \; ,
\end{equation}
with $\Theta(x)$ being the usual step-function.  
Note that the remaining integration over 
$\theta$ can also be done analytically but the results are rather 
unwieldy so I do not list them here.

\section*{Static Limit}

In the limit $\omega\rightarrow 0$ it is possible to obtain 
analytic expressions 
for all four structure functions.  The results for $m_\alpha$ and 
$m_\beta$ defined
in Eq.(\ref{massdef}) are
\begin{eqnarray}
m_\alpha^2&=&-\frac{m_D^2 \sqrt{1+\xi}}{2 k_{x}^2 \sqrt{\xi}}%
\left(k_z^2 \rm{arctan}{\sqrt{\xi}}-\frac{k_{z} k^2}{\sqrt{k^2+\xi k_{x}^2}}%
\rm{arctan}\frac{\sqrt{\xi} k_{z}}{\sqrt{k^2+\xi k_{x}^2}}\right) \; , \\
m_\beta^2&=&m_{D}^2 
\frac{(\sqrt{\xi}+(1+\xi)\rm{arctan}{\sqrt{\xi}})(k^2+\xi k_x^2)}
{2  \sqrt{\xi} \sqrt{1+\xi} (k^2+ \xi k_x^2)} \nonumber \\
&&+m_D^2\frac{\xi k_z\left(%
k_z \sqrt{\xi} + \frac{k^2(1+\xi)}{\sqrt{k^2+\xi k_{x}^2}} %
\rm{arctan}\frac{\sqrt{\xi} k_{z}}{\sqrt{k^2+\xi k_{x}^2}}\right)}{%
2  \sqrt{\xi} \sqrt{1+\xi} (k^2+ \xi k_x^2)}\, ,
\end{eqnarray}
with similar results for $m_\gamma^2$ and $m_\delta^2$.

\bibliographystyle{JHEP}

\addtocontents{toc}{\protect\contentsline {chapter}{\numberline {}Bibliography}{93}}
\bibliography{books,TFTSlac,lattice,astro}
\end{document}